\documentclass[a4paper,aps,preprint,nofootinbib,floatfix,superscriptaddress,prd]{revtex4-2}
\usepackage[T1]{fontenc} % if needed
\usepackage{graphicx,color,amsmath,amsfonts,multirow}
\usepackage{subcaption}
\usepackage{amssymb}
\usepackage{amsmath}
\usepackage{hyperref}
\usepackage{bm}% bold math
\usepackage{verbatim}
\usepackage{setspace}
\usepackage{float}
\usepackage{appendix}
\usepackage{latexsym}
\usepackage{tikz}
\usepackage[compat=1.1.0]{tikz-feynman}
\usepackage{caption}
\usepackage{subcaption}
\usepackage{feynmp}
\usepackage{tikzsymbols}
\usepackage{array}% For table's line thickness change
\usepackage{pifont} % For special character using commend \ding{}
\usepackage{bbold}
\usepackage{pifont}
\usepackage{array}

\usepackage{natbib}
\usepackage{graphicx}
\def\be{\begin{equation}}
\def\ee{\end{equation}}
\newcommand{\bea}{\begin{eqnarray}}
\newcommand{\eea}{\end{eqnarray}}
\newcommand{\bers}{\begin{eqnarray*}}
\newcommand{\nn}{\nonumber}

\numberwithin{equation}{section}

\def\vev#1{\left\langle #1\right\rangle}

\begin{document} 
%\maketitle
\preprint{arXiv:2408.14942}
%\preprint{Prepared for submission to Phys.~Rev.~D}
%\preprint{~}
%\journal{Physical Reviews D}
\title{Neutrino Masses and Mixing in an Axion Model}
\author{Sin Kyu Kang }
\email{skkang@snut.ac.kr}
\affiliation{School of Natural Science, Seoul National University of Science and Technology, Seoul 01811, Korea}
\author{ Hiroshi Okada}
\email{hiroshi3okada@htu.edu.cn}
\affiliation{Department of Physics, Henan Normal University, Xinxiang 453007, China}
\date{\today}

\begin{abstract}
We propose a novel framework that simultaneously addresses three critical issues: tiny neutrino masses and their mixing patterns, dark matter, and the strong CP problem. Our model extends the Peccei-Quinn (PQ) symmetry by incorporating modular $S_3$ symmetry, which plays a central role in explaining the observed neutrino mixing structure.
The field content includes two vector-like colored fermions and three colored scalars as $S_3$ singlets, an isospin doublet inert scalar and a singlet PQ scalar, each assigned appropriate modular weights. We show that such an extension, together with a suitable assignment of modular weights to the fields, can lead to holomorphic modular forms of Yukawa interactions, which can be derived from a superpotential.
Furthermore, we explore an extension of the model to include non-holomorphic Yukawa interactions in the non-supersymmetric framework and show that the results are distinct from the holomorphic case.
Tiny neutrino masses are generated radiatively through colored mediators, while the KSVZ-type axion appears to dynamically resolve the strong CP problem. We investigate the phenomenology of lepton flavor violation and the muon $g-2$ anomaly within this framework. Additionally, we explore the axion's properties and its role as dark matter.
\end{abstract}
                              
\maketitle
\newpage

\section{Introduction}
Despite the considerable success of the Standard Model (SM), it falls short in explaining the tiny neutrino masses~\cite{Kajita:2016cak,McDonald:2016ixn} and cosmological dark matter (DM)~\cite{Bertone:2004pz,Planck:2018vyg}.
To address these shortcomings, new physics beyond the SM is necessary.
The most widely accepted mechanism for generating tiny neutrino masses is the seesaw mechanism, while the weakly interacting massive particle (WIMP) has long been considered as a strong candidate for DM.
Another enduring issue in the SM is the strong charge-parity (CP) problem in Quantum Chromodynamics (QCD), which involves the unexpectedly small value of the CP-violating phase parameter  $\bar{\theta}$.
The Peccei-Quinn (PQ) mechanism~\cite{PhysRevLett.38.1440,PhysRevD.16.1791} offers an elegant solution by introducing a new symmetry that, when spontaneously broken, results in a pseudo-Goldstone boson called the axion~\cite{PhysRevLett.40.223,PhysRevLett.40.279}. 
This mechanism effectively nullifies the CP-violating phase, providing a dynamic solution to the problem. Axion models, such as the KSVZ\cite{Kim:1979if,Shifman:1979if} and  DFSZ~\cite{Zhitnitsky:1980tq,Dine:1981rt} types, vary in whether SM quarks or exotic quarks carry the PQ charge.

While most extensions of the SM address these problems separately, there have been efforts to explore their connections.
The scotogenic scenario  simultaneously accommodates neutrino mass generation and a DM candidate, with neutrino masses arising radiatively from the exchange of DM states~\cite{Tao:1996vb,Ma:2006km,Hirsch:2013ola}.
On the other hand, axions produced non-thermally in the early Universe via the so-called misalignment mechanism~\cite{Preskill:1982cy,Abbott:1982af,Dine:1982ah}, can  serve as excellent alternatives~\cite{DiLuzio:2020wdo,Adams:2022pbo} to the WIMP DM~\cite{Arcadi:2017kky}.
There have also been investigations into embedding  realize  the DFSZ or KSVZ axion within the type-I seesaw model to address both issues coherently
\cite{Mohapatra:1982tc,Langacker:1986rj,Shin:1987xc, He:1988dm}.

Recently, a novel idea has been proposed in which  neutrino masses are generated at the quantum level through colored mediators that also resolve the strong CP problem. This new class of KSVZ-type axion models address key issues simultaneously:  tiny neutrino masses, DM and the strong CP problem \cite{Batra:2023erw}.

In this work, we extend the model to account for the observed neutrino mixing inferred from neutrino oscillation experiments by imposing $S_3$ modular symmetry on leptons and scalars \cite{deAdelhartToorop:2011re,Feruglio:2017spp,Ding:2023htn,Kobayashi:2018vbk,Kobayashi:2018wkl,Kobayashi:2019rzp}, known as the minimal non-Abelian discrete flavor symmetry \cite{Altarelli:2010gt,Ishimori:2010au, Ishimori:2012zz, Hernandez:2012ra, King:2013eh}. We introduce two vector-like fermions that transform as singlets under $S_3$, along with an isospin doublet inert scalar in the SM, both carrying nonzero modular weights. 
Neutrino masses are generated radiatively, offering an attractive scenario that  simultaneously explains the smallness of neutrino masses and provides a DM candidate. 
Additionally, we will examine several phenomenological aspects, including lepton flavor violations (LFVs), and the muon anomalous magnetic moment, and will discuss strategy for searching for axion DM.

{Modular symmetry was originally introduced in the context of supersymmetry (SUSY) where 
the holomorphicity of modular forms arises naturally \cite{Feruglio:2017spp}. More recently, inspired by modular invariant theory based
 on automorphic forms \cite{Ding:2020zxw}, a non-supersymmetric formulation of the modular flavor symmetry has been proposed in Ref. \cite{Ding:2024inn}. 
In this work, we consider both SUSY and non-SUSY frameworks to explain lepton mixing and compare their resulting predictions.
While both approaches yield nearly identical physical predictions for neutrino masses and mixing,
the corresponding parameter space required to fit experimental data differ significantly.
In SUSY framework, we suppress SUSY-related effects by assuming a high SUSY breaking scale and a small value of $\tan\beta$.
Some remarks on this treatment will be provided below.}
%
%
%In most studies, supersymmetry (SUSY) is used to protect modular symmetry and explain neutrino mixing. 
%
%
%In contrast, we propose a novel approach that achieves similar protection without SUSY by introducing a $U(1)$ PQ symmetry, an additional $Z_2$ symmetry, and appropriate modular weight assignments. This framework enables the construction of holomorphic Yukawa couplings and scalar potential terms, preserving the modular structure and suppressing undesired terms even in the absence of SUSY.
%Alternatively,  the model can be embedded within a SUSY framework to protect the holomorphic modular form. But, such an extension introduces additional contributions to neutrino masses and mixing through sperpartner effects. 
%To address this, we may consider  the decoupling limit of SUSY by assuming a large SUSY breaking scale, effectively rendering the SUSY effects negligible.
%In both approaches, the resulting physical predictions for neutrino masses and mixing are nearly the same.

This paper is organized as follows. Section II introduces an axion framework designed to address tiny neutrino masses and mixing, dark matter and the strong CP problem. We incorporate modular $S_3$ symmetry to specifically explain neutrino mixing within the model.
In Section III, we demonstrate the leptonic Yukawa structures based on the proposed framework from Section II.
We also show how tiny neutrino masses are radiatively generated through a two-loop mechanism, followed by the results of a numerical analysis.
Section IV explores the phenomenology of lepton flavor violations and the anomaly of the lepton $g-2$.
In section V, we discuss the properties of the axion and its role as dark matter within the framework.
The conclusion and remarks are provided in section VI. Additionally, reviews of {two-loop integral of the neutrino masses and the modular $S_3$ symmetry are provided in the Appendices.}

%%%%%%%%%%%%%%%%%%%%%%%%%%%%%%%%%%%%%%%%%%%%%%%%%%%%%%%%%%%%%%%%%%%%%%%%%%%%%
\section{Framework}
%%%%%%%%%%%%%%%%%%%%%%%%%%%%%%%%%%%%%%%%%%%%%%%%%%%%%%%%%%%%%%%%%%%%%%%%%%%%%
%
The model we consider is an extension of the original KSVZ model~\cite{Kim:1979if,Shifman:1979if}.
As in the KSVZ model, we introduce vectorlike fermions $\Psi_{L_i}$ and $\Psi_{R_i}$ which are triplets under the fundamental representation of SU(3)$_c$, singlets under SU(2)$_L$, and with $Y=0$. To explain the observed neutrino mixing by imposing a flavor symmetry, we introduce two generations of the vector-like fermions.
A complex scalar singlet $\sigma$ is also  introduced to break a U($1$)$_{\text{PQ}}$ symmetry spontaneously, providing masses to those exotic fermions. The phase of $\sigma$ corresponds to the axion field $a$. 
The differing PQ charges of $\Psi_{L_i,R_i}$  ensure the anomalous axion-gluon coupling necessary to resolve the strong CP problem.
Note that nonzero PQ charge $\omega$ is assigned to $\Psi_{L_i}$, while $\Psi_{R_i}$ carries no PQ charge.

As proposed in \cite{Batra:2023erw},  the vector-like fermion $\Psi_{L,R}$ fields
% in the SU($3)_c$ complex representation $(p,q)$ with $p>q = 0,1,2, \cdots$ 
can serve as mediators for neutrino mass generation at the two-loop level. 
To facilitate this process, two additional scalars, $\eta$ and $\chi$ with the same SU($3)_c$ triplet representation, are introduced. These scalars contribute to the generation of neutrino masses at the two-loop level.
While $\Psi_{L,R}$ and $\chi$ are hyperchargeless and transform as SU($2)_L$ singlets, {$\eta\equiv[\eta^+,\eta^0]^T$} has $Y=1/2$ and transforms as a SU($2)_L$ doublet. 
As mentioned, the U($1$)$_{\text{PQ}}$ symmetry is spontaneously broken by non-zero vacuum of the complex scalar singlet $\sigma$ with nonzero PQ charge $\omega$, which also generates the masses of $\Psi_{L,R}$.
Since the SU(3)$_c$ invariant term arising from $ \bm{3} \otimes \bm{3}\otimes \bm{3}$ is antisymmetric, we introduce three $\chi_i$(with $i=1,2,3)$ to allow trilinear term that is essential for generating neutrino masses at the two-loop level, and
two copies of the colored fermion $\Psi_i$.  And we note that terms involving $\overline{\Psi^c_{R_i}} \Psi_{R_i}$ and $\overline{\Psi^c_{L_i}} \Psi_{L_i}$ are not allowed.
TABLE \ref{tab:fields} lists all the new fields and their transformation properties under the SM and PQ symmetries.

To elucidate the neutrino mixing and masses observed in neutrino oscillation experiments, we enforce modular $S_3$ symmetry on the leptons and scalars. The leptons $\bar{L}_{L_2}\equiv (\overline{L_{L_\mu}},\overline{L_{L_\tau}})^T$ and $e_{R_2}\equiv (e_{R_\mu},e_{R_\tau})^T$
are organized into doublets (${\bm 2}$) under $S_3$, while the remaining leptons and scalars
%except for $\chi_i$
are singlets (${\bm 1})$ under $S_3$.
Specifically, $\Psi_{L_1}$ and $\Psi_{R_1}$ transform as $\bm{1}$, while  $\Psi_{L_2}$ and $\Psi_{R_2}$ transform as $\bm{1}^{\prime}$ under $S_3$.
The scalar fields $(\chi_1,\chi_2,\chi_3)$ are assigned to be $({\bm 1}^{\prime},{\bm 1},{\bm 1})$ under $S_3$.
%While $\chi_1$ transforms as $\bm{1}^{'}$ , $\chi=(\chi_2, \chi_3)$ transforms as $\bm{2}$ under $S_3$.
The modular weights $k$ assigned to the fields are listed  in TABLE~\ref{tab:fields}, and those assigned to the modular forms of Yukawa couplings can be found in TABLE~\ref{tab:couplings}.
We also impose $Z_2$ symmetry to prevent unwanted terms such as $(\eta^\dag H)(\chi_2\chi_3)$ from being generated in the Lagrangian and scalar potential {where $H$ is the SM Higgs boson}. 
Then, as will become evident later, neutrino masses are generated through {solely} the two-loop diagram shown in FIG. \ref{fig:1}.
%%%

{ 
%As highlighted in the introduction, 
Imposing the modular $S_3$ symmetry, supplemented by PQ symmetry and an additional 
$Z_2$ symmetry together with the assignment of modular weight, we may construct holomorphic forms for Yukawa couplings and scalar potential terms, which can be derived from superpotential when we consider SUSY framework.
This construction reflects that even non-SUSY framework is as robust as its SUSY counterpart by prohibiting undesirable terms and maintaining consistency at tree level. 
While holomorphicity of yukawa structure can be protected by SUSY, non-holomorphic terms can be arisen at loop level, which are suppressed compared to leading contributions arisen at tree level.
%Furthermore, for completeness, we may extend the model to a SUSY framework, assuming superpartner masses are assumed to be extremely heavy, which causes SUSY effects to decouple while preserving the modular symmetry structure.
Then, both approaches yield nearly identical results for neutrino masses and mixing.}

\subsection{holomorphic case}
Based on these symmetries, we can construct a renormalizable Yukawa Lagrangian for lepton sector as follows:
\begin{align}
-{\cal L}_{Lepton} &=
\alpha_\ell (Y^{(2)}_{\bf 2}\otimes\overline {L_{L_2}}\otimes e_{R_2})_{\bf1}H
+\beta_\ell (Y^{(4)}_{\bf 2}\otimes\overline{ L_{L_2}}\otimes e_{R_e})_{\bf1}H\nn\\
&+\gamma_\ell (Y^{(2)}_{\bf 2}\otimes\overline{ L_{L_e}}\otimes e_{R_2})_{\bf1}H
+\delta_\ell (Y^{(4)}_{\bf 1}\otimes\overline{ L_{L_e}}\otimes e_{R_e})_{\bf1}H\nn\\
%%%
&+\alpha_\nu (Y^{(6)}_{\bf 2}\otimes\overline{ L_{L_2}}\otimes \Psi_{R_2})_{\bf1}\tilde\eta
+\beta_\nu (Y^{(6)}_{\bf 1'}\otimes\overline{ L_{L_e}}\otimes \Psi_{R_2})_{\bf1}\tilde\eta
%+\beta_\nu (Y^{(6)}_{\bf 2}\otimes\bar L_{L_e}\otimes N_{R_2})_{\bf1}\tilde\eta
%+\gamma_\nu (Y^{(6)}_{\bf 1}\otimes\bar L_{L_2}\otimes N_{R_2})_{\bf1}\tilde\eta
\nn\\
%%%
&+\rho_\nu (Y^{(6)}_{\bf 1}\otimes\overline{ L_{L_e}}\otimes \Psi_{R_1})_{\bf1}\tilde\eta
+\sigma_\nu (Y^{(6)}_{\bf 2}\otimes\bar L_{L_2}\otimes \Psi_{R_1})_{\bf1}\tilde\eta\nn\\
%%%
&+ \xi_0 (Y^{(6)}_{\bf 1} \otimes \Psi_{R_1}\otimes \Psi_{R_2})_{\bf1^{\prime}}\chi_1
+ \xi_1 (Y^{(6)}_{\bf 1} \otimes \Psi_{R_2} \otimes \Psi_{R_1})_{\bf1^{\prime}}\chi_1 \nn \\
&+\zeta_1 (\overline{ \Psi_{L_1}}\otimes \Psi_{R_1})_{\bf1}\sigma
+\zeta_2 (\overline{ \Psi_{L_2}}\otimes \Psi_{R_2})_{\bf1}\sigma + {\rm h.c.}, \label{eq:lag-lep}
\end{align}
where $\tilde\eta\equiv i\sigma_2 \eta^*$, $\sigma_2$ being second Pauli matrix,
$\zeta_{1(2)}=c_{1(2)}/(-i \tau+i \bar{\tau})^2$~\footnote{The invariance under the modular $S_3$ symmetry is discussed in Appendix~\ref{sec:realization}.}.
For simplicity, %from now on 
we omit color and SU(2)$_L$ indices.
{This leptonic Lagrangian can be derived from a superpotential, so it can be embedded in the SUSY framework.}

% \begin{widetext}
\begin{center} 
\begin{table}[tb]%[tbc]
%\begin{tiny}
\begin{tabular}{|c||c|c|c|c|c|c|c|c||c|c|c|c|c|c||}\hline\hline  
&\multicolumn{8}{c||}{ Fermions} & \multicolumn{6}{c||}{Bosons} \\\hline
  & ~$\overline{L_{L_e}}$~& ~$\overline{ L_{L_2}}$~ & ~$e_{R_e}$~& ~$e_{R_2}$~& ~$\Psi_{L_1}$~ 
  & ~$\Psi_{L_2}$~ & ~$\Psi_{R_1}$~ 
  & ~$\Psi_{R_2}$~ &  ~$H$~  & ~$\eta^*~$ &$~\sigma$ ~& $\chi_1$ & $\chi_2 $ & $\chi_3$~
  \\\hline 
 %%%
$SU(3)_c$ &  $\bm{1}$ & $\bm{1}$   & $\bm{1}$  & $\bm{1}$ &  $\bm{3}$ &  $\bm{3}$ &  $\bm{3}$ &  $\bm{3}$ & $\bm{1}$  & $\bm{3}$ & 
 $\bm{1}$&$\bm{3}$ & $\bm{3}$  & $\bm{3}$  \\ \hline
 $SU(2)_L$ & $\bm{2}$  & $\bm{2}$  & $\bm{1}$ & $\bm{1}$   & $\bm{1}$  & $\bm{1}$& $\bm{1}$   & $\bm{1}$ & $\bm{2}$ & $\bm{2}$ &  $\bm{1}$ & $\bm{1}$ &  $\bm{1}$  &  $\bm{1}$  \\ \hline 
 %%%
$U(1)_Y$ & $\frac12$ & $\frac12$ & $-1$  & $-1$& $0$ & $0$  &$0$ & $0$  & $\frac12$  & -$\frac12$ &0&0 &0 & 0   \\\hline
 %%%
$U(1)_{\rm PQ}$ & $0$ &  $0$ & $0$ & $0$ & $w$ & $w$ & $0$ & $0$ & $0$ & $0$ & $w$ & $0$ &$0$ & $0$ \\ \hline
 $S_3$ & $\bm1$ & $\bm2$ & $\bm1$ & $\bm2$ & $\bm1$ & $\bm1'$ & $\bm1$ & $\bm1'$ &$\bm1$ & $\bm1$& $\bm1$ & $\bm1'$& $\bm1$ & $\bm1$    \\\hline
$Z_2$ & $1$ & $1$ & $1$ & $1$ & $-1$ & $-1$ & $-1$ & $-1$ &$1$ & $-1$& $1$ & $1$& $-1$ & $-1$    \\\hline
 $-k$ & $-2$ & $-2$ & $-2$ & $0$ & $-2$ & $-2$ & $-2$ & $-2$ &$0$ & $-2$ & $0$ & $-2$ &$-2$& $-2$\\\hline
% $Z_2$ & $+$ & $+$ & $+$ & $+$ & $-$ & $-$ &  $-$ & $-$ &$+$ & $-$ & $+$ & $+$  \\\hline
\end{tabular}
\caption{Field contents of fermions and bosons
and their charge assignments under $SU(3)_c\otimes SU(2)_L\otimes U(1)_Y\otimes S_{3}\otimes Z_2\otimes  U(1)_{\rm PQ}$ in the lepton and boson sector, 
where $-k$ is the number of modular weight %, $a=1,2,3$, 
and the quark sector is the same as the SM. $L_{L_2}\equiv(L_\mu, L_\tau)$, and $e_{R_2}\equiv(e_{R_\mu},e_{R_\tau}$).}
\label{tab:fields}
% \end{tiny}
\end{table}
\end{center}
%\end{widetext}

% \begin{widetext}
\begin{center} 
\begin{table}[tb]%[tbc]
%\begin{tiny}
\begin{tabular}{|c||c|c|c|c|c|c|c||}\hline\hline  
 &\multicolumn{7}{c||}{Couplings}  \\\hline
  & ~$Y^{(4)}_{\bf1}$~& ~$Y^{(6)}_{\bf1}$~ & ~$Y^{(6)}_{\bf1'}$~ & ~$Y^{(8)}_{\bf1}$~ & ~$Y^{(2)}_{\bf2}$~& ~$Y^{(4)}_{\bf2}$~& ~$Y^{(6)}_{\bf2}$~  \\\hline 
 $S_3$ & ${\bf1}$ & ${\bf1}$& ${\bf1^{\prime}}$ & ${\bf1}$ &${\bf2}$ & ${\bf2}$ & ${\bf2}$    \\\hline
 $-k$ & $4$ & $6$ & $6$ &  $8$ & $2$ & $4$& $6$   \\\hline
\end{tabular}
\caption{Modular weight assignments for Yukawa interaction.}
\label{tab:couplings}
% \end{tiny}
\end{table}
\end{center}
%\end{widetext}
%%%%%%%%%%%%%%%%%%%%%%%%%%%%%%%%%%%%%%%%%%%%%%%%%%%%%%%%%%%%%%%%%%%%%%%%%
{
\subsection{Non-holomorphic case}
The non-holomorphic modular flavor symmetry extends the original holomorphic modular invariance by incorporating polyharmonic Maaß forms~\cite{Ding:2024inn, Qu:2024rns}. While holomorphic modular forms are restricted to carry non-negative integer weights, non-holomorphic modular forms may possess negative integer weights, thereby broadening the scope of modular flavor model building.
% offering greater flexibility in constructing modular-invariant structures.
In this work, we extend the framework by introducing non-holomorphic modular form known as the polyharmonic Maaß form of 
$SL(2,Z)$.
%as an extension to the lepton-sector Lagrangian given in Eq.~\ref{eq:lag-lep}. 
The minimal extension involves the inclusion of the polyharmonic Maaß form of $Y^{(2)}_{{\bf 1}}$ which serves as a modular Yukawa coupling that transforms as a singlet under $S_3$ and carries modular weight 2. The explicit form of $Y^{(2)}_{{\bf 1}}$ will be given in Sec.\ref{yukawa}.
The resulting non-holomorphic Yukawa term added to the Lagrangian Eq.(\ref{eq:lag-lep}) takes the following form,
\begin{align}
&\sigma'_\ell  Y^{(2)}_{\bf 1}\otimes (\overline {L_{L_2}}\otimes e_{R_2})_{\bf1}H + {\rm h.c.}
\equiv \sigma_\ell (L_{L_\mu} e_{R_\mu} + L_{L_\tau} e_{R_\tau}) H  + {\rm h.c.}, \label{eq:lag-lep_nonholo}
\end{align}
where $\sigma_\ell$ is an effective coupling constant arising from the modular form.
}

%%%%%%%%%%%%%%%%%%%%%%%%%%%%%%%%%%%%%%%%%%%%%%%%%%%%%%%%%%%%%%%%%%%%%%%%%%%%%

The scalar potential, invariant under $SU(3)_c\otimes SU(2)_L\otimes U(1)_Y\otimes S_{3}\otimes Z_2\otimes  U(1)_{\rm PQ}$, is given by
\begin{align}
{\cal V} = & -\mu_H^2 |H|^2 +
%|Y^{(4)}_{\bf1}|
\mu^2_\eta |\eta|^2+\mu^2_{\chi_i} |\chi_i|^2+\mu^2_\sigma |\sigma|^2 \nn \\
&+Y^{(6)}_{\bf1^{\prime}}\mu_{3} \chi_1 \chi_2 \chi_3 
+Y^{(4)}_{{\bm 1}} \kappa_{k} \eta^\dagger H \chi_k
%+ \lambda H^\dagger \eta \chi \chi 
+ \text{h.c.} \nn \\
&+ \frac14 \lambda_H|H|^4
 +\frac14 \lambda_\sigma |\sigma|^4 
+|Y^{(8)}_{\bf1}|\left( \frac14\lambda_\eta |\eta|^4
+\frac14 \lambda_{\chi_i} |\chi_i|^4  
%\right. \nn \\ &\left.
+\lambda_{\eta\chi_i} |\eta|^2|\chi_i|^2\right)\nn\\
&+|Y^{(4)}_{\bf1}|\lambda_{\eta\sigma} |\eta|^2|\sigma|^2+\lambda_{H\sigma} |H|^2|\sigma|^2
\nn \\
&+|Y^{(4)}_{\bf1}|(\lambda_{\chi_i\sigma} |\chi_i|^2|\sigma|^2
+\lambda_{H\chi_i}|H|^2|\chi_i|^2) \nn \\
&+|Y^{(4)}_{\bf1}|(\lambda_{H\eta}|H|^2|\eta|^2+\lambda_{H\eta}'  |H^\dag\eta|^2),
%+\frac14\lambda_{H\eta}'' [Y^{(4)}_{\bf1}(H^\dag\eta)^2+ {\rm h.c.}].
 \label{eq:pot}
\end{align}
where $i=1,2,3$, $k=2,3$, $\mu^2_\eta\equiv\tilde\mu^2_\eta/(-i\tau+i\bar\tau)^2$, and $\mu^2_{\chi_i}\equiv\tilde\mu^2_{\chi_i}/(-i\tau+i\bar\tau)^2$.
To preserve the SU($3)_c$ symmetry the colored scalars $\eta$ and $\chi_{1,2,3}$ must not acquire a vacuum expectation value {(VEV)}, so that the only {VEVs} are $\vev{\sigma} = v_\sigma/\sqrt{2}$ breaking U($1$)$_{\text{PQ}}$, and $\vev{H} =[0, v_H/\sqrt{2}]^T\ (v_H/\sqrt{2}\simeq 174$~GeV) triggering electroweak (EW) symmetry breaking. 
After EW symmetry breaking, mixing masses between $\eta$ and $\chi_i$ are generated, which result in a mixing between $\eta^{0}$ and $\chi_i(=2,3)$. The squared-mass matrix is given by
\begin{align}
M_{\eta\chi}^2=
\left[\begin{array}{ccc}
\mu^2_{\eta} &   Y^{(4)}_{\bf1}\kappa_{2} v^{\prime}
& Y^{(4)}_{\bf1}\kappa_{3}  v^{\prime} \\ 
Y^{(4)}_{\bf1}\kappa_2 v^{\prime}  & \mu^2_{\chi_2} & 0 \\ 
Y^{(4)}_{\bf1}\kappa_3 v^{\prime} & 0 &  \mu^2_{\chi_3} 
\end{array}\right],
\end{align}
where $v^{\prime}=v_H/\sqrt{2}$, and we have ignored the contributions from $v_{\sigma}$ by assuming the coupling constants associated with $\sigma$ are very small.
Then,  the scalar fields $(\eta^0, \chi_i)$ are expressed in terms of the mass eigenstates denoted by $S_i$ as follows,
\begin{eqnarray}
\left(
\begin{array}{c}
\eta^{0}\\
\chi_2\\
\chi_3
\end{array}
 \right)
=U\cdot 
\left(
\begin{array}{c}
S_1\\
S_2 \\
S_3
\end{array}
 \right),\label{eq:umat}
\end{eqnarray}
where $U$ is a $3\times 3$ unitary matrix that diagonalizes the matrix $M_{\eta\chi}^2$.
The masses of $S_i$ are denoted as $m_{S_i}$ and the mass of $\chi_1$ is denoted as $m_{\chi_1}$.

{It is worthwhile to notice that the Yukawa structures presented above may receive potential corrections.
As discussed in \cite{Criado:2018thu}, the primary sources of such corrections include higher dimensional operators, renormalization group (RG) evolution and SUSY breaking effects.
The higher dimensional operators, though in principle allowed, are strongly constrained by the imposed symmetries and can be 
safely suppressed by assuming a large cutoff scale (i.e., the scale of new physics). 
As a result, their impacts on the predicted flavor structures can be negligible.
RG corrections arising from the running of parameters from the high energy scale down to the electroweak scale can modify the predictions, but their effects become negligibly small in SUSY case if we assume a low $\tan\beta$ and a sufficiently high SUSY breaking scale. 
Even in non-SUSY case, RG effects can be kept under control by assuming a strongly hierarchical neutrino mass spectrum, masses of TeV scale mediators 
and no large Yukawa couplings in the lepton sectors.
The SUSY breaking effects can also be ignorable under the same conditions of low $\tan\beta$ and high SUSY breaking scale, as discussed in \cite{Criado:2018thu}.
Therefore, in this work, we will neglect these corrections and focus on the leading-order predictions based on the modular symmetry framework.}

%%%%%%%%%%%%%%%%%%%%%%%%%%%%%%%%%%%%%%%%%%%%%%%%%%%%%%%%%%%%%%%%%%%%%%%%%%%%
\section{Leptonic Yukawa Structures and Radiative Neutrino mass generation}\label{yukawa}
%%%%%%%%%%%%%%%%%%%%%%%%%%%%%%%%%%%%%%%%%%%%%%%%%%%%%%%%%%%%%%%%%%%%%%%%%%%%%
Now, let us demonstrate how the lepton mixing matrix can be predicted by imposing a modular $S_3$ symmetry in the lepton sector.
The  modular forms with the lowest weight 2, $Y^{(2)}_{\bf2}\equiv (y_1,y_2)^{T}$, transforming
as a doublet of $S_3$ is written in terms of Dedekind eta-function  $\eta(\tau)$ and its derivative \cite{Novichkov:2019sqv} as follows:
%%%%%%%%%a %%%%%%%%%%%%%%
\begin{eqnarray} 
\label{eq:Y-S3}
y_1(\tau) &=& \frac{i}{4\pi}\left( \frac{\eta'(\tau/2)}{\eta(\tau/2)}  +\frac{\eta'((\tau +1)/2)}{\eta((\tau+1)/2)}  
- \frac{8\eta'(2\tau)}{\eta(2\tau)}  \right), \nonumber \\
y_2(\tau) &=& \frac{\sqrt3 i}{4\pi}\left( \frac{\eta'(\tau/2)}{\eta(\tau/2)}  -\frac{\eta'((\tau +1)/2)}{\eta((\tau+1)/2)}  
 \right) \label{Yi}.
\end{eqnarray}
%%%%%%%%%%%%%%%%%%%%%
%
Then, any couplings of higher weight are constructed by multiplication rules of $S_3$ {as presented in Appendix~\ref{s3_rules}},
and we find the following couplings~\cite{Novichkov:2019sqv}:
%%%%%%%%%%%%%%%%%%%%%%%%%%
\begin{align}
&Y^{(4)}_{\bf1}=y^2_1+y^2_2,\quad
Y^{(6)}_{\bf1}=3y^2_1y_2-y^3_2,\quad
Y^{(6)}_{\bf1'}=y_1^3-3y_1y_2^2,\nn\\
&Y^{(4)}_{\bf2}
\equiv
\left[\begin{array}{c}
Y^{(4)}_{{\bf2},1} \\ 
Y^{(4)}_{{\bf2},2}  \\ 
\end{array}\right]
=
\left[\begin{array}{c}
2y_1y_2 \\ 
y^2_1-y_2^2  \\ 
\end{array}\right],\quad
Y^{(6)}_{\bf2}
\equiv
\left[\begin{array}{c}
Y^{(6)}_{{\bf2},1} \\ 
Y^{(6)}_{{\bf2},2}  \\ 
\end{array}\right]
=
\left[\begin{array}{c}
y^3_1+y_1y_2^2 \\ 
y^3_2+y_1^2 y_2 \\ 
\end{array}\right]. 
\end{align}
%%%%%%%%%%%%%%%%%%%
{Note that the modular forms given above are holomorphic. 

In contrast, the polyharmonic Maaß form $Y^{(2)}_{{\bf 1}}$ is constructed from the modified Eisenstein series $\hat{E}_2(\tau)$~\cite{Qu:2024rns}, 
which is a well-known example of a non-holomorphic modular form. 
It is explicitly defined as
\begin{eqnarray}
Y^{(2)}_{{\bf 1}}(\tau)=\hat{E}_2(\tau)=1-\frac{3}{\pi y}-24 \sum^{\infty}_{n=1}\sigma_1(n) q^n,
\end{eqnarray}
where $y={\rm Im}\tau$, $q=e^{2\pi i \tau}$ and $\sigma_1(n)=\sum_{d|n}d$ denotes the sum of the positive divisors of the integer $n$.
The non-holomorpicity of $\hat{E}_2(\tau)$ arises explicitly from its dependence on $y$.
Its corresponding q-expansion is given by
\begin{eqnarray}
\hat{E}_2(\tau)=1-\frac{3}{\pi y}-24 q - 72 q^2-96 q^3-168 q^4-144 q^5-\cdot\cdot\cdot.
\end{eqnarray}
}

The structure of Yukawa couplings is determined by the modular symmetry. 
After the {EW} spontaneous symmetry breaking,  the charged lepton mass matrix is given by
\begin{align}
{\rm Holomorphic\ case}:
m_\ell&= \frac {v_H}{\sqrt{2}}
\left[\begin{array}{ccc}
\delta_\ell Y_{\rm1}^{(4)} & \gamma_\ell y_1 & \gamma_\ell y_2 \\ 
\beta_\ell Y^{(4)}_{{\bf2},1} & \alpha_\ell y_2 &  \alpha_\ell y_1 \\ 
\beta_\ell Y^{(4)}_{{\bf2},2} & \alpha_\ell y_1 & -\alpha_\ell y_2 \\ 
\end{array}\right],\\
%%%
{\rm Non-holomorphic\ case}:
m_\ell&= \frac {v_H}{\sqrt{2}}
\left[\begin{array}{ccc}
\delta_\ell Y_{\rm1}^{(4)} & \gamma_\ell y_1 & \gamma_\ell y_2 \\ 
\beta_\ell Y^{(4)}_{{\bf2},1} & \alpha_\ell y_2 +\sigma_\ell &  \alpha_\ell y_1 \\ 
\beta_\ell Y^{(4)}_{{\bf2},2} & \alpha_\ell y_1 & -\alpha_\ell y_2 +\sigma_\ell \\ 
\end{array}\right].
\end{align}
%where {$\langle H\rangle\equiv [0, v_H/\sqrt2]^T$}.
Then the charged lepton mass eigenstates and corresponding mass-squared eigenvalues can be obtained by diagonalizing
the hermitian matrix $m_\ell m^{\dagger}_\ell$ as
{
$%|D_\el|^2\equiv 
V_{e_L}^\dag m_\ell m^\dag_\ell V_{e_L}$
$={\rm diag.}(|m_e|^2,|m_\mu|^2,|m_\tau|^2)$, where $V_{e_L}$ denotes the unitary matrix transforming left-handed charged leptons into
the mass eigenstates.
%In our numerical analysis presented  below, the free parameters $\alpha_\ell,\beta_\ell,\gamma_\ell$ are dteremined so as to reporduce
%the observed charged lepton masses.
%$\{\alpha_\ell, \beta_\ell, \gamma_\ell\}$ are determined in order to fit the three observed charged-lepton masses by the following relations:
%\begin{align}
%&{\rm Tr}[m_\ell m_\ell^\dag] = |m_e|^2 + |m_\mu|^2 + |m_\tau|^2,\quad
% {\rm Det}[m_\ell m_\ell^\dag] = |m_e|^2  |m_\mu|^2  |m_\tau|^2,\nn\\
%&({\rm Tr}[m_\ell m_\ell^\dag)^2 -{\rm Tr}[(m_\ell m_\ell^\dag)^2] =2( |m_e|^2  |m_\mu|^2 + |m_\mu|^2  |m_\tau|^2+ |m_e|^2  |%m_\tau|^2 ).\label{eq:l-cond},
%\end{align}
%where $\delta_\ell$ is an input parameter that is free.
%%%%
}

%%%%%%%%%%%%
The Dirac Yukawa matrix is given by
\begin{align}
y_D &=%\frac1{2}
\left[\begin{array}{cc}
\rho_\nu  Y^{(6)}_{\bf1} & \beta_\nu  Y^{(6)}_{\bf1'}
%\beta_\nu   Y^{(6)}_{\bf2,1} &% \beta_\nu   Y^{(6)}_{\bf2,2} 
 \\ 
\sigma_\nu Y^{(6)}_{{\bf2},1} 
 &%\gamma_\nu  Y^{(6)}_{\bf1}+
- \alpha_\nu   Y^{(6)}_{{\bf2},2}     
% &\alpha_\nu   Y^{(6)}_{\bf2,1}    
\\ 
\sigma_\nu Y^{(6)}_{{\bf2},2}  & \alpha_\nu   Y^{(6)}_{{\bf2},1}  
% & %\gamma_\nu  Y^{(6)}_{\bf1} -\alpha_\nu   Y^{(6)}_{\bf2,2}   
  \\ 
\end{array}\right].
\label{eq:yD}
\end{align}
%where $Y^{(6)}_{\bf2}\equiv[Y^{(6)}_{{\bf2},1},Y^{(6)}_{{\bf2},2}]^T$.
Since the scalar $\eta$ can not have a {nonzero VEV},
%nontrivial vacuum, 
the light neutrino masses can not be generated at tree level.

The mass matrix for the heavy colored fermions $(\Psi_{L_1},\Psi_{L_2},\Psi_{R_1},\Psi_{R_2})$ is given by
\begin{align}
M_{\Psi}=
\frac{v_{\sigma}}{\sqrt2}
\left[\begin{array}{cccc}
0 & 0 & \zeta_1   & 0 \\
0 & 0 & 0 & \zeta_2  \\
\zeta_1   & 0 & 0 & 0 \\
0 & \zeta_2    & 0& 0  \\
\end{array}\right].\label{eq:mn}
\end{align}
The mass eigenvalues of $\Psi_{1,2}$ are found by $M_{1,2}\equiv \frac{v_\sigma}{\sqrt2}\zeta_{1,2}$.

%
%%%%%%%%%%%%%%%%%%%%%
\begin{figure}[ht]
\centering
%\begin{subfigure}[b]{0.30 \textwidth}
%    \centering
    \begin{tikzpicture}
\begin{feynman}
\vertex (cc);
    \vertex [left =1.5cm of cc] (a);
    \vertex [above =1.5cm of cc] (c);
    \vertex [right=1.5cm of cc] (b);
    \vertex [above left =1.08cm and 1.08cm of cc] (d);
    \vertex [above right=1.08cm and 1.08cm of cc] (e);
    \vertex [above left=1.5cm of d] (i1) {$H$};
    \vertex [left=1.5cm of a] (i2) {$\bar{\nu}$};
    \vertex [above right=1.5cm of e] (f1) {$H$};
    \vertex [right=1.5cm of b] (f2) {$\nu$};
    \diagram* {
        (a) -- [ anti charged scalar,  bend left, edge label=$\eta$] (d) --[ charged scalar, bend left, edge label=$\chi$](c),
    (c)--[anti charged  scalar, bend left, edge label=$\chi$](e) --[ charged scalar, bend left, edge label=$\eta$](b),
% -- [scalar, bend left, edge label=$\chi$] (c),
        (a) -- [ anti fermion, edge label=$\Psi_R$] (cc) -- [ fermion, edge label=$\Psi_R$] (b),
 %       (c) -- [  scalar,  bend left, edge label=$\chi$] (e) -- [ scalar, bend left, edge label=$\Phi$] (b),
%(a) -- [ scalar, quarter right,  edge label'=$\Phi_1^*$] (d) -- [  scalar,quarter right, edge label'=$\Phi_2$] (b),
 (cc)--[charged scalar , edge label'=$\chi$](c),
        (a) -- [anti fermion] (i2),
        (b) -- [anti fermion] (f2),
        (d) -- [anti charged scalar] (i1),
        (e) -- [anti charged scalar] (f1),
    };
% (c) -- [anti charged scalar, edge label'=$\Phi_2^{*}$] (f1),
%a -- [charged scalar, quarter left, edge label=$\eta$] (t1) -- [anti charged scalar, quarter left, edge label=$\eta$] (c),
\end{feynman}
\end{tikzpicture}
   \caption{Feynman diagram for two-loop neutrino mass.}
   \label{fig:1}
%\end{subfigure}
\end{figure}
%%%%%%%%%%

%%%%%%%%%%%%%%%%%%%%%%%%%
The Yukawa and scalar interactions described in Eqs.\eqref{eq:lag-lep} and \eqref{eq:pot} can generate Majorana neutrino masses at the two-loop level, as illustrated by the diagram in FIG.~\ref{fig:1}. 
A distinctive aspect of neutrino mass generation is that the masses of the neutrinos are mediated by the colored particles $\Psi$, $\eta$, and $\chi$, all of which transform under the same SU($3)_c$ representation.
While $\eta$ is a SU(2$)_L$ doublet, $\Psi$ and $\chi$ are SU(2$)_L$ singlets. 
The coupling between $L$ and $\Psi$ necessitates that  $\eta$ has a hypercharge $Y=1/2$. In our scenario $\Psi$ and $\chi$ carry no hypercharge, which ensures the Majorana nature of light neutrinos. 

The Lagrangian terms contributing to neutrino masses arisen at two loop are written as
\begin{eqnarray}
{\cal L}&\supset &[\alpha_{\nu}(Y^{(6)}_{\bm{2},1}\bar{\nu}_{\tau}-Y^{(6)}_{\bm{2},2}\bar{\nu}_{\mu})+\beta_{\nu}Y^{(6)}_{\bm{1}^{\prime}}\bar{\nu}_e]\Psi_{R_2}\tilde{\eta}+[\rho_{\nu}Y^{(6)}_{\bm{1}}\bar{\nu}_e+\sigma_{\nu}
(Y^{(6)}_{\bm{2},1}\bar{\nu}_{\mu}+Y^{(6)}_{\bm{2},2}\bar{\nu}_{\tau})]\Psi_{R_1}\tilde{\eta} \nonumber  \\
&+&[\xi_0Y^{(6)}_{\bm{1}}\overline{\Psi_{R_1}^C} \Psi_{R_2}+\xi_1Y^{(6)}_{\bm{1}}\overline{\Psi_{R_2}^C} \Psi_{R_1}]\chi_1
+Y^{(6)}_{\bm{1}}\mu_3 \chi_1\chi_2\chi_3+Y^{(4)}_{\bm{1}}(\kappa_2 \chi_2+\kappa_3\chi_3)\eta^{\dagger}H. \nn \\
\end{eqnarray}
Then, the combinations of the couplings for neutrino mass generation are denoted as follows;
\begin{eqnarray}
{\cal Y}^{1}&=&{\cal C}^1\cdot \left(
\begin{array}{ccc}
(\beta_{\nu}Y^{(6)}_{\bm{1}^{\prime}}\cdot  \rho^{\ast}_{\nu}Y^{(6)^{\ast}}_{\bm{1}}) & (\beta_{\nu}Y^{(6)}_{\bm{1}^{\prime}}\cdot \sigma^{\ast}_{\nu}Y^{(6)^{\ast}}_{\bm{2},1}) &(\beta_{\nu}Y^{(6)}_{\bm{1}^{\prime}} \cdot \sigma^{\ast}_{\nu}Y^{(6)^{\ast}}_{\bm{2},2}) \\
(\beta_{\nu}Y^{(6)}_{\bm{1}^{\prime}}\cdot \sigma^{\ast}_{\nu}Y^{(6)^{\ast}}_{\bm{2},1} )
& -(\alpha_{\nu}Y^{(6)}_{\bm{2},2} \cdot \sigma^{\ast}_{\nu}Y^{(6)^{\ast}}_{\bm{2},1}) & -(\alpha_{\nu}Y^{(6)}_{\bm{2},2} \cdot \sigma^{\ast}_{\nu}Y^{(6)^{\ast}}_{\bm{2},2}) \\
(\beta_{\nu}Y^{(6)}_{\bm{1}^{\prime}} \cdot \sigma^{\ast}_{\nu}Y^{(6)^{\ast}}_{\bm{2},2}) & -(\alpha_{\nu}Y^{(6)}_{\bm{2},2} \cdot \sigma^{\ast}_{\nu}Y^{(6)^{\ast}}_{\bm{2},2}) & (\alpha_{\nu}Y^{(6)}_{\bm{2},1} \cdot \sigma^{\ast}_{\nu}Y^{(6)^{\ast}}_{\bm{2},2})
\end{array}
\right), ~~~\nn \\
\\
{\cal Y}^{2}&=&{\cal C}^2\cdot \left(
\begin{array}{ccc}
(\rho_{\nu}Y^{(6)}_{\bm{1}}\cdot \beta^{\ast}_{\nu}Y^{(6)^{\ast}}_{\bm{1}^{\prime}}) & -(\rho_{\nu}Y^{(6)}_{\bm{1}} \cdot \alpha^{\ast}_{\nu}Y^{(6)^{\ast}}_{\bm{2},2}) & (\rho_{\nu}Y^{(6)}_{\bm{1}} \cdot \alpha^{\ast}_{\nu}Y^{(6)^{\ast}}_{\bm{2},1})
\\
-(\rho_{\nu}Y^{(6)}_{\bm{1}} \cdot \alpha^{\ast}_{\nu}Y^{(6)^{\ast}}_{\bm{2},2}) &
-(\sigma_{\nu}Y^{(6)}_{\bm{2},1} \cdot \alpha^{\ast}_{\nu}Y^{(6)^{\ast}}_{\bm{2},2})
& (\sigma_{\nu}Y^{(6)}_{\bm{2},1} \cdot \alpha^{\ast}_{\nu}Y^{(6)^{\ast}}_{\bm{2},1}) \\
(\rho_{\nu}Y^{(6)}_{\bm{1}} \cdot \alpha^{\ast}_{\nu}Y^{(6)^{\ast}}_{\bm{2},1})
 & (\sigma_{\nu}Y^{(6)}_{\bm{2},1} \cdot \alpha^{\ast}_{\nu}Y^{(6)^{\ast}}_{\bm{2},1}) & (\sigma_{\nu}Y^{(6)}_{\bm{2},2}\cdot \alpha^{\ast}_{\nu}Y^{(6)^{\ast}}_{\bm{2},1})
\end{array}
\right),
\nn
\end{eqnarray}
where ${\cal C}^1\equiv \mu_3 \kappa_2 \kappa_3 \xi_1 ~(Y^{(6)}_{\bm{1}} \cdot Y^{(4)}_{\bm{1}})^2$ and ${\cal C}^2\equiv \mu_3 \kappa_2 \kappa_3 \xi_0 ~(Y^{(6)}_{\bm{1}} \cdot Y^{(4)}_{\bm{1}})^2$.
The light neutrino masses derived from the two-loop diagram are given by
\begin{eqnarray}
(m_{\nu})_{\alpha \beta}=\frac{N_c}{4(4\pi)^2} [{\cal Y}^{1}  L_{12}+{\cal Y}^{2} L_{21}],
%(m_{\nu})_{\alpha \beta}=\frac{N_c}{(16\pi^2)^2} Y^{j}_{a\alpha} (Y_{\chi})^k_{ab} Y^{l}_{b\beta} \mu_{jkl} L^{jkl}_{ab},
\end{eqnarray}
where 
$N_c(=3)$ is the color factor. The loop functions $L_{lm}$ are \cite{Aoki:2014cja}~\footnote{{A brief derivation of the two-loop integral is presented in Appendix~~\ref{neut-int}.}}
\begin{eqnarray}
L_{lm}&=& \int^1_0 dx dy dz \frac{\delta(x+y+z-1)}{(1-y)}
 \left[\sum_{i=1}^{3}  U_{1i}^2 U_{2i}U_{3i}
{\cal I}\left(
\frac{m^2_{S_i}}{M_m^2},\frac{m^2_{S_i \Psi_l}}{M^2_m}
\right) \right. \nn \\
&& \left. 
-\sum_{i,j=1(i\neq j )}^3 U_{1i}U_{2i} U_{1j}U_{3j}{\cal I}\left(\frac{m^2_{S_i}}{M_m^2},\frac{m^2_{S_j \Psi_l}}{M^2_m}\right)
\right],\label{eq:neut-int1}
\end{eqnarray}
with
\begin{eqnarray}
{\cal I}(a,b)=\frac{a^2 \ln a}{(1-a)(a-b)}-\frac{b^2 \ln b}{(1-b)(b-a)}, ~~{\rm and}~~
m^2_{S_i\Psi_j}=\frac{ x m^2_{S_i}+y M^2_{j}+z m^2_{\chi_1}}{y(1-y)}. ~~~\label{eq:neut-int2}
% m^2_{H_j S}=\frac{ u m^2_H+v M^2_J+w m^2_S}{w(1-w)}
\end{eqnarray}
%where $M_k$ denotes the mass of $\Psi_{R_k}$.

The neutrino mass matrix $m_{\nu}$ is diagonalized by a unitary matrix $U_{\nu}$ as $U_{\nu}^\dag m_{\nu} U^{*}_{\nu}={\rm diag}[D_{\nu_1} ,D_{\nu_2},D_{\nu_3}]$. %where the lightest neutrino mass is zero. 
Note here that, with only two copies of $\Psi (n_{\Psi} = 2)$, one of the three light neutrinos is predicted to be massless.
%%%
Therefore, there is one Majorana phase and we define diag[$1,e^{\alpha_{21}/2},1$] with $\alpha_{21}\neq0$.
%%%
% two-one component to be nonzero $\alpha_{21}\neq0$.
%%%
Because of vanishing the lightest neutrino mass, the other two neutrino mass eigenvalues are given in terms of two neutrino mass squared differences that are experimentally observed. 
\begin{align}
& ({\rm NH}):\ D_{\nu_2}\approx \sqrt{\Delta m^2_{\rm sol}},\quad D_{\nu_3}\approx \sqrt{|\Delta m^2_{\rm atm}|}, \\
& ({\rm IH}):\ D_{\nu_1}\approx \sqrt{|\Delta m^2_{\rm atm}-\Delta m^2_{\rm sol}|},\quad D_{\nu_2}\approx \sqrt{|\Delta m^2_{\rm atm}|},
\end{align}
where  NH and IH are respectively normal hierarchy and inverted hierarchy of neutrino masses. 
Since $\Delta m^2_{\rm sol}/\Delta m^2_{\rm atm} <<1$ from the experiments, the sum of neutrino masses $\sum D_\nu$ is approximately given by
\begin{align}
& ({\rm NH}):\ \sum D_\nu\approx \sqrt{|\Delta m^2_{\rm atm}|}\simeq 50\ {\rm meV}, \\
& ({\rm IH}):\  \sum D_\nu \approx 2  \sqrt{|\Delta m^2_{\rm atm}|}\simeq 100\ {\rm meV}.
\end{align}
Thus, the upper bound from the minimal cosmological model
$\Lambda$CDM $+\sum D_{\nu}$; $\sum D_{\nu}\le$ 120 meV~\cite{Vagnozzi:2017ovm, Planck:2018vyg},
is already satisfied.
But recent combined data of DESI and CMB gives more stringent upper bound; $\sum D_{\nu}\le$ 72 meV~\cite{DESI:2024mwx}.
{
If this experimental result is confirmed,  IH would be disfavored.}
Then, the PMNS mixing matrix is given by
$U_{\rm PMNS}=V^{\dagger}_{eL}U_{\nu}$. From the standard parameterization of $U_{\rm PMNS}$, 
the mixing angles are presented in terms of the entries of $U_{\rm PMNS}$ as follows:
\begin{eqnarray}
s^2_{12}=\frac{|(U_{\rm PMNS})_{12}|^2}{1-|(U_{\rm PMNS})_{13}|^2},\quad  s^2_{13}=|(U_{\rm PMNS})_{13}|^2,\quad
s^2_{23}=\frac{|(U_{\rm PMNS})_{23}|^2}{1-|(U_{\rm PMNS})_{13}|^2},
\end{eqnarray}
where we have used short-hand notations $s^2_{12}(c^2_{12}),\ s^2_{13}(c^2_{13}),\ s^2_{23}(c^2_{23})$ for $\sin^2\theta_{12}(\cos^2\theta_{12})$, $\sin^2\theta_{13}(\cos^2\theta_{13})$, $\sin^2\theta_{23}(\cos^2\theta_{23})$, respectively.
As a constraint, we take into account the neutrinoless double beta decay whose amplitude is proportional to
the effective neutrino mass given by
\begin{align}
& ({\rm NH}):
\langle m_{ee}\rangle =|
%m_{\nu_1}\cos^2\theta_{12}\cos^2\theta_{13}+
D_{\nu_2}s^2_{12} c^2_{13}e^{i\alpha_{21}}+
D_{\nu_3}s^2_{13} e^{-2i\delta_{\rm CP}}|,\\
% e^{i(\alpha_{31}-2\delta_{\rm CP})}|
& ({\rm IH}):
\langle m_{ee}\rangle =
%m_{\nu_1}\cos^2\theta_{12}\cos^2\theta_{13}+
|D_{\nu_1} c^2_{12} c^2_{13}+D_{\nu_2} s^2_{12} c^2_{13}e^{i\alpha_{21}} |.
\end{align}
The most stringent bound on $\langle m_{ee}\rangle$ comes from KamLAND-Zen experiments whose
upper bound is found as $\langle m_{ee}\rangle<(36-156)$ meV at 90 \% confidence level \cite{KamLAND-Zen:2016pfg}. 
To constrain the model parameters,  we will use the bound on $\langle m_{ee}\rangle$ from KamLAND-Zen experiment when our $\langle m_{ee}\rangle$ is nearby this bound.

%%%%%%%%%%%%%%%%%%%%%%%%%%%%%%%%%%%%%%%%%%%%%%%%%%%%%%%%%%%%%%
\subsection{Numerical results}
%%%%%%%%%%%%%%%%%%%%%%%%%%%%%%%%%%%%%%%%%%%%%%%%%%%%%%%%%%%%%%
%
\begin{figure}
\centering
\includegraphics[width=0.60\textwidth]{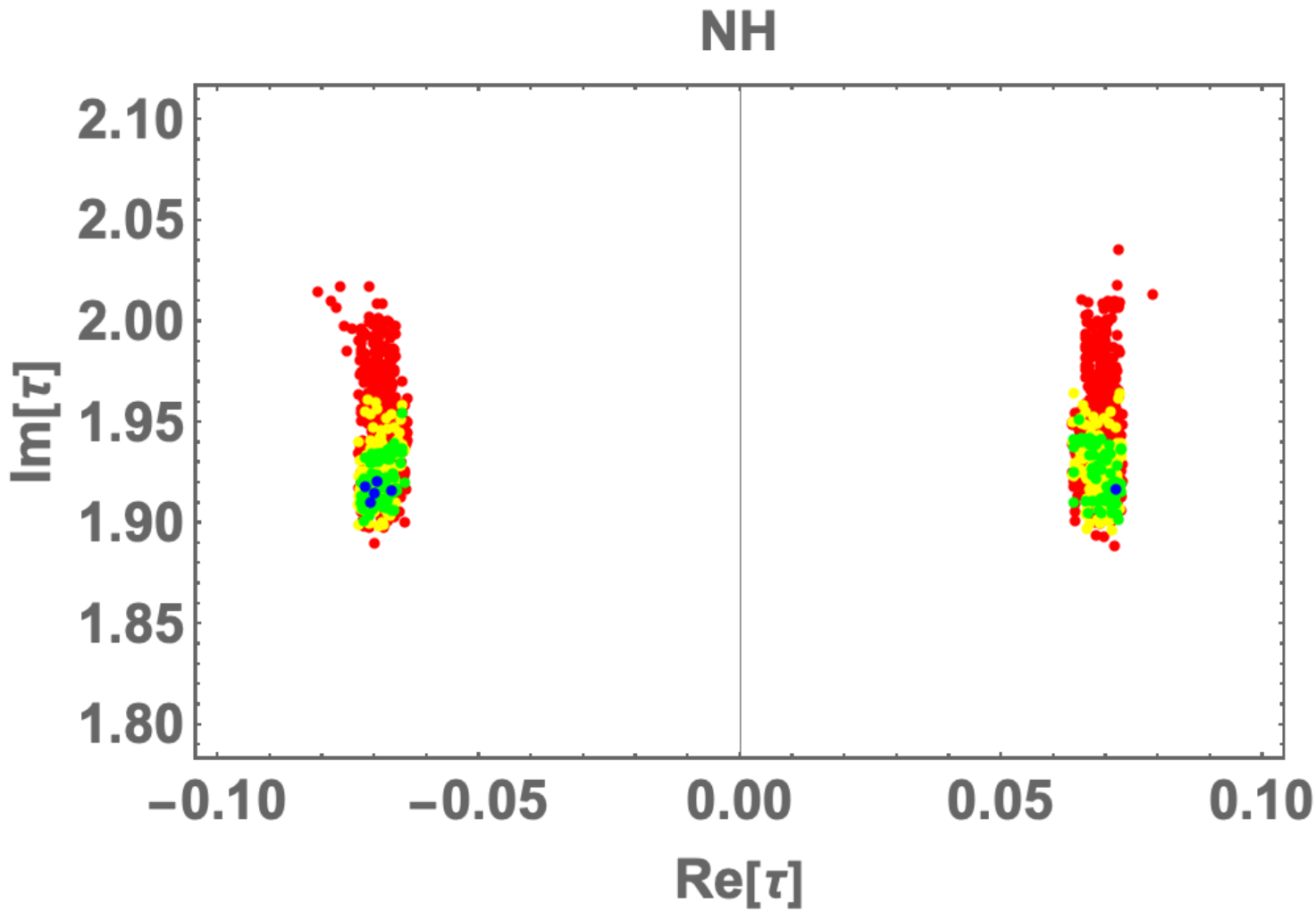}
\caption{Scatter plot for Re[$\tau$] {\it vs.} Im[$\tau$].
The red, yellow, green and blue points correspond to $3\sigma-5\sigma$, $2\sigma-3\sigma$, $1\sigma-2\sigma$, and $0\sigma-1\sigma$, respectively. 
}
\label{fig:modulus}
\end{figure}
  \begin{figure}
\centering
\includegraphics[width=0.45\textwidth]{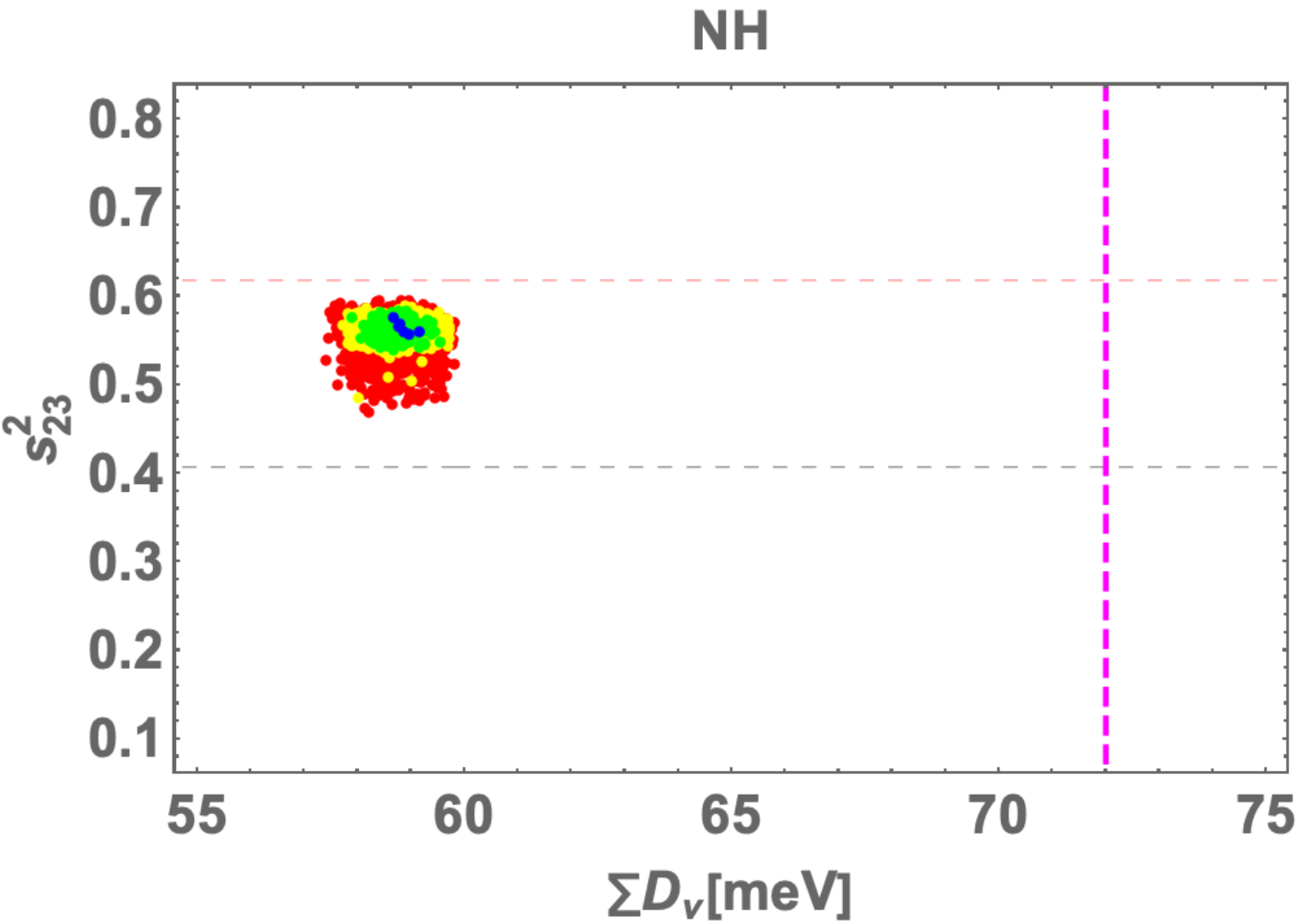}
\includegraphics[width=0.47\textwidth]{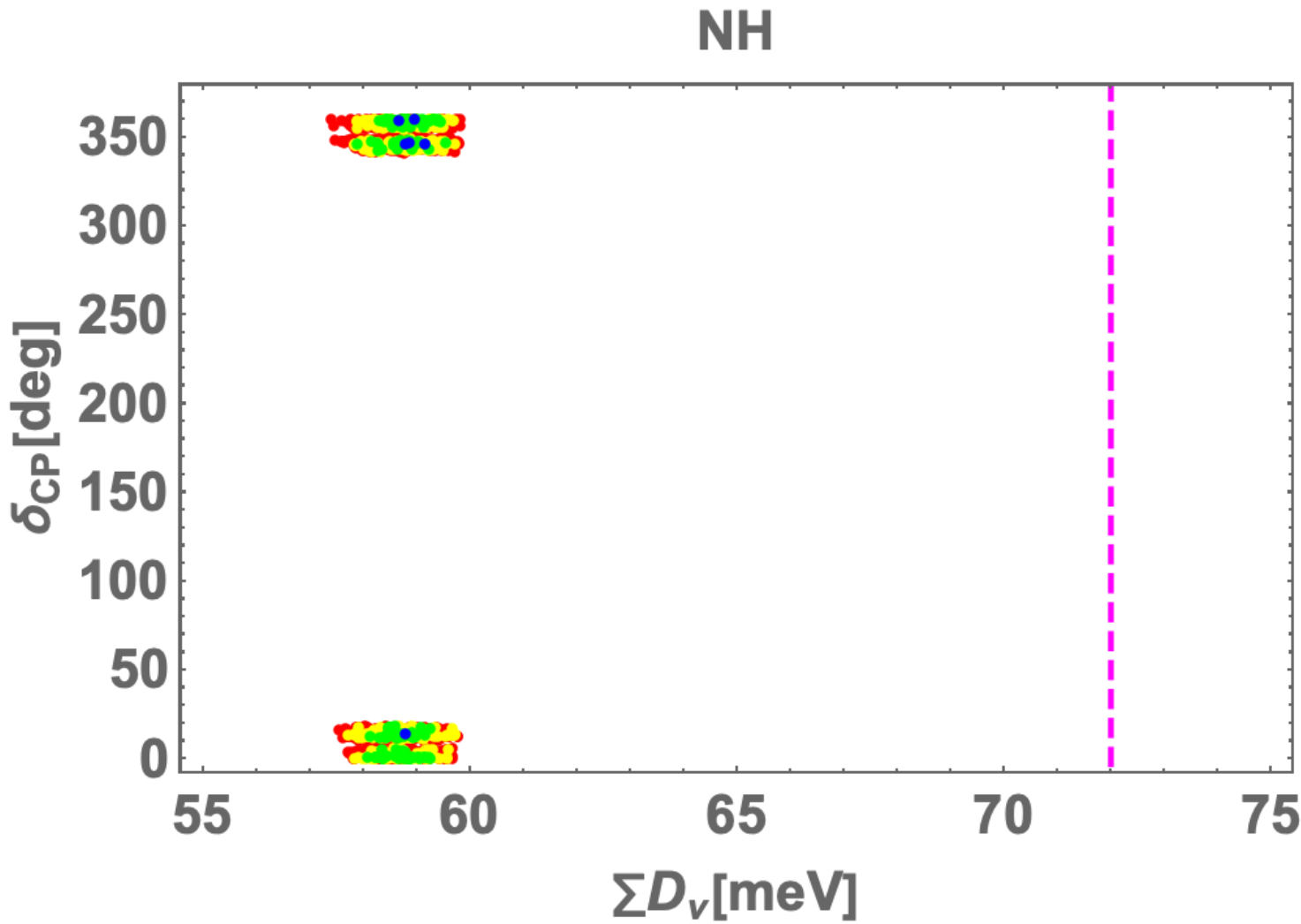}
\includegraphics[width=0.45\textwidth]{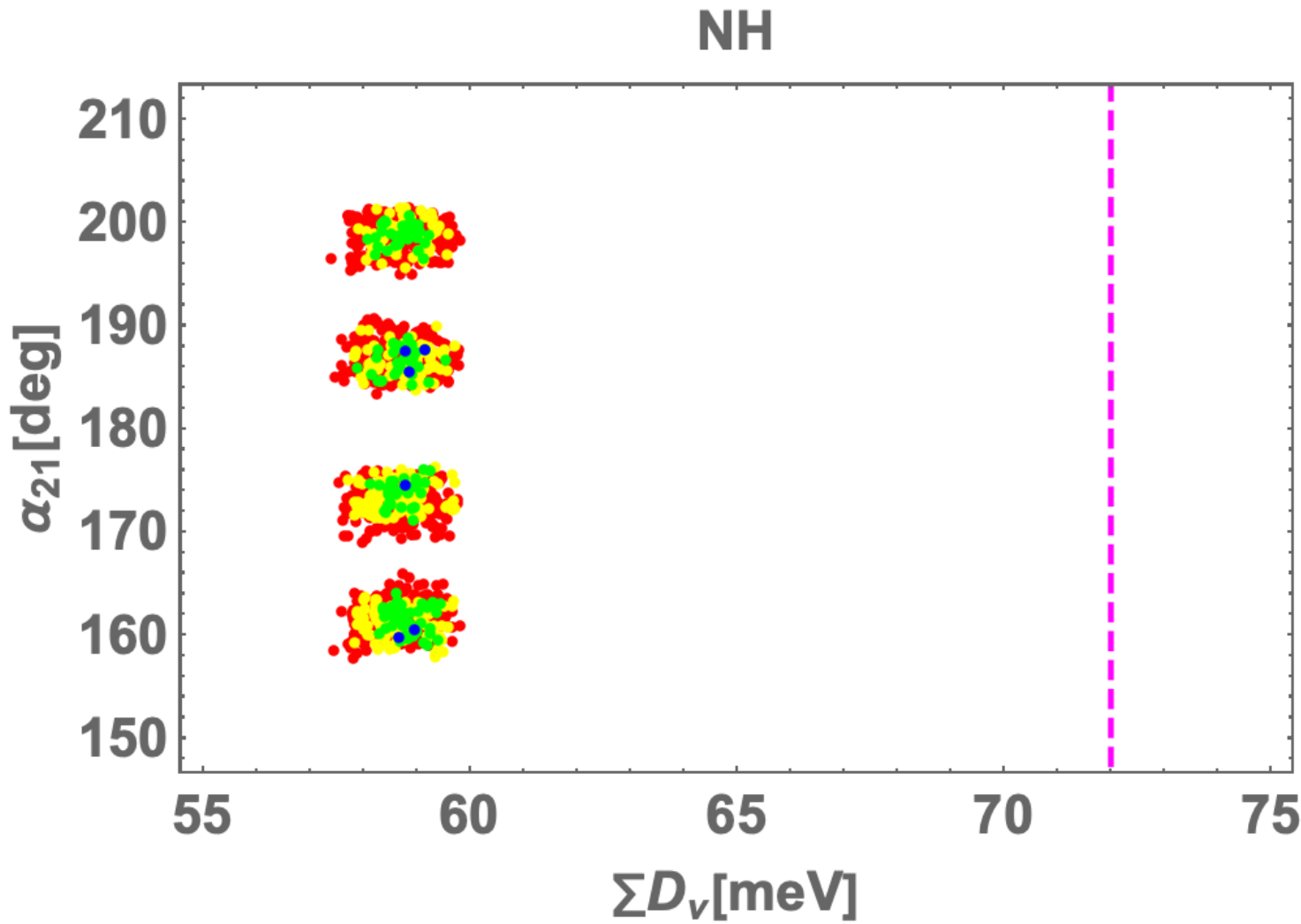}
      \includegraphics[width=0.47\textwidth]{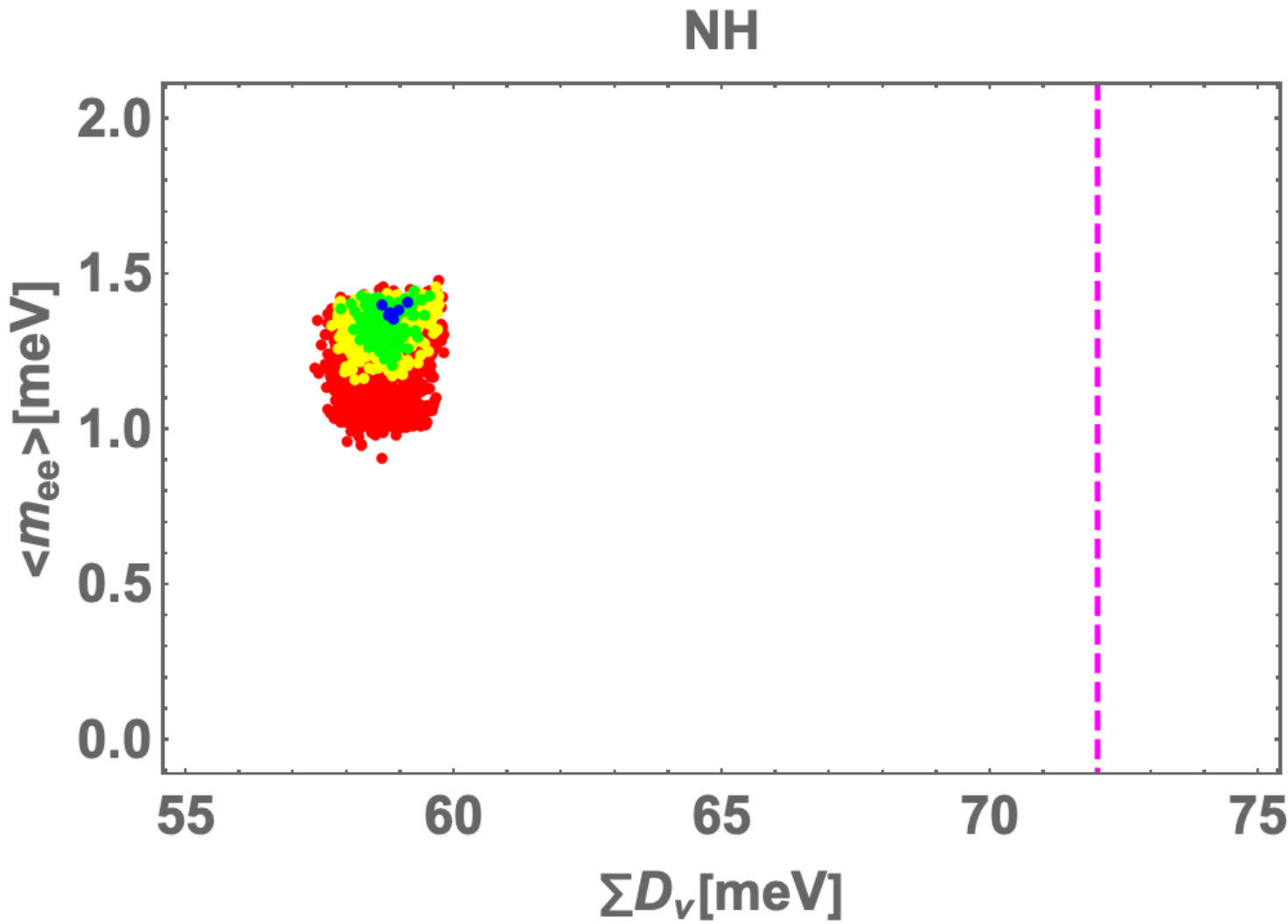}

\caption{Plots for (a) $s^2_{23}$ (b) $\delta_{\rm CP}$ (c) $\alpha_{21}$ (d)$\langle m_{ee}\rangle$ {\it vs.} $\sum D_{\nu}$.
The colors of the points carry the same meanings as  in FIG.\ref{fig:modulus}.}
\label{fig:sum}
\end{figure}
  \begin{figure}
\centering
\includegraphics[width=0.45\textwidth]{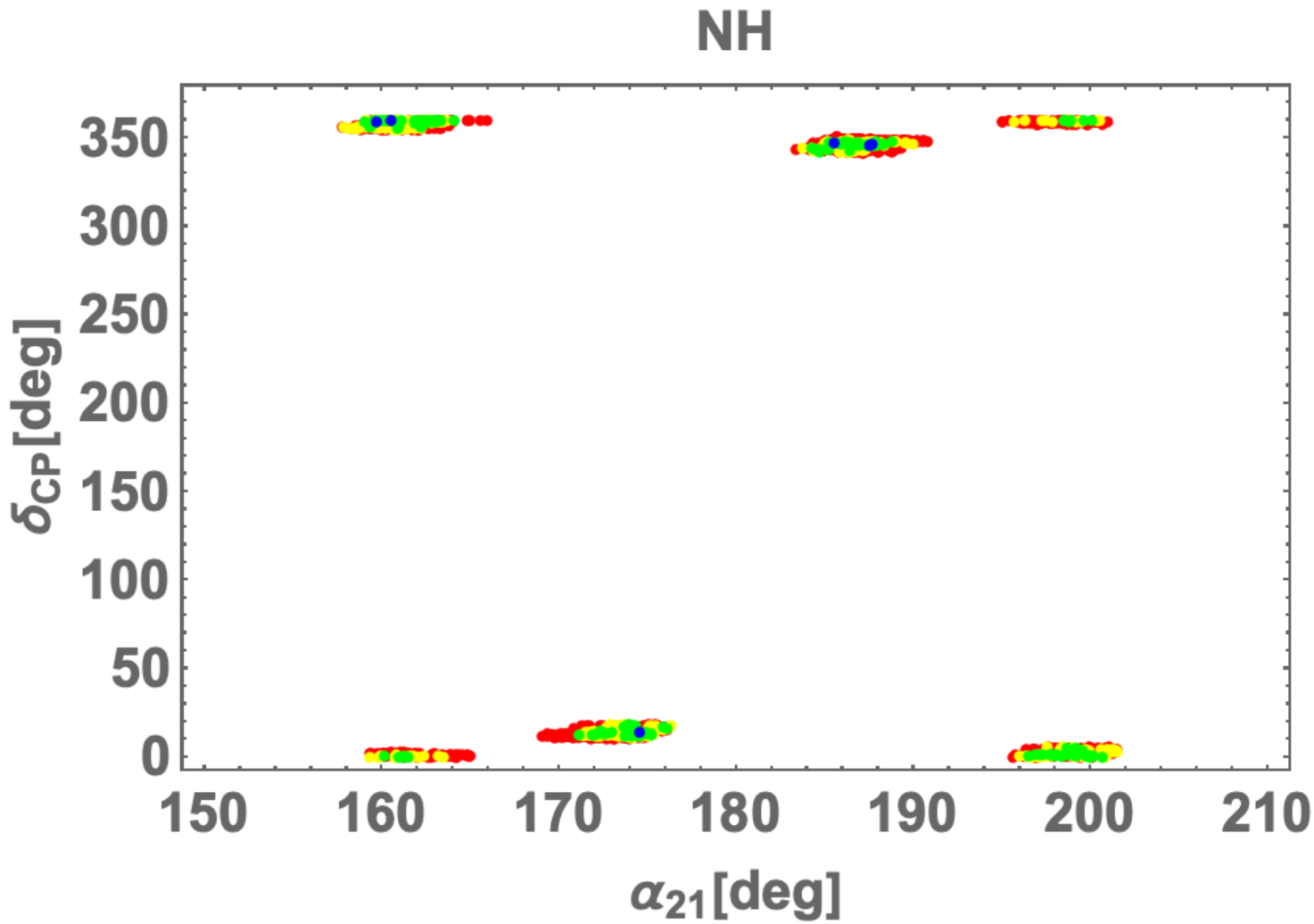}
\caption{Plot for the allowed regions of the parameter space $(\alpha_{21}, \delta_{\rm CP})$.
The colors of the points carry the same meanings as  in FIG.\ref{fig:modulus}.}
\label{fig:phase}
\end{figure}

 For our numerical analysis, at first, we naively consider the experimentally allowed regions for  the mass-squared differences and mixing parameters within the 3$\sigma$ ranges in Nufit 5.0~\cite{Esteban:2020cvm} as follows:
\begin{align}
({\rm NH}):&|\Delta m^2_{\rm atm}| =[2.431-2.598]\times 10^{-3} ~{\rm eV}^2,~
\Delta m^2_{\rm sol}= [6.82-8.04]\times 10^{-5} ~{\rm eV}^2, \\
&s^2_{12}=[0.269-0.343], ~~s^2_{23}=[0.407-0.618], ~~
s^2_{13}=[0.02034-0.02430].\nonumber \\
%%%
({\rm IH}):&|\Delta m^2_{\rm atm}| =[2.412-2.583]\times 10^{-3} ~{\rm eV}^2,~
\Delta m^2_{\rm sol}= [6.82-8.04]\times 10^{-5} ~{\rm eV}^2, \\
&s^2_{12}=[0.269-0.343], ~~s^2_{23}=[0.411-0.621], ~~
s^2_{13}=[0.02053-0.02436].\nonumber
\end{align}
However since the {$\chi^2$} of $s_{23}$ is far from Gaussian form, we perform our  {$\chi^2$} analysis more precisely, making the use of the reliable five data of Nufit5.0;
[$\Delta m^2_{\rm atm}$, $\Delta m^2_{\rm sol}$, $s^2_{12}$, $s^2_{23}$, $s^2_{13}$].
Note that Dirac CP phase is considered as our output parameter, since wide region is allowed at 3$\sigma$. 
Based on the formulae for neutrino masses, mixing, charged lepton masses, and effective neutrino mass, we perform a parameter scan of the model.
We randomly select data points that satisfy the
experimental results for neutrino oscillation parameters within $5\sigma$ range and to bounds on experiments such as neutrinoless double beta decay.
As explained earlier, this model predicts that the lightest neutrino is massless. Therefore, we set $D_{\nu_1(\nu_3)}=0$, considering {the NH(IH)} for neutrino mass spectrum, and retain  only the nonzero Majorana phase $\alpha_{21}$.
%%%

%%%%%%%%%%%%%%%%%%%%%%%
\subsubsection{Holomorphic case}
%%%%%%%%%%%%%%%%%%%%%%%
For the complex dimensionless parameters $\beta_{\nu}/\rho_{\nu}, \sigma_{\nu}/\rho_{\nu}$ and a real parameter $\alpha_{\nu}/\rho_{\nu}$, we scan over the absolute range $[10^{-5}-10^{5}]$. Since $\rho_{\nu}$ is an overall factor in the  neutrino mass matrix, its specific value does not need to be fixed.
As mentioned earlier, the free parameter for the charged lepton mass matrix is $\delta_l$, which we scan over the range $[0.01-100]$.
We note that the values of $\delta_l$ beyond 100 violate perturbativity
%%%
but the total values are within the perturbative limit due to multiplication of modular coupling $Y^{(4)}_{\bf 1}$ as can be seen in Eq.~(\ref{eq:lag-lep}).
%%%
We also scan the relevant mass parameters 
$m_{\eta^\pm}, M_{1,2}$ %m_\eta, m_\chi, m_\psi$
 in the range $[10^{3}-10^{5}]$ GeV.
Note here that $m_{S_i}\ (i=1,2,3)$ are not needed to be fixed since there masses can be involved in overall factors of the neutrino mass matrix. 
Through a $\chi^2$ analysis, we find that the best-fit values of the parameters are obtained at $\chi_{\rm min} = 3.84$. 
The corresponding best-fit parameter values are listed in TABLE III.
%
%As already mentioned in the previous subsection, 
IH of the holomorphic case is disfavored by the neutrino oscillation data, and we concentrate on the case of NH.

%%%%%%%%%%%%%%%%%%
\begin{table}[tb]
    \setlength\tabcolsep{0.2cm}
    \begin{tabular}{c|c||c|c||c|c}
%        \toprule
\hline
        parameter    &  BF & parameter & BF & parameter & BF \\ \hline \hline
          $\tau$ &  $-0.072+1.918 i $ & $m_{\eta^\pm}$ & 1200.73 GeV &$M_1$ &1171.5 GeV \\ \hline
       $M_2$  &23870 GeV &  $\sigma_\nu/\rho_\nu$ & $-0.0034-0.0362 i$ &
 $\alpha_\nu/\rho_\nu$ &   $-1.705$ \\ \hline
 $\beta_\nu/\rho_\nu$ &   $-0.027-1.959 i$ &
$\delta_l$ & $-3.742+20.796 i$ & $\alpha_l$ & 0.0023 \\ \hline
$\beta_l$ & 60.298 & $\gamma_l$ & 0.504 & $$ &  \\ \hline
$s_{12}$ & 0.5317 &$s_{23}$ & 0.7519 &$s_{13}$ & 0.1504 \\ \hline
$\Delta m^2_{\rm sol}$ & $7.353\times 10^{-5}~{\rm eV}^2$ &  $\Delta m^2_{\rm atm}$ & $2.52\times 10^{-3}~{\rm eV}^2$ &
$\delta_{\rm CP}$ & $345.951^{\circ}$ \\ \hline
$\alpha_{21}$ & $187.503^{\circ}$ &$\sum D_{\nu}$ & $5.88\times 10^{-2}~{\rm eV}$ & $\langle m_{ee}\rangle$ 
& $1.37\times 10^{-3}~ {\rm eV}$ \\
%        \bottomrule
\hline
    \end{tabular}
    \caption{\label{tab:BF}%
      Best-fit (BF) parameter values in the NH case corresponding to $\chi_{\rm min} =3.84$.}
\end{table}
%%%%%%%%%%%%%%

FIG. \ref{fig:modulus} shows scatter plot of  the real and imaginary parts of the modulus $\tau$ constrained within $5\sigma$.
The red, yellow, green and blue points correspond to $3\sigma-5\sigma$, $2\sigma-3\sigma$, $1\sigma-2\sigma$, and $0\sigma-1\sigma$, respectively. 
From the plot, we see that the allowed regions of the modulus $\tau$ within the $5\sigma$ range confined to a relatively narrow space as 
$0.062 \lesssim |{\rm Re}[\tau]|\lesssim 0.082$ and  $1.88 \lesssim{\rm Im}[\tau]\lesssim 2.04$.
%, and $ 0.062 \lesssim {\rm Re}[\tau]\lesssim 0.08$ and  $1.88 \lesssim{\rm Im}[\tau]\lesssim 2.04$.

FIG. \ref{fig:sum} represents scatter plots of (a) $s^2_{23}$ (b) $\delta_{\rm CP}$ (c) $\alpha_{21}$ (d) $\langle m_{ee}\rangle$ in terms of
$\sum D_{\nu}$. The color scheme for the data points is consistent with that in FIG. \ref{fig:modulus}.
The pink vertical dashed lines indicate the upper limit on the sum of neutrino masses from combined data of DESI and CMB.
From the plots, the allowed range for $\sum D_{\nu}$ up to $5\sigma$ is between 57.2 meV and 60 meV.
In panel (a), it is evident that larger values of $s^2_{23}$ are favored.
For the Dirac CP phase $\delta_{\rm CP}$,  the allowed regions are approximately $0\sim~20^{\circ}$ and $335^{\circ}\sim350^{\circ}$.
In contrast, for the Majorana CP phase $\alpha_{21}$, the allowed regions are approximately  $157^{\circ}\sim 166^{\circ}$, $168^{\circ}\sim 176^{\circ}$, $183^{\circ}\sim 191^{\circ}$, and $195^{\circ}\sim 202^{\circ}$. The effective mass $\langle m_{ee}\rangle$ has an allowed region of $0.8{\rm meV}\sim 1.5{\rm meV}$,
% within the  $5 \sigma$ C.L., 
which is significantly below the current experimental bound  from the current KamLAND-Zen experiment \cite{KamLAND-Zen:2016pfg}.
FIG. \ref{fig:phase} shows the correlation between $\alpha_{21}$ and $\delta_{\rm CP}$. Six distinct allowed regions can be identified in the plot.
%

%%%%%%%%%%%%%%%%%%%%%%%
\subsubsection{Non-holomorphic case}
%%%%%%%%%%%%%%%%%%%%%%%
In the non-holomorphic case, the charged lepton mass matrix includes an additional free parameter, $\sigma_\ell$, as introduced in Eq.~(\ref{eq:lag-lep_nonholo}).
%the previous section.
We scan $\sigma_\ell$ over the absolute range [$10^{-5}-10^5$], keeping all other parameters and their respective scan ranges identical to those in the holomorphic case.

\begin{figure}
\centering
\includegraphics[width=0.60\textwidth]{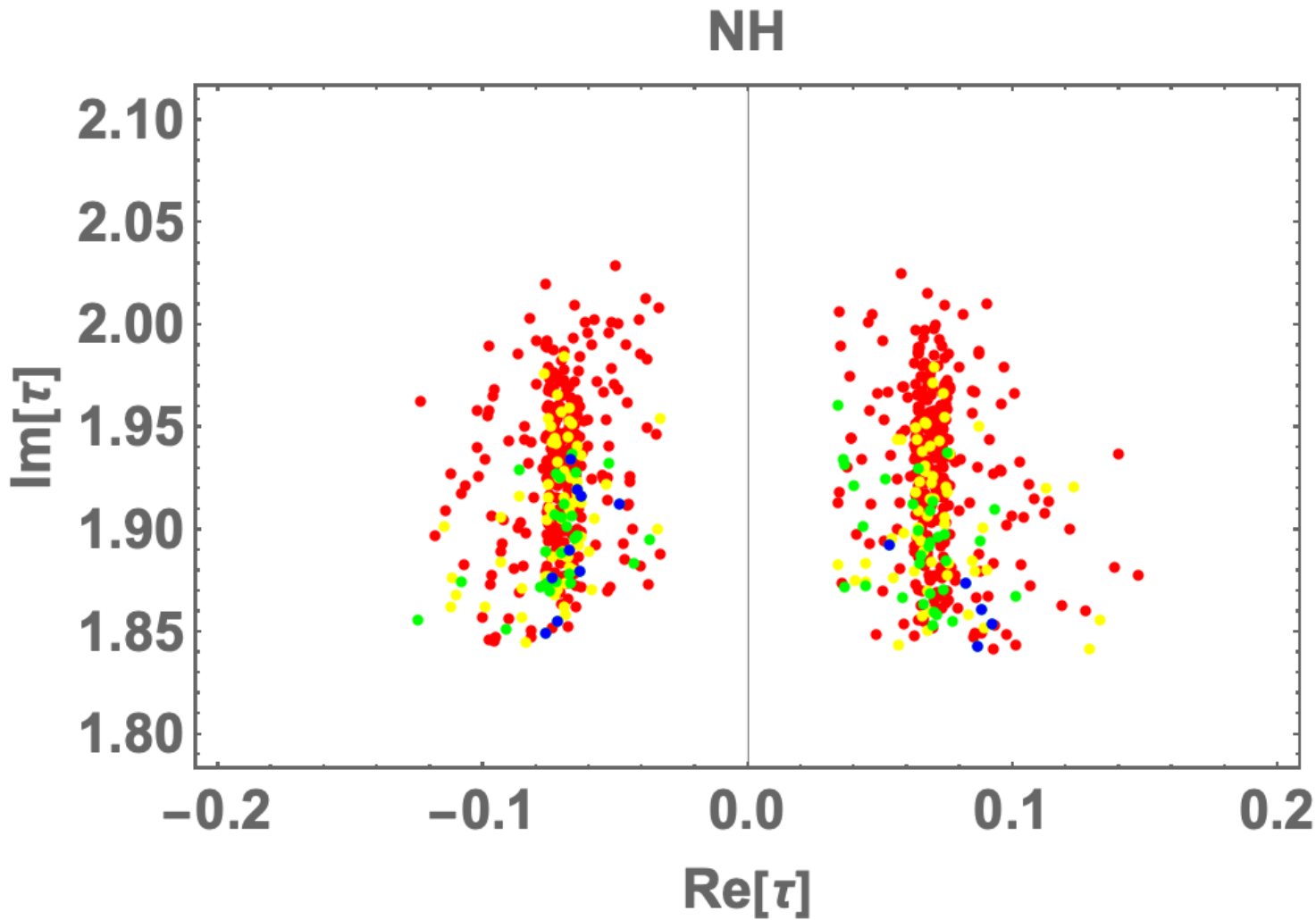}
\caption{Scatter plot for Re[$\tau$] {\it vs.} Im[$\tau$].
The red, yellow, green and blue points correspond to $3\sigma-5\sigma$, $2\sigma-3\sigma$, $1\sigma-2\sigma$, and $0\sigma-1\sigma$, respectively. 
}
\label{fig:modulus_no}
\end{figure}
  \begin{figure}
\centering
\includegraphics[width=0.45\textwidth]{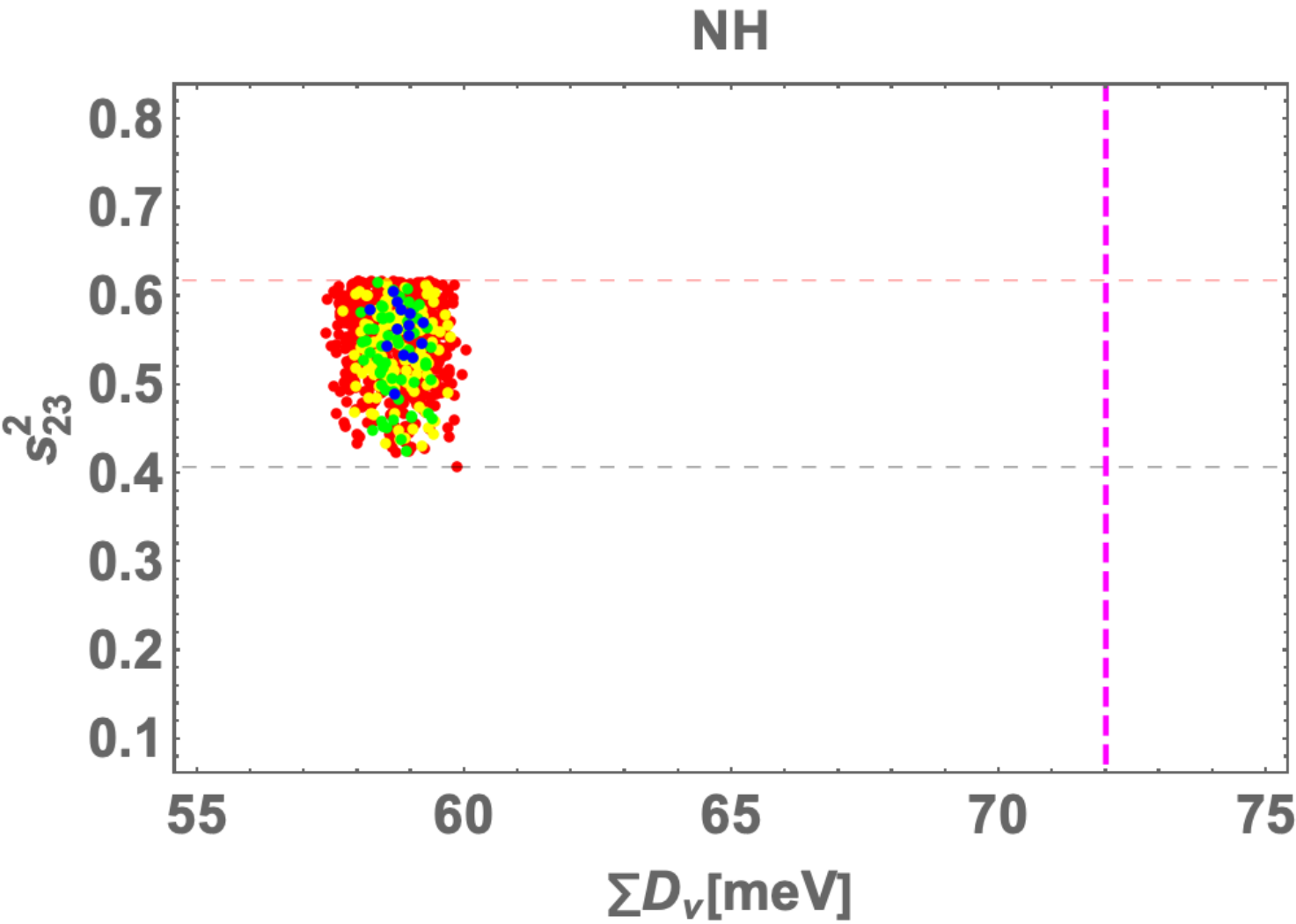}
\includegraphics[width=0.47\textwidth]{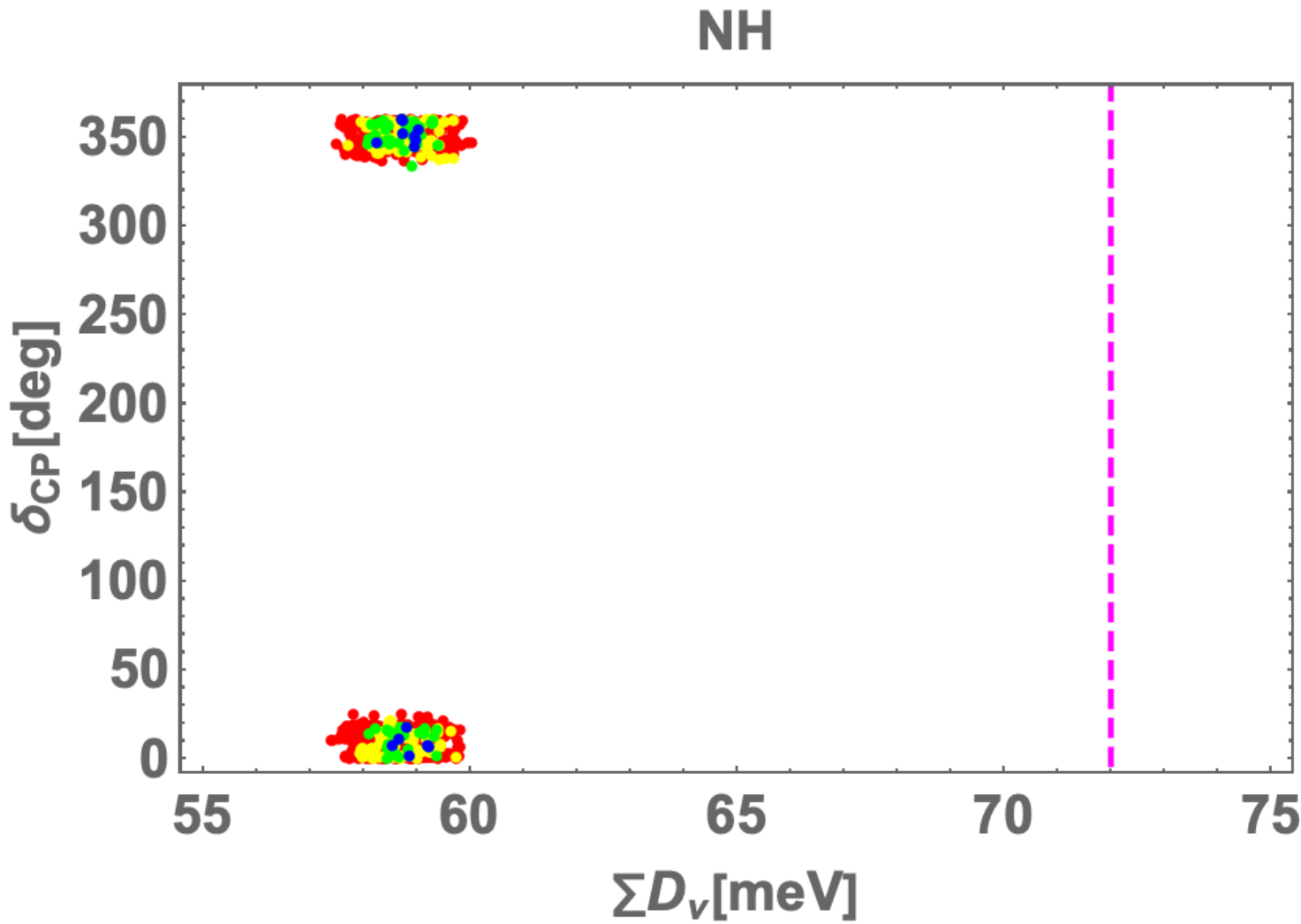}
\includegraphics[width=0.45\textwidth]{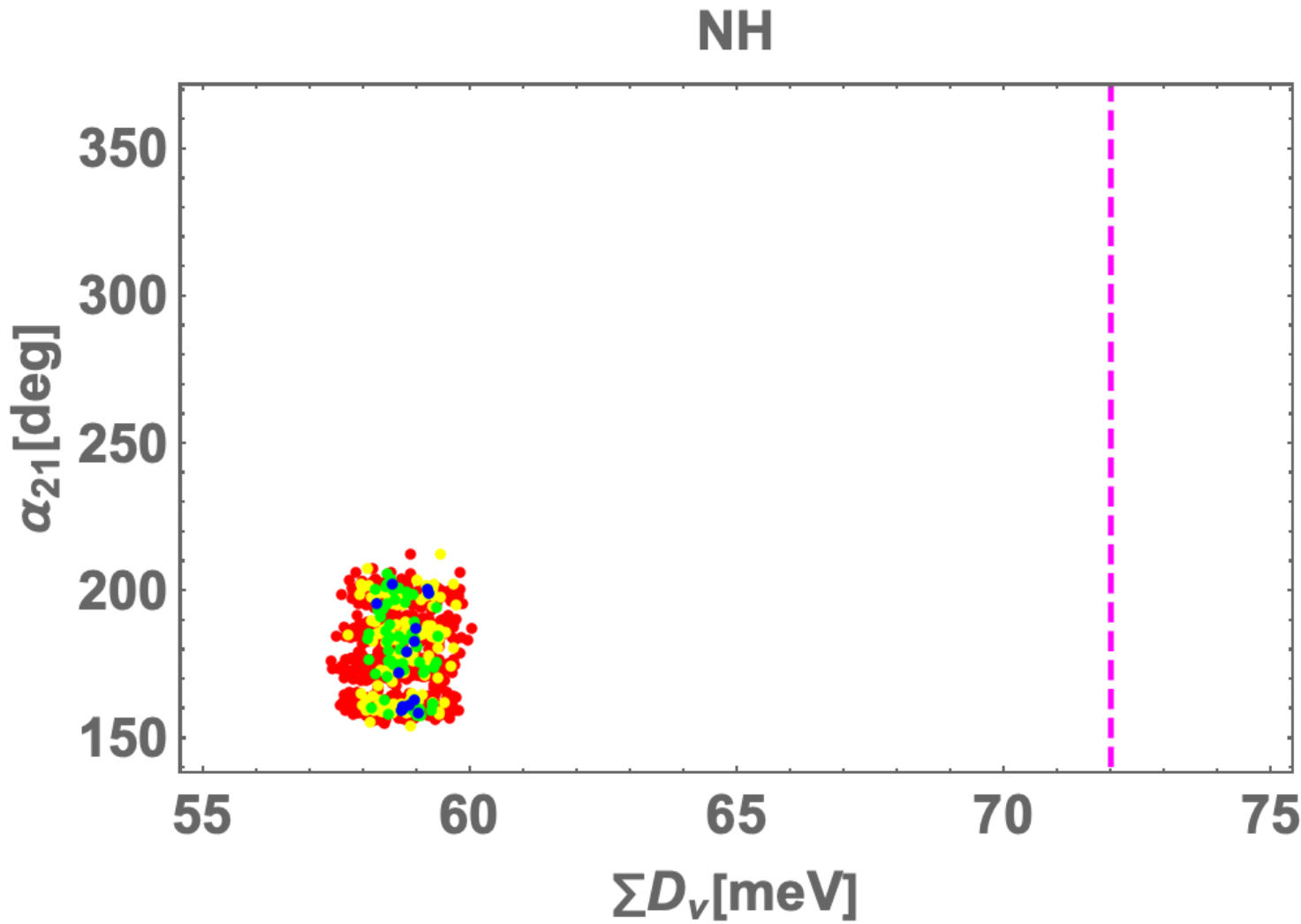}
      \includegraphics[width=0.47\textwidth]{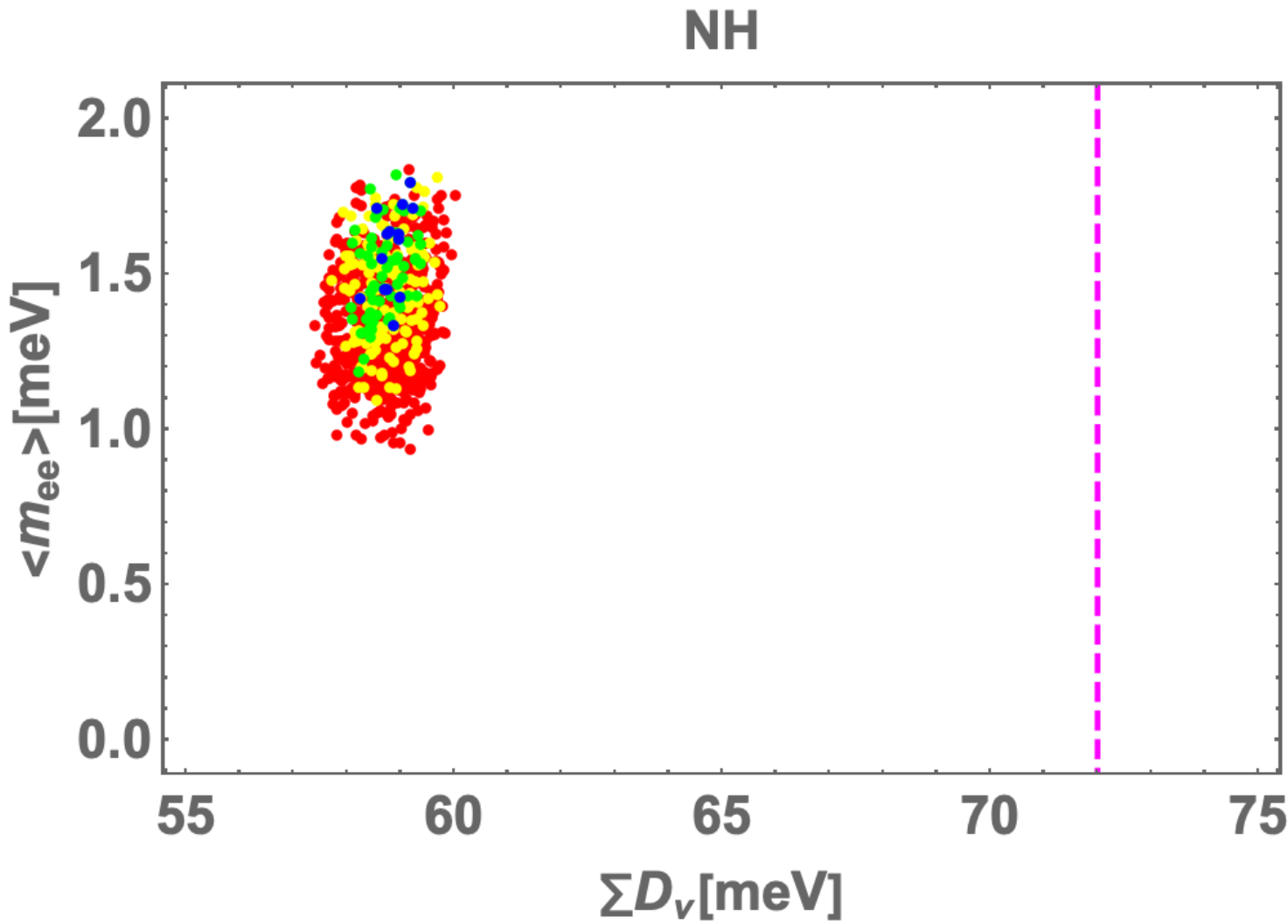}

\caption{Plots for (a) $s^2_{23}$ (b) $\delta_{\rm CP}$ (c) $\alpha_{21}$ (d)$\langle m_{ee}\rangle$ {\it vs.} $\sum D_{\nu}$.
The colors of the points carry the same meanings as in FIG.\ref{fig:modulus}.}
\label{fig:sum_no}
\end{figure}
  \begin{figure}
\centering
\includegraphics[width=0.45\textwidth]{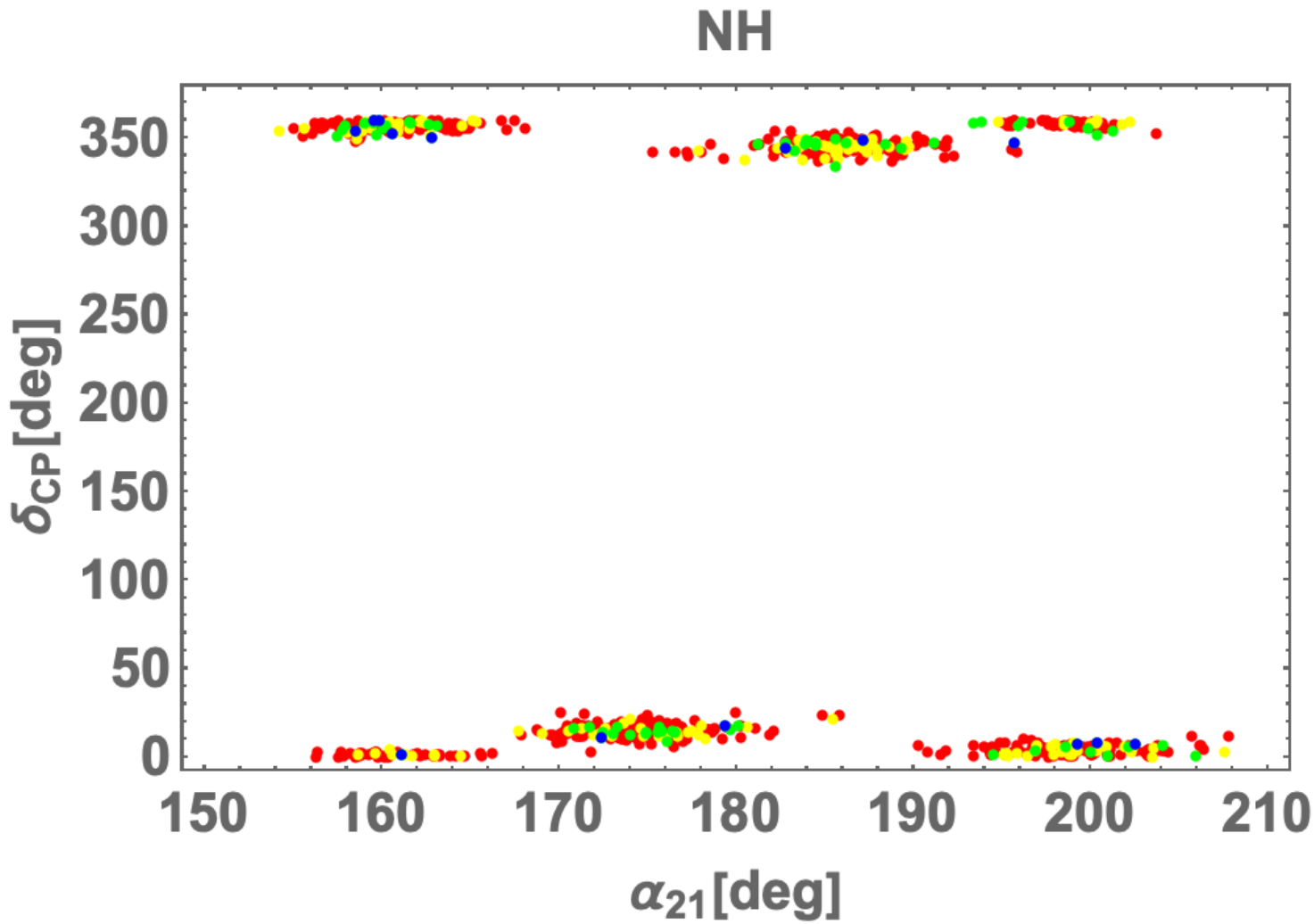}
\caption{Plot for the allowed regions of the parameter space $(\alpha_{21}, \delta_{\rm CP})$.
The colors of the points carry the same meanings as in FIG.\ref{fig:modulus}.}
\label{fig:phase_no}
\end{figure}
%{\bf NO case:}\\
From a $\chi^2$ analysis for the NH case, we find that the best-fit values of the parameters are obtained at $\chi_{\rm min} = 3.042$. 
The corresponding best-fit parameter values are listed in TABLE IV.
%
%
%
%%%%%%%%%%%%%%%%%%
\begin{table}[tb]
    \setlength\tabcolsep{0.2cm}
    \begin{tabular}{c|c||c|c||c|c}
%        \toprule
\hline
        parameter    &  BF & parameter & BF & parameter & BF \\ \hline \hline
          $\tau$ &  $-0.088 + 1.861 i $ & $m_{\eta^\pm}$ & 1127.68 GeV &$M_1$ &1439.18 GeV \\ \hline
       $M_2$  &26885.8 GeV &  $\sigma_\nu/\rho_\nu$ & $-0.0028 + 0.0426 i$ &
 $\alpha_\nu/\rho_\nu$ &   $-2.060$ \\ \hline
 $\beta_\nu/\rho_\nu$ &   $-0.03219 + 2.375 i$ &
$\delta_l$ & $-3.178 - 19.53 i$ & $\alpha_l$ & 0.0021 \\ \hline
$\beta_l$ & 60.707  & $\gamma_l$ & 0.500 & $\sigma_\ell\times10^4$ & $-2.338- 2.566 i$  \\ \hline
$s_{12}$ & 0.5390 &$s_{23}$ & 0.7645 &$s_{13}$ & 0.1488 \\ \hline
$\Delta m^2_{\rm sol}$ & $7.229\times 10^{-5}~{\rm eV}^2$ &  $\Delta m^2_{\rm atm}$ & $2.530\times 10^{-3}~{\rm eV}^2$ &
$\delta_{\rm CP}$ & $17.875^{\circ}$ \\ \hline
$\alpha_{21}$ & $179.363^{\circ}$ &$\sum D_{\nu}$ & $0.0588~{\rm eV}$ & $\langle m_{ee}\rangle$ 
& $1.63\times 10^{-3}~ {\rm eV}$ \\
%        \bottomrule
\hline
    \end{tabular}
    \caption{\label{tab:BF_no}%
      Best-fit (BF) parameter values in the NH case corresponding to $\chi_{\rm min}=3.042$.}
\end{table}
%%%%%%%%%%%%%%
%
FIG. \ref{fig:modulus_no} shows the scatter plot of  the real and imaginary parts of the modulus $\tau$ constrained within $5\sigma$ region.
The color legends are the same as the  holomorphic case. 
From the plot, we see that the allowed region for the modulus $\tau$
%within the $5\sigma$ range
is confined to a relatively narrow range,
$0.025 \lesssim |{\rm Re}[\tau]|\lesssim 0.145$ and  $1.83 \lesssim{\rm Im}[\tau]\lesssim 2.04$.
%, and $ 0.062 \lesssim {\rm Re}[\tau]\lesssim 0.08$ and  $1.88 \lesssim{\rm Im}[\tau]\lesssim 2.04$.

FIG. \ref{fig:sum_no} represents the scatter plots of (a) $s^2_{23}$ (b) $\delta_{\rm CP}$ (c) $\alpha_{21}$ (d) $\langle m_{ee}\rangle$ in terms of
$\sum D_{\nu}$. The color scheme for the data points is consistent with that in FIG. \ref{fig:modulus}.
The vertical pink dashed lines indicate the upper limit on the sum of neutrino masses derived from the combined data of DESI and CMB. 
From the plots, we find that the allowed range for $\sum D_{\nu}$ up to $5\sigma$ lies between 57.2 meV and 60 meV.
Panel (a) represents a preference for larger values of $s^2_{23}$.
For the Dirac CP phase $\delta_{\rm CP}$,  the allowed regions are approximately $0\sim~20^{\circ}$ and $330^{\circ}\sim360^{\circ}$.
In contrast, for the Majorana CP phase $\alpha_{21}$, the allowed region lies approximately  in the region of $150^{\circ}\sim 210^{\circ}$.
The effective mass $\langle m_{ee}\rangle$ is constrained to the range $0.9{\rm meV}\sim 1.83{\rm meV}$,
% within the  $5 \sigma$ C.L., 
 which is 
significantly below the current experimental bound  from the current KamLAND-Zen experiment \cite{KamLAND-Zen:2016pfg}.
FIG. \ref{fig:phase_no} shows the correlation between $\alpha_{21}$ and $\delta_{\rm CP}$. 
Six allowed regions could be identified in the plot.
\begin{figure}
\centering
\includegraphics[width=0.60\textwidth]{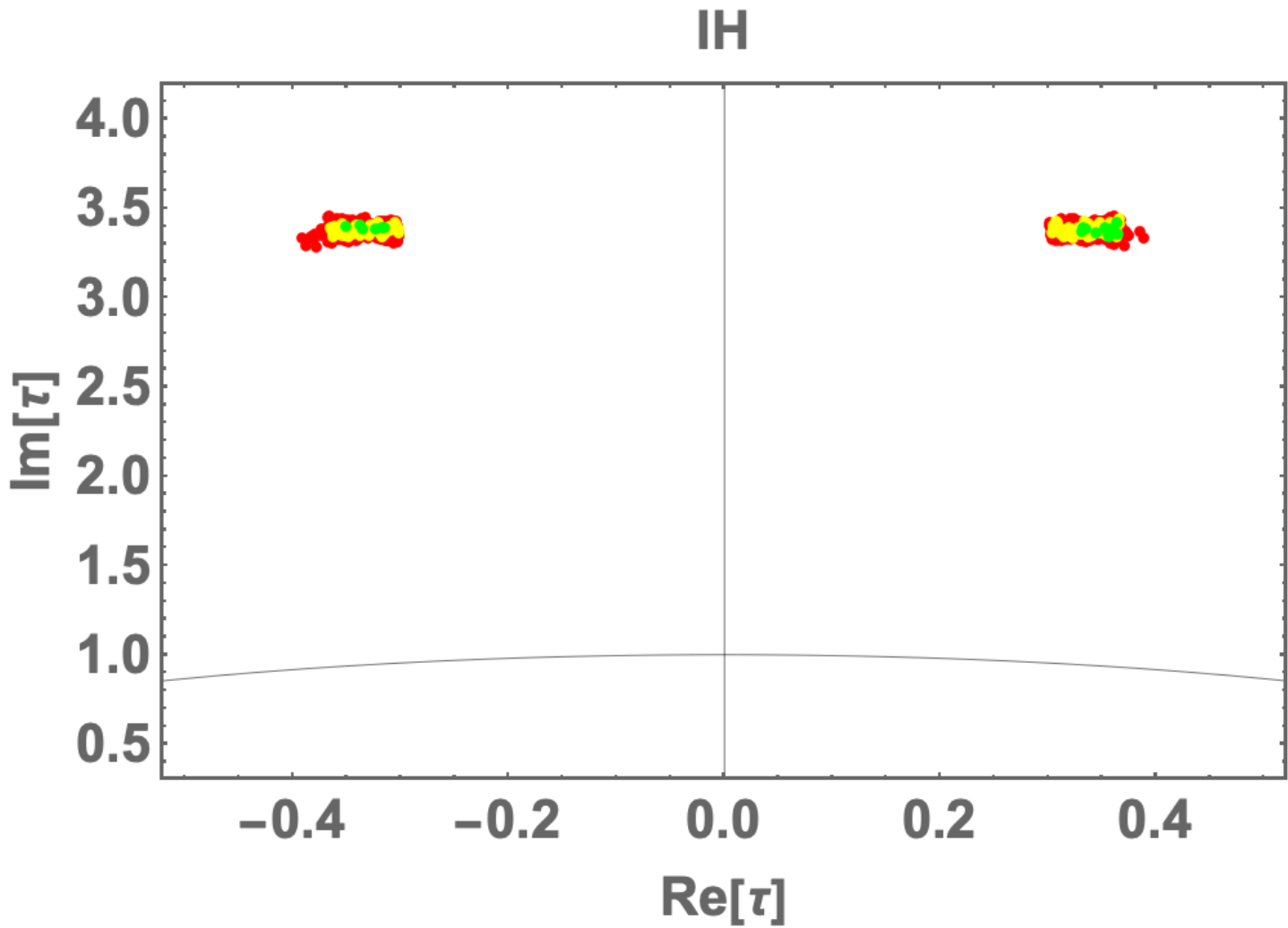}
\caption{Scatter plot for Re[$\tau$] {\it vs.} Im[$\tau$].
The red, yellow, green and blue points correspond to $3\sigma-5\sigma$, $2\sigma-3\sigma$, $1\sigma-2\sigma$, and $0\sigma-1\sigma$, respectively. 
}
\label{fig:modulus_io}
\end{figure}
  \begin{figure}
\centering
\includegraphics[width=0.45\textwidth]{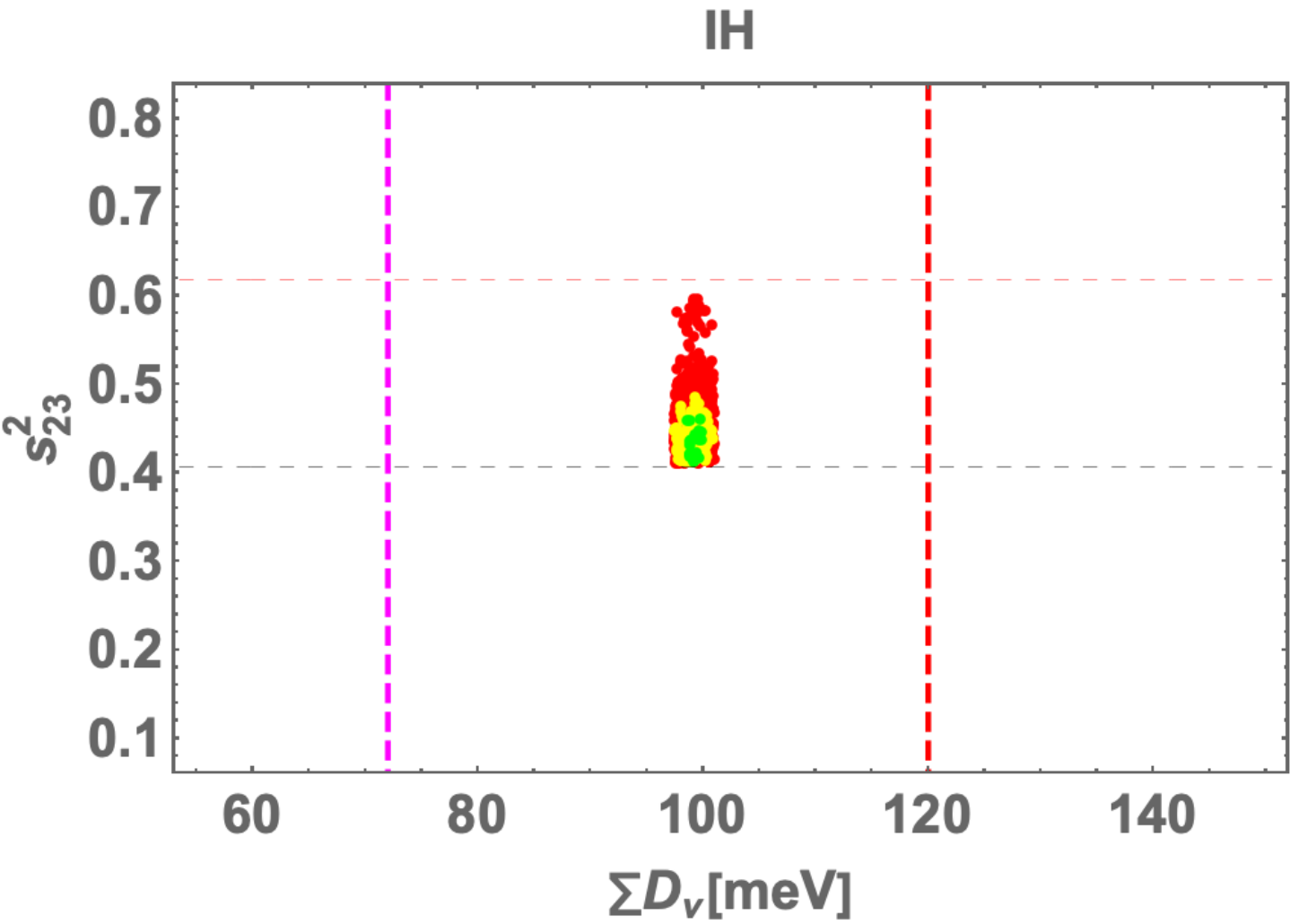}
\includegraphics[width=0.47\textwidth]{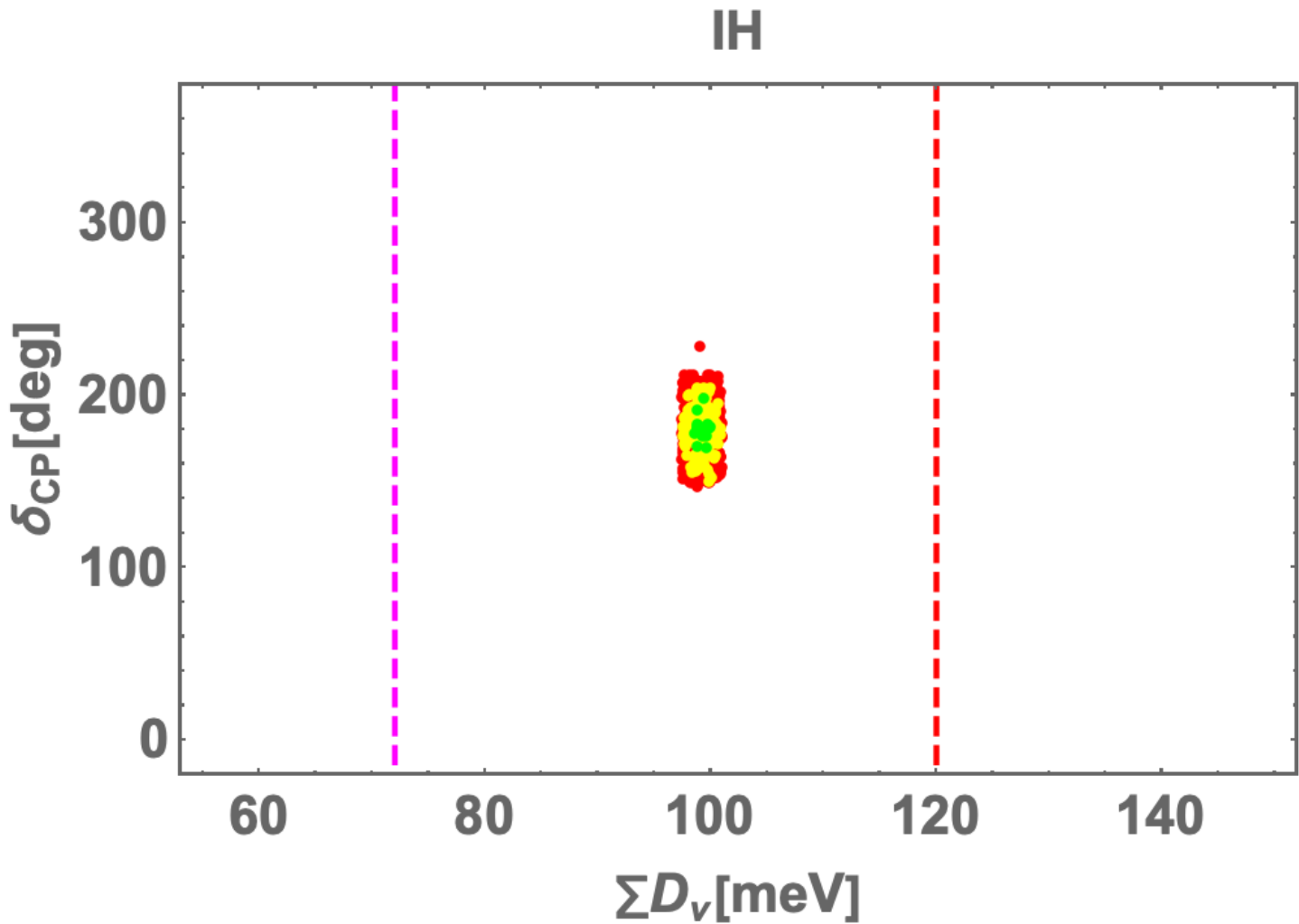}
\includegraphics[width=0.45\textwidth]{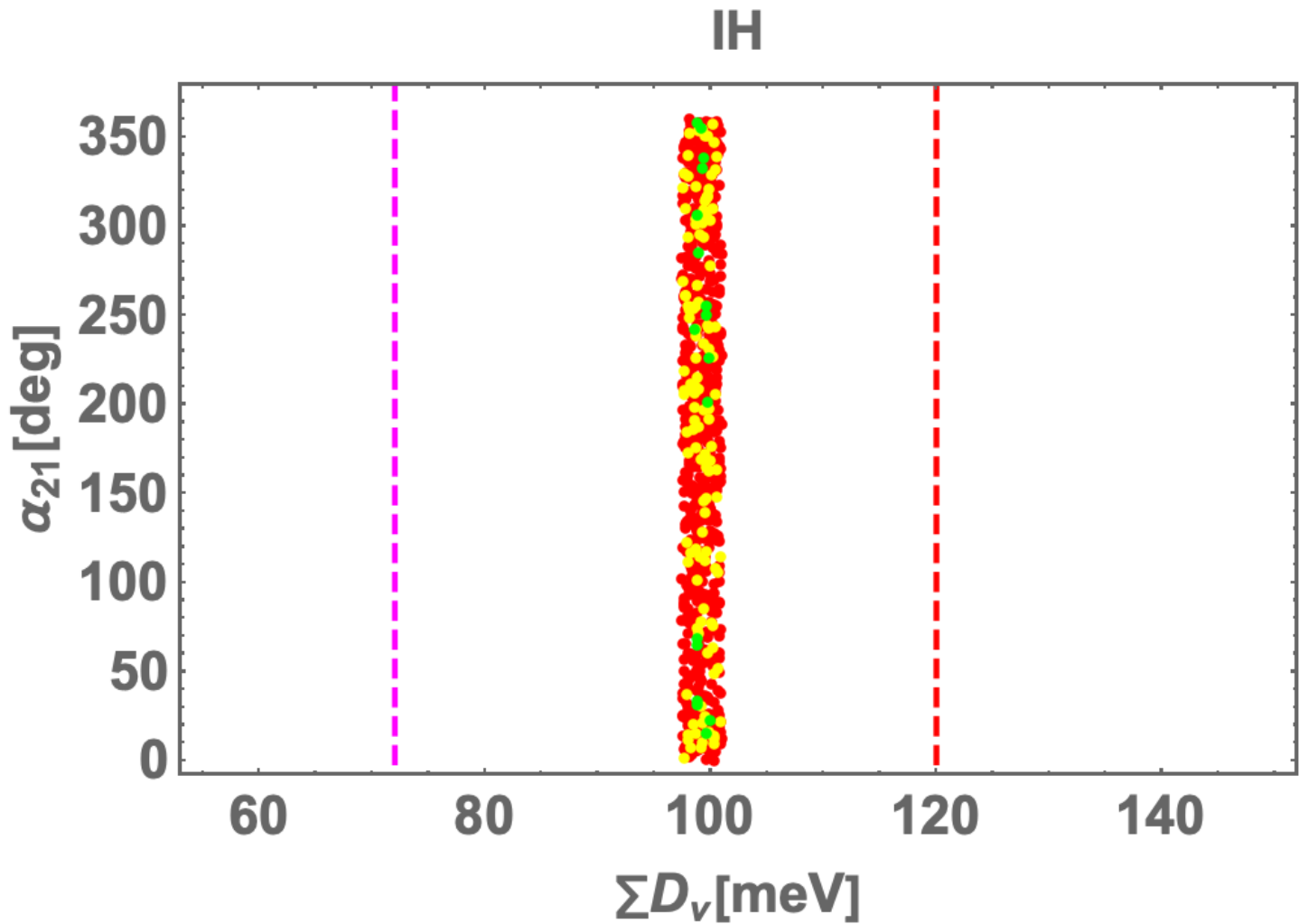}
      \includegraphics[width=0.47\textwidth]{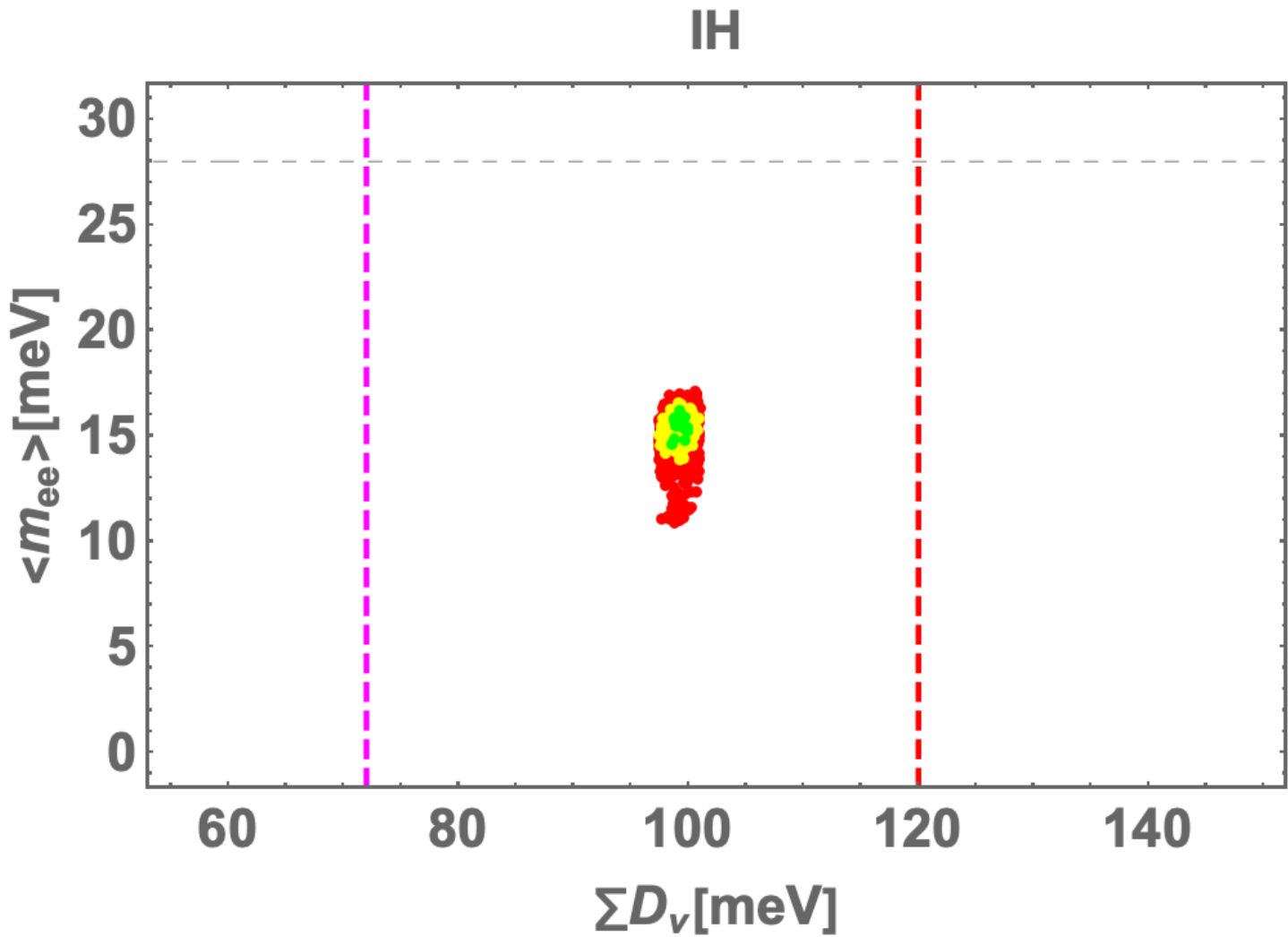}

\caption{Plots for (a) $s^2_{23}$ (b) $\delta_{\rm CP}$ (c) $\alpha_{21}$ (d)$\langle m_{ee}\rangle$ {\it vs.} $\sum D_{\nu}$.
The colors of the points carry the same meanings as  in FIG.\ref{fig:modulus}.}
\label{fig:sum_io}
\end{figure}
  \begin{figure}
\centering
\includegraphics[width=0.45\textwidth]{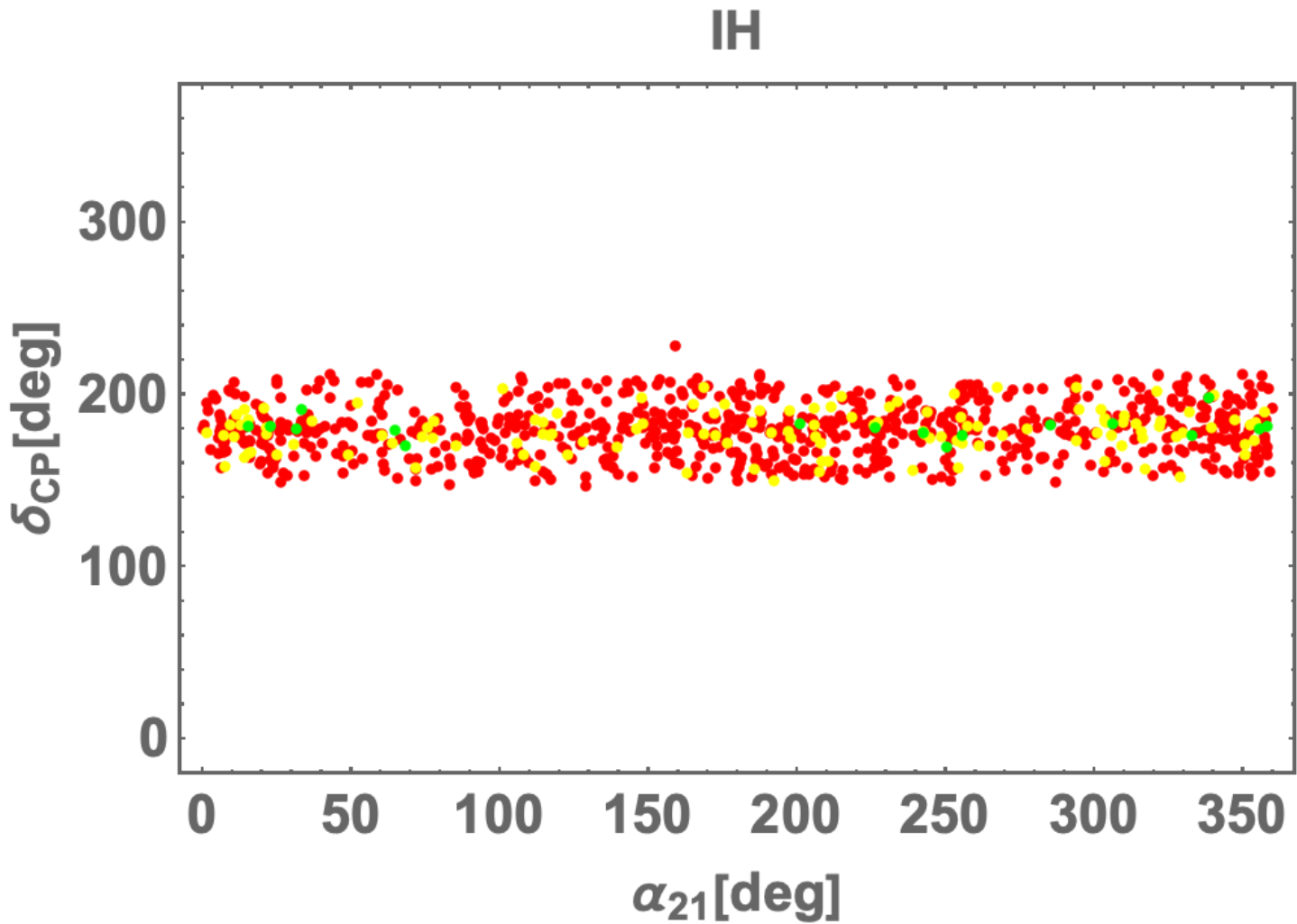}
\caption{Plot for the allowed regions of the parameter space $(\alpha_{21}, \delta_{\rm CP})$.
The colors of the points carry the same meanings as  in FIG.\ref{fig:modulus}.}
\label{fig:phase_io}
\end{figure}
%{\bf IO case:}\\
From a $\chi^2$ analysis for the IH case, we find that the best-fit values of the parameters are obtained at $\chi_{\rm min} = 7.82429$. 
The corresponding best-fit parameter values are listed in TABLE V.
%
%
%
%%%%%%%%%%%%%%%%%%
\begin{table}[tb]
    \setlength\tabcolsep{0.2cm}
    \begin{tabular}{c|c||c|c||c|c}
%        \toprule
\hline
        parameter    &  BF & parameter & BF & parameter & BF \\ \hline \hline
          $\tau$ &  $0.344 + 3.36 i $ & $m_{\eta^\pm}$ & 7598.89 GeV &$M_1$ &9345.7 GeV \\ \hline
       $M_2$  &2463.32 GeV &  $\sigma_\nu/\rho_\nu$ & $0.0850 - 0.0593 i$ &
 $\alpha_\nu/\rho_\nu$ &   $-0.3047$ \\ \hline
 $\beta_\nu/\rho_\nu$ &   $0.1057 - 0.3162 i$ &
$\delta_l$ & $-0.4228 - 0.4928 i$ & $\alpha_l$ & 0.0022 \\ \hline
$\beta_l$ & 63.991  & $\gamma_l$ & 0.465 & $\sigma_\ell$ & $-0.002617 - 0.01258 i$  \\ \hline
$s_{12}$ & 0.5798 &$s_{23}$ & 0.6612 &$s_{13}$ & 0.1514 \\ \hline
$\Delta m^2_{\rm sol}$ & $7.476\times 10^{-5}~{\rm eV}^2$ &  $\Delta m^2_{\rm atm}$ & $2.482\times 10^{-3}~{\rm eV}^2$ &
$\delta_{\rm CP}$ & $182.212^{\circ}$ \\ \hline
$\alpha_{21}$ & $285.241^{\circ}$ &$\sum D_{\nu}$ & $0.0989~{\rm eV}$ & $\langle m_{ee}\rangle$ 
& $0.0155~ {\rm eV}$ \\
%        \bottomrule
\hline
    \end{tabular}
    \caption{\label{tab:BF_io}%
      Best-fit (BF) parameter values in the IH case corresponding to $\chi_{\rm min} =7.82429$.}
\end{table}
%%%%%%%%%%%%%%
%
FIG. \ref{fig:modulus_io} shows the scatter plot for  the real and imaginary parts of the modulus $\tau$ constrained.
The color legends are the same as the  holomorphic case. 
From the plot, we see that the allowed region of the modulus $\tau$ is confined to a relatively narrow region,
$0.3 \lesssim |{\rm Re}[\tau]|\lesssim 0.4$ and  $3.25 \lesssim{\rm Im}[\tau]\lesssim 3.50$.
%, and $ 0.062 \lesssim {\rm Re}[\tau]\lesssim 0.08$ and  $1.88 \lesssim{\rm Im}[\tau]\lesssim 2.04$.

FIG. \ref{fig:sum_io} represents the scatter plots of (a) $s^2_{23}$ (b) $\delta_{\rm CP}$ (c) $\alpha_{21}$ (d) $\langle m_{ee}\rangle$ in terms of
$\sum D_{\nu}$. The color scheme for the data points is consistent with that in FIG. \ref{fig:modulus}.
The vertical pink(red) dashed lines indicate the upper limit on the sum of neutrino masses from combined data of DESI and CMB (cosmological observations).
From the plots, the allowed range for $\sum D_{\nu}$ %up to $5\sigma$ 
lies between 93 meV and 102 meV.
Panel (a) shows a tendency toward smaller values of $s^2_{23}$.
For the Dirac CP phase $\delta_{\rm CP}$,  the allowed region is approximately  $130^{\circ}\sim220^{\circ}$.
In contrast, for the Majorana CP phase $\alpha_{21}$, the allowed region spans the full phase range.
The effective mass $\langle m_{ee}\rangle$ lies between $10{\rm meV}$ and $17{\rm meV}$,
% within the  $5 \sigma$ C.L., 
which is still below the current experimental bound  from the current KamLAND-Zen experiment \cite{KamLAND-Zen:2016pfg} whose line is indicated by the gray dotted one. 
FIG. \ref{fig:phase_io} shows the correlation between $\alpha_{21}$ and $\delta_{\rm CP}$. 
Unlike the NH case, no clear correlation is observed.

\section {Lepton flavor violation and lepton $g-2$}
Lepton flavor violating (LFV) processes are induced  by the Yukawa interactions $\overline{L_{L_i}}\tilde{\eta} \Psi_{R_j}$
, and the relevant Lagrangian is given  by
\begin{eqnarray}
{\cal L}\supset
(\beta_{\nu}Y^{(6)}_{{\bf1}^{\prime}}\bar{e}
-\alpha_{\nu}Y^{(6)}_{{\bf 2},2}\bar{\mu}+
\alpha_{\nu}Y^{(6)}_{{\bf 2},1}\bar{\tau})\tilde{\eta} \Psi_{R_2}+
(\rho_{\nu}Y^{(6)}_{\bf 1}\bar{e}+
\sigma_{\nu}Y^{(6)}_{{\bf 2},1}\bar{\mu}+
\sigma_{\nu}Y^{(6)}_{{\bf 2},2}\bar{\tau})\tilde{\eta} \Psi_{R_1} .
\end{eqnarray}
The matrix of Yukawa couplings for the interactions described above is given by Eq.(\ref{eq:yD}).
We consider the LFV processes $\ell_i\to\ell_j\gamma$ ${((l_1,l_2,l_3)\equiv (e, \mu, \tau))}$, with $\mu\to e \gamma$ providing the most stringent constraint
on the model parameter.
%{\color{blue}$\ell_1\equiv e$, $\ell_2\equiv \mu$, and $\ell_3\equiv \tau$}.
The branching ratio of the  LFV processes $\ell_i\to\ell_j\gamma$ is given by ~\cite{Baek:2016kud, Lindner:2016bgg}
\begin{align}
&{\rm BR}(\ell_i\to\ell_j\gamma)\approx\frac{48\pi^3\alpha_{e} C_{ij}}{G_F^2 (4\pi)^4}
\left|\sum_{\alpha=1}^{2}y_{D_{j\alpha}} y^\dag_{D_{\alpha i}} F(M_{\alpha},m_{\eta^\pm})\right|^2,\\
%%%
%%%
&F(m_a,m_b)\approx\frac{2 m^6_a+3m^4_am^2_b-6m^2_am^4_b+m^6_b+12m^4_am^2_b\ln\left(\frac{m_b}{m_a}\right)}{12(m^2_a-m^2_b)^4},
\end{align}
where $C_{21}=1$, $C_{31}=0.1784$, $C_{32}=0.1736$, {$\alpha_{e}\approx1/137$ is the fine structure constant}, 
%$\alpha_{em}(m_Z)=1/128.9$, 
and $G_F=1.166\times10^{-5}$ GeV$^{-2}$.
The current experimental upper bounds are given by~\cite{TheMEG:2016wtm, Aubert:2009ag,Renga:2018fpd}
\begin{align}
{\rm BR}(\mu\to e\gamma)\lesssim 4.2\times10^{-13},\quad 
{\rm BR}(\tau\to e\gamma)\lesssim 3.3\times10^{-8},\quad
{\rm BR}(\tau\to\mu\gamma)\lesssim 4.4\times10^{-8}. \label{eq:lfvs-cond}
\end{align}

Our numerical analysis reveals that the best fit values predict the branching ratios for ${\rm Br}(l_i\rightarrow l_j \gamma)$ as follows:
{
\begin{align}
&{\rm Holomorphic\ case(NH\ only)}:\nn\\
&{\rm BR}(\mu\to e\gamma)\lesssim 5.97\times10^{-25},\quad 
{\rm BR}(\tau\to e\gamma)\lesssim 1.55\times10^{-23},\quad
{\rm BR}(\tau\to\mu\gamma)\lesssim 1.19\times10^{-26},\\
%%%%%%%%%
&{\rm Non-holomorphic\ case}:\nn\\
&{\rm NH}:\ {\rm BR}(\mu\to e\gamma)\lesssim 9.60\times10^{-25},\quad 
{\rm BR}(\tau\to e\gamma)\lesssim 2.10\times10^{-23},\quad
{\rm BR}(\tau\to\mu\gamma)\lesssim 2.27\times10^{-26},\\
&{\rm IH}:\ {\rm BR}(\mu\to e\gamma)\lesssim 2.63\times10^{-31},\quad 
{\rm BR}(\tau\to e\gamma)\lesssim 2.77\times10^{-25},\quad
{\rm BR}(\tau\to\mu\gamma)\lesssim 3.14\times10^{-32},
\label{eq:lfvs-ourmodel}
\end{align}
}%
which are significantly below the current experimental bounds stated in Eq.~(\ref{eq:lfvs-cond}).

In a similar way as in the above LFVs, we can also formulate the muon and electron $g-2$ as follows:
\begin{align}
\Delta a_{l}\approx -\frac{m^2_{l}}{(4\pi)^2}
\sum_{\alpha=1}^{2} \left(y_{D_{l\alpha}} y^\dag_{D_{\alpha l}}\right) F({M_\alpha},m_{\eta^\pm}),
\label{g-2}
\end{align}
where $l=e,\mu$.
If $m_{\eta^\pm}=M_{1,2}=100$ GeV, Eq.(\ref{g-2}) is simplified as follows:
\begin{align}
&|\Delta a_{e}|= 6.86\times 10^{-15}\times \left(\sum_{\alpha=1}^{2}y_{D_{e\alpha}} y^\dag_{D_{\alpha e}}\right)\lesssim 8.8\times 10^{-13}, \nn \\
&|\Delta a_{\mu}|= 2.95\times 10^{-10}\times \left(\sum_{\alpha=1}^{2}y_{D_{\mu \alpha}} y^\dag_{D_{\alpha \mu}}\right)\lesssim 2.61 \times 10^{-9},
\label{g-2-const}
\end{align}
where the constraints are set so that the new contributions do not exceed the current discrepancy
between the experimental values and the SM predictions. These constraints can be expressed in
terms of the Yukawa upper bounds as $y_{D}\simeq 4.61 (1.21)$ for electron (muon) $g-2$ , which are weaker than the constraint from
$\mu \rightarrow e \gamma$.

%%%%%%%%%%%%%%%%%%%%%%%%%%%%%%%%%%%%%%%%%%%%%%%%%%%%%%%%%%%%%%%%%%%%%%%%%%%%%
\section{Implications for Invisible Axion and Dark Matter}
%%%%%%%%%%%%%%%%%%%%%%%%%%%%%%%%%%%%%%%%%%%%%%%%%%%%%%%%%%%%%%%%%%%%%%%%%%%%%
%
\subsection{Invisible Axion}
Decomposing the complex scalar field in polar coordinate,
 $\sigma = (v_\sigma + \rho) \exp(i a /v_\sigma)/\sqrt{2}$, 
the field $a$ corresponds to the axion, while the radial mode has a mass term as well as cubic and quartic self-interactions in the scalar potential.
% contains the axion $a$ and the radial mode $\rho$.
In the PQ broken phase, we see the fermion $\Psi_i$ gets massive with mass $M_{i}=\zeta_{i} v_\sigma/\sqrt{2}$ in Eq.~(\ref{eq:mn}).
%$m_{\Psi_i}=Y_{\Psi_i} v_\sigma/\sqrt{2}$.
The PQ symmetry is spontaneously broken at a scale $f_{\text{PQ}} = \langle \sigma \rangle= v_\sigma/\sqrt{2}$, leading to the axion decay constant
\begin{equation}
    f_a = \frac{f_{\text{PQ}}}{N} = \frac{v_{\sigma}}{\sqrt{2} N} \; ,
    \label{eq:axionfa}
\end{equation}
where $N$ is the color anomaly factor.
To be viable, the axion solution to the strong CP problem requires a nonvanishing anomaly factor $N$ to ensure an axion-gluon coupling. 
As expected, in this model, $N$ depends on the multiplicity of the colored fermions $n_\Psi$ and on the $\Psi_L$ PQ charge $\omega$. Since they are $SU(2)$ singlets, we get
\begin{eqnarray}
N &=& 2 \, n_\Psi \, \omega \, (2 n \pm 1) \; T(1,0) \nn \\
   &=& 2 \omega
\label{eq:Nmodel}
\end{eqnarray}
with $T(1,0)$ the Dynkin index of the SU($3)_c$ representation ${\bf 3}\sim (1,0)$, which is 1/2.
By selecting the minimal multiplicity and setting $\omega=\frac{1}{2}$, we find $N=1$, consistent with the original KSVZ axion model
~\cite{Kim:1979if,Shifman:1979if}.

The axion is coupled to a colored fermion $\Psi_i$ via an axial-vector current, 
\begin{eqnarray}
J_a^5=\frac{\omega}{2N}\left(\frac{\partial_{\mu} a}{f_a}\right)\sum_{i=1}^2\overline{\Psi_i}\gamma^{\mu}\gamma^5 \Psi_i.
\end{eqnarray}
The derivative coupling of the axion is just a current coupling to ABJ anomaly for $\Psi_i$ and we can obtain
a Lagragian term with the coupling of $a$ to the dual field strength tensor of the gluons $G_{\mu\nu}$,
\begin{align}
{\cal L}=N\frac{a}{f_a}\frac{g^2_s}{32\pi^2}G_{\mu\nu}\tilde{G}_{\mu \nu},
\end{align}
where $\tilde{G}_{\mu \nu}\equiv \epsilon_{\mu \nu \rho \sigma}G^{\rho\sigma}/2$.

The QCD axion mass derived from a non-perturbative potential at next-to-leading order is given by~\cite{GrillidiCortona:2015jxo}
\begin{equation}
    m_a = 5.70(7) \left(\frac{10^{12} \text{GeV}}{f_a}\right) \mu \text{eV} \; .
    \label{eq:axionmass}
\end{equation}
This relation between $m_a$ and $f_a$ is a model-independent prediction of the QCD axion if the only explicit breaking of the PQ symmetry is by nonperturbative QCD effects. 
%The typical value of $f_{PQ}$ is around $10^{12}$ GeV, so that axions are responsible for the observed DM relic density \cite{Preskill:1982cy,Abbott:1982af,Dine:1982ah}.
In subsec.\ref{axion-DM}, we will briefly demonstrate how the DM relic density constrains $f_a$, leading to an estimation of the axion mass $m_a$.

Axions have been extensively searched for in numerous experiments, primarily through the axion-photon coupling  as described in the effective Lagrangian given by
\begin{eqnarray}
{\cal L}=\frac{g_{a\gamma\gamma}}{4}a F_{\mu\nu}\tilde{F}^{\mu\nu},
\end{eqnarray}
where $F_{\mu\nu}$ represents  the photon field strength and $\tilde{F}_{\mu\nu}$ denotes its dual.
Consequently, the strength of the axion-photon coupling is crucial for experimental axion searches.
In this model, since the only chiral fermions with non-zero PQ charges are $\Psi_i$ and they are $SU(2)_L$ singlets,
the ratio of the electromagnetic and color anomaly factors, $E/N$ to be zero, which results in the axion-to-photon coupling $g_{a\gamma\gamma}$ to be  
{
\begin{eqnarray}
g_{a\gamma\gamma}&=&\frac{\alpha_e}{2\pi f_a}\left[ \frac{E}{N}-1.92(4)\right]
= -1.92(4) \frac{\alpha_e}{2\pi f_a}. \label{eq:agg}
\end{eqnarray}
%where $\alpha_e$ is the fine structure constant, approximately  $1/137$.
}
Using the value of $f_a$ constrained by the relic density discussed in Subsec.\ref{axion-DM}, we can determine $g_{a\gamma\gamma} (\sim -8\times 10^{-15}~\mbox{GeV}^{-1})$.
This result can be interpreted as a lower bound to $g_{a\gamma\gamma}$.
Axions can couple to nucleons as $c_N (\partial_{\mu} a/2 f_a) \bar{N} \gamma^{\mu}\gamma_5 N$.
For KSVZ-like models, the axion couplings to proton and neutron are given respectively by
$c_p=-0.47(3)$ and $c_n=-0.02(3)$ \cite{GrillidiCortona:2015jxo}.
By fixing $f_a$ as above, we can also determine the coupling strength $c_N/(2 f_a)$.
The results of the axion parameter presented in this work are consistent with the current 
indirect astrophysical and cosmological observations, as well as constraints from laboratory searches~(for reviews see Refs.~\cite{DiLuzio:2020wdo,Adams:2022pbo}).
%, constrain the axion parameter space due to its couplings to photons, nucleons and electrons.

%%%%%%%%%%%%%%%%%%%%%%%%%%%%%%%%%%%%%%%%%%%%%%%%%%%%%%%%%%%%%%%%%%%%%%%%%%%%%
\subsection{ Axion Dark Matter } \label{axion-DM}
%%%%%%%%%%%%%%%%%%%%%%%%%%%%%%%%%%%%%%%%%%%%%%%%%%%%%%%%%%%%%%%%%%%%%%%%%%%%%
%
Axions are a promising dark matter (DM) candidate due to their naturally light mass, weak coupling with ordinary matter, cosmological stability, and potential for nonthermal production in the early Universe.
Axion DM can be produced through the misalignment mechanism ~\cite{Preskill:1982cy,Abbott:1982af,Dine:1982ah}, which results in
a relic density given by~\cite{DiLuzio:2020wdo}
\begin{equation}
   \Omega_{a, mis} h^2 \simeq  \Omega_\text{CDM} h^2 \langle \theta_0^2 \rangle \left(\frac{f_a}{ 2\times 10^{11} \ \text{GeV}} \right)^{\frac{7}{6}} \; ,
   \label{eq:relica}
\end{equation}
where the free parameter $\langle \theta_0^2\rangle$ denotes the initial average misalignment angle squared, which
depends on whether PQ symmetry is broken before or after inflation. In the latter case,  $\langle \theta_0^2\rangle$ is estimated to be  $2.15^2$  based on the periodic axion potential including anharmonicities \cite{Visinelli:2009zm,DiLuzio:2020wdo}.
The observed cold DM~(CDM) relic abundance obtained by Planck is
$\Omega_{\text{CDM}} h^2 = 0.1200 \pm 0.0012$~\cite{Planck:2018vyg}. 
Then, assuming the axion generated through the misalignment constitutes $100\%$ of CDM, we get $f_a=2.7\times 10^{11}$ GeV from Eq.~(\ref{eq:relica}).
Consequently, the axion mass is estimated to be $ 21.1 \mu$eV from Eq.~(\ref{eq:axionmass}).
{If the PQ symmetry is broken after inflation, the formation and decay of cosmic strings and domain walls also contribute significantly to the relic axion abundance. 
Incorporating these contributions, the total relic density is estimated to be \cite{Kawasaki:2014sqa},
\begin{equation}
   \Omega_{a, tot} h^2 =(1.6\pm0.4) \times 10^{-2}\times \left(\frac{f_a}{10^{10} \ \text{GeV}}\right)^{(6+n)/(4+n)},
\end{equation}
where QCD scale is taken to be 400 MeV and $n=6.68$.
The result implies $\Omega_{a, tot} h^2\simeq 3 \Omega_{a, mis} h^2$.
This additional contributions effectively reduce the value of the axion decay constant $f_a$ required to match the observed dark matter density, leading to a correspondingly higher axion mass.
Assuming that $ \Omega_{a, tot} h^2$ accounts for $100\%$ of CDM,  the value of $f_a$ is shifted to $(4.6-7.2) \times 10^{10}$ GeV, with the corresponding axion mass in the range $(0.8-1.3)\times 10^{-4}$ eV.
For these values of $f_a$, the axion-to-photon coupling $g_{a\gamma\gamma}$ is predicted to be approximately $-(3\sim 5) \times 10^{-14}~\mbox{GeV}^{-1}$ as given by Eq.~(\ref{eq:agg}).
These parameter values fall within the permissible region of the hadronic parameter space, as discussed in  \cite{DiLuzio:2020wdo}. }
%%%%%%%%%%%%%%%%%%%%%%%
%Recent numerical simulations indicate that these topological defects can enhance the relic density by a factor of O(1–10) [see, e.g., Ref. [Kawasaki et al., arXiv:1412.0789]]. This additional contribution would imply a lower value of the axion decay constant $f_a$ needed to match the observed dark matter density and correspondingly a higher axion mass.}
%
%Then, assuming the axion constitutes $100\%$ of CDM, we get $f_a=2.7\times 10^{11}$ GeV from Eq.~(\ref{eq:relica}).
%Consequently, the axion mass is estimated to be $ 21.1 \mu$eV from Eq.~(\ref{eq:axionmass}).
%We note that this estimation is simply made without accounting for the potential effects from topological defects  that may have been formed fat inflation.
%For this value of $f_a$, the axion-to-photon coupling $g_{a\gamma\gamma}$ is predicted to be $-8\times 10^{-15}$ from Eq.~(\ref{eq:agg}).
%These parameter values fall within the permissible region of the hadronic parameter space, as discussed in  \cite{DiLuzio:2020wdo}

In the pre-inflationary scenario, we assume that the masses of exotic fermions and scalars exceed the reheating temperature of the Universe, i.e. 
$(M_{1,2}, {m_{\chi_1}}, \ m_{S_{1,2,3}})~ \gtrsim T_\text{RH} ~\gtrsim 4.7 \ \text{MeV}$,
%$m_{\Psi,\eta,\chi} > T_\text{RH} \gtrsim 4.7 \ \text{MeV}$, 
where the lower limit on $T_\text{RH}$ is set by Big Bang nucleosynthesis~\cite{deSalas:2015glj}.
As a result, the abundance of stable baryonic or charged relics,  as well as topological defects, would be erased during inflation.
In this scenario, the axion imprints itself on primordial fluctuations, reflected in the cosmic microwave background anisotropies and large-scale structure. The isocurvature fluctuations generated are constrained by cosmic microwave background data~\cite{Beltran:2006sq}, resulting in an upper bound on the inflationary scale $H_I$~\cite{DiLuzio:2016sbl}:
\begin{equation}
   H_I 
\lesssim 
 \frac{0.9\times 10^7}{\Omega_a h^2/\Omega_\text{CDM} h^2} \left(\frac{\langle \theta_0\rangle}{\pi} \frac{f_a}{ 10^{11} \ \text{GeV}} \right) \; \text{GeV} \; \lesssim
2.0\times 10^{7}\; \text{GeV},
   \label{eq:Iso}
\end{equation}
where the numerical value for the upper bound is derived, assuming axions constitute all DM,  from the threshold value of $\langle \theta_0 \rangle$ as adopted
from the results in \cite{Borsanyi:2016ksw}.

Axion DM can also be produced through kinetic misalignment mechanism \cite{Co:2019jts}.
In the kinetic misalignment mechanism, the axion field has a significant initial velocity $\dot{a}_0$ or kinetic energy.
Thus, the axion starts its evolution with both a misalignment angle $\theta_0$ and a non-zero velocity $\dot{a}_0$.
Once the axion field starts oscillating, it behaves like CDM
%cold dark matter
 with a relic density that can be estimated.
In the kinetic misalignment scenario, the relic density can be modified due to the initial kinetic energy. 
If the kinetic energy density  at a temperature is larger than the potential barrier, kinetic misalignment occurs and the axion
oscillations are delayed until the kinetic energy is below that of the potential \cite{Co:2019jts}.
By considering the effects of inflation on kinetic misalignment and parametric resonance, the authors in \cite{Co:2020dya} obtained
$4\times 10^8 < f_a/[{\rm GeV}] < 10^{11}$ for consistent production of axion DM.
For the result of $f_a$, the axion mass is estimated to be $57~\mu{\rm eV}\lesssim m_a \lesssim14.25~ {\rm meV}$  from Eq.~(\ref{eq:axionmass}), and  the axion-to-photon coupling is predicted to be $-2.2\times 10^{-14}\lesssim g_{a\gamma\gamma} \times[{\rm GeV}] \lesssim -5.6 \times 10^{-12}$  from Eq.~(\ref{eq:agg}).

Future helioscope and haloscope experiments, including IAXO, ADMX, MADMAX, and CAPP, are expected to probe unexplored regions of the parameter space for the axion-photon coupling $|g_{a\gamma\gamma}|$ and the axion mass $m_a$
significantly advancing our ability to detect or constrain axion DM models.
As is known, the region of $|g_{a\gamma\gamma}|$ around $10^{-12}\sim 10^{-11}~{\rm GeV}^{-1}$ will be probed
at IAXO \cite{Armengaud:2014gea}. The full ladscape of QCD axion models for masses $1~\mu{\rm eV}\lesssim m_a \lesssim 100~\mu{\rm eV}$ and
$50~\mu{\rm eV}\lesssim m_a \lesssim 120~\mu{\rm eV}$
will be probed at ADMX \cite{Stern:2016bbw} and MADMAX \cite{Beurthey:2020yuq}, respectively.
In addition, CAPP aims to achieve a sensitivity to $|g_{a\gamma\gamma}|$ as low as $10^{-16} ~{\rm GeV}^{-1}$, with a focus
on the axion mass ranging from approximately $1 \mu{\rm eV}$ to $10 \mu{\rm eV}$, and extending their search to  higher mass ranges up to $30 \mu{\rm eV}$ \cite{Semertzidis:2019gkj}.
{Therefore, the predicted values of $g_{a\gamma\gamma}$ and $m_a$ in our model may fall within the reach of upcoming experimental searches.}

{Axions can also be probed indirectly via astrophysical and cosmological observations, particularly through the decay or conversion of
 axion dark matter into photons under strong magnetic fields or in dense environments \cite{DiLuzio:2020wdo,OHare:2024nmr}. 
These indirect searches cover a wide mass range, from ultralight axions ($m_a \lesssim 10^{-14}~\mathrm{eV}$) to masses above the MeV scale. In the context of our model, we focus on the region $m_a \sim {\cal O}(10\text{--}100)~\mu\mathrm{eV}$, which is predicted with axion dark matter.
%well-motivated theoretically and relevant to the predictions above.
%
Axion to photon conversion in the magnetospheres of neutron stars, particularly near the Galactic Center, has been probed by the Breakthrough Listen project using radio surveys in the C-band. These studies constrain the coupling to $g_{a\gamma\gamma} \lesssim 10^{-11}~\mathrm{GeV}^{-1}$ for axion masses between $15~\mu\mathrm{eV}$ and $35~\mu\mathrm{eV}$~\cite{Foster:2022fxn}. 
Similarly, radio data from the Bullet Cluster (1E 0657-55.8) yield an upper bound on the axion-photon coupling of $g_{a\gamma\gamma} \sim 10^{-12} \text{--} 10^{-11}~\mathrm{GeV}^{-1}$ for $m_a \sim 10\text{--}30~\mu\mathrm{eV}$~\cite{Chan:2021gjl}.
Axions may also influence stellar evolution by enhancing cooling processes such as the Primakoff effect. Observations of horizontal branch stars in globular clusters place a strong upper limit of $g_{a\gamma\gamma} \lesssim 6.6 \times 10^{-11}~\mathrm{GeV}^{-1}$~\cite{Ayala:2014pea}. This is consistent with bounds from the CAST helioscope, which similarly constrains $g_{a\gamma\gamma}$ for $m_a \lesssim 0.02~\mathrm{eV}$~\cite{CAST:2017uph}.
In addition, recent constraints derived from SN 1987A gamma ray data have improved limits on axion-photon couplings by accounting for axion to photon conversion in the magnetic field of the supernova progenitor itself. These analyses yield limits of
$g_{a\gamma\gamma} \lesssim \text{few} \times 10^{-11}~\mathrm{GeV}^{-1} \quad \text{for} \quad 10^{-6} \lesssim m_a \lesssim 10^{-3}~\mathrm{eV},$
representing an order-of-magnitude improvement over earlier bounds based solely on conversion in the Galactic magnetic field~\cite{Manzari:2024jns}. In our model, the predicted coupling lies well below those bounds.}

%%%%%%%%%%%%%%%%%%%%%%%%%%%%%%%%%%%%%%%%%%%%%%%%%%%%%%%%%%%%%%%%%%%%%%%%%%%%%
\section{Conclusions}
%%%%%%%%%%%%%%%%%%%%%%%%%%%%%%%%%%%%%%%%%%%%%%%%%%%%%%%%%%%%%%%%%%%%%%%%%%%%%
%
In this paper, we have constructed a predictive axion model with modular $S_3$ symmetry to address a connection among seemingly unrelated issues: tiny neutrino masses and mixing, dark matter and the strong CP problem. This was achieved within a novel class of KSVZ axion schemes, containing exotic colored fermions and scalars which act as neutrino-mass mediators at the two-loop level.
The radiative generation of neutrino masses at the two-loop level accounts for their smallness, while the neutrino mixing arises from the modular
$S_3$ symmetry that we consider. 
{We have demonstrated that the combination of PQ symmetry and $Z_2$ symmetry, integrated with modular $S_3$ and the appropriate assignment of modular weights to the fields, can effectively constrain the forms of Yukawa interactions and the scalar potential.
This framework is applicable to both SUSY (holomorphic) and non-SUSY (non-holomorphic) scenarios, and ensures the robustness of the resulting predictions. 
We have presented the numerical results illustrating how the predictions and physical  implications for neutrino masses and mixing differ between the holomorphic and non-holomorphic cases.
In the holomorphic case, only NH is favored by the neutrino oscillation data , whereas  in the non-holomorphic case, both NH and IH are consistent with experimental data.
Under reasonable conditions, we have ignored the subleading corrections such as  those from higher-dimensional operators, RG running, and SUSY-breaking effects, and focused on the leading-order predictions.}
%This model can also be embedded in a SUSY framework to protect the holomorphic modular forms. In the decoupling limit of SUSY with a large SUSY-breaking scale, the physical predictions for neutrino masses and mixing remain nearly identical, highlighting the framework's flexibility and robustness in both SUSY and non-SUSY contexts.}
This scenario also predicts lepton flavor violations with a small rate and neutrinoless double beta decay, with the amplitude proportional to the effective neutrino mass being well below the current experimental bounds. Since the lightest neutrino mass vanishes, the sum of neutrino masses is written in terms of two experimental values $\Delta m^2_{\rm atm}$ and $\ \Delta m^2_{\rm sol}$. That leads $\sum D_\nu\simeq 50(100)$ meV in the NH(IH) case which is in good agreement with the recent combined DESI and CMB data (cosmological observations). 
%
%Due to the potential dangers posed by colored relics, we have explored the scenario where the PQ symmetry is broken before inflation in the context of axion dark matter. With an initial misalignment angle of $\theta_0 \sim \mathcal{O}(1)$, axions could account for the entire CDM budget if the axion decay constant $f_a$ is approximately $2.7 \times 10^{11}$ GeV.
%This parameter region is currently being investigated by haloscope experiments.In our scenario, the axion DM is correlated with neutrinos in a specific manner, where their interactions are determined by the $S_3$ symmetry that is known as the minimal group in non-Abelian discrete flavor symmetries.

{Furthermore, we have demonstrated that the axion emerging from our framework not only resolves the strong CP problem but also serves as a viable dark matter candidate across multiple cosmological scenarios, including post-inflationary string-induced contributions and kinetic misalignment production. The resulting axion decay constant and mass, constrained by the relic density, predict an axion-photon coupling in the range 
$|g_{a\gamma\gamma}|\sim 10^{-15}\text{--} 10^{-12}~{\rm GeV}^{-1}$, which lies within the reach of upcoming experimental searches such as IAXO, ADMX, MADMAX, and CAPP. Moreover, these predictions remain consistent with existing astrophysical and cosmological bounds. Thus, our model offers a highly testable and unified explanation of neutrino masses, dark matter, and the strong CP problem—linking axion phenomenology and lepton flavor structure through a common modular origin.}

%%%%%%%%%%%%%%%%%%%%%%%%%%%%%%%%%%%%%%%%%%%%%%%%%%%%%%%%%%%%%%%%%%%%%%
\begin{acknowledgments}
 SKK was supported by the National Research Foundation of Korea (NRF) grant funded by the Korea government (MSIT) (No.2023R1A2C1006091).
 HO is supported by Zhongyuan Talent (Talent Recruitment Series) Foreign Experts Project. 
\end{acknowledgments}
%%%%%%%%%%%%%%%%%%%%%%%%%%

\appendix
{
\section{Two-loop integral}
\label{neut-int}
Let us briefly show how the two-loop integrals in Eqs.~(\ref{eq:neut-int1} and \ref{eq:neut-int2}) are derived.
The relevant two-loop integral takes the form
\begin{align}
\int\frac{d^4k_2}{(2\pi)^4}\int\frac{d^4k_1}{(2\pi)^4}
\frac{{/\hspace{-2mm}k_1/\hspace{-2mm}k_2}}{[k_1^2 -M^2_{\ell}][k_2^2 -M^2_{m}][k_1^2 -m^2_{S_i}][k_2^2 -m^2_{S_j}][(k_1+k_2)^2 -m^2_{\chi_1}]}.
\end{align}
We begin by calulating the $k_1$ integral:
\begin{align}
& \int\frac{d^4k_1}{(2\pi)^4} 
\frac{{/\hspace{-2mm}k_1}}{[k_1^2 -M^2_{\ell}] ][k_1^2 -m^2_{S_i}] [(k_1+k_2)^2 -m^2_{\chi_1}]}
=\frac{i}{(4\pi)^2}\int \frac{[dx]_3}{y-1} \frac{{/\hspace{-2mm}k_2}}{k^2_2 + D_1^2},\label{eq:k1int}\\
%%%
& (D_{\ell}^{i1})^2=\frac1{y(y-1)}(x m^2_{S_i} + y M^2_\ell + z m^2_{\chi_1}) ,  \label{eq:a3}
\end{align}
where $[dx]_3\equiv dxdydz\delta(1-x-y-z)$.
Substituting Eq.~(\ref{eq:k1int}) into the remaining $k_2$ integral, we obtain
\begin{align}
& \int\frac{d^4k_2}{(2\pi)^4}
\frac{k_2^2}{[k_2^2 +  (D_{\ell}^{i1})^2] [k_2^2 -M^2_{m}] [k_2^2 -m^2_{S_j}] } \nn\\
&= - \frac{i}{(4\pi)^2} 
% \int[dx']_3  \ln(x' \zeta^{i1}_{\ell m} + y' + z' \xi^j_m),
\left[ 
\frac{(\zeta^{i1}_{\ell m} )^2\ln[\zeta^{i1}_{\ell m}] }{(-1+\zeta^{i1}_{\ell m} )(\zeta^{i1}_{\ell m} - \xi^j_m)}
-
\frac{(\xi^j_m)^2\ln[\xi^j_m] }{(-1+\xi^j_m )(\zeta^{i1}_{\ell m} - \xi^j_m)}
  \right],  \label{eq:a4} \\
%%%
\zeta^{i1}_{\ell m} &\equiv - \frac{ (D_{\ell}^{i1})^2}{M^2_m}, \quad  \xi^j_m \equiv \frac{m^2_{S_j}}{M^2_m}. \label{eq:a5}
\end{align}
Notice here that the logarithmic divergence vanishes when we sum over the indices $i,j$ due to the unitarity of the mixing matrix of $U$ in Eq.~(\ref{eq:umat}).
Using Eqs.~(\ref{eq:k1int})-\ref{eq:a5}), we can obtain the expressions given in Eqs.~(\ref{eq:neut-int1} and \ref{eq:neut-int2}).
}

\section{$S_3$ symmetry}
\label{s3_rules}
Due to its three dimensional representation can be decomposed as $\bm{3}_S = \bm{2}\oplus \bm{1}_S$ or
$\bm{3}_A = \bm{2}\oplus \bm{1}_A$ .
{
The multiplication rules among doublets and singlets are given by~\cite{Novichkov:2019sqv} as follows;
\begin{align}
& \begin{pmatrix}
   a_1 \\
    a_2
\end{pmatrix}_{\bm{2}} \otimes \begin{pmatrix}
   b_1 \\
    b_2
\end{pmatrix}_{\bm{2}}
=(a_1 b_1+a_2 b_2)_{\bm{1}_S}\oplus
(a_1 b_2-a_2 b_1)_{\bm{1}_A}\oplus
\begin{pmatrix}
   a_1b_2 + a_2 b_1 \\
   a_1 b_1 - a_2 b_2 
\end{pmatrix}_{\bm{2}},\\
%%%
&\begin{pmatrix}
   a_1 \\
    a_2
\end{pmatrix}_{\bm{2}} \otimes (y')_{\bm 1_S} 
=
\begin{pmatrix}
   a_1 y' \\
    a_2y'
\end{pmatrix}_{\bm{2}} ,\quad
\begin{pmatrix}
   a_1 \\
    a_2
\end{pmatrix}_{\bm{2}} \otimes (y')_{\bm 1_A} 
=
\begin{pmatrix}
-   a_2 y' \\
    a_1 y'
\end{pmatrix}_{\bm{2}} ,\\
%%%
 & (x)_{\bm 1_S}\otimes (y)_{\bm 1_A} = (xy)_{\bm 1_A}, 
 \quad
(x)_{\bm 1_A}\otimes (y)_{\bm 1_A} = (xy)_{\bm 1_S}.
\end{align}
}

 %%%%%%%%%%%%%%%%%%%%%%%%%%%%%%%%%%%%%%%%%%%%%%%%%%%%%%%%%%%
%
\section{Modular $S_3$ symmetry} 
\label{sec:realization}
The modular group $\bar\Gamma$ is the group of linear fractional transformation
$\gamma$ acting on the modulus  $\tau$, 
belonging to the upper-half complex plane as:
\begin{equation}\label{eq:tau-SL2Z}
\tau \longrightarrow \gamma\tau= \frac{a\tau + b}{c \tau + d}\ ,~~
{\rm where}~~ a,b,c,d \in \mathbb{Z}~~ {\rm and }~~ ad-bc=1, 
~~ {\rm Im} [\tau]>0 ~ ,
\end{equation}
 which is isomorphic to  $PSL(2,\mathbb{Z})=SL(2,\mathbb{Z})/\{I,-I\}$ transformation.
This modular transformation is generated by $S$ and $T$, 
\begin{eqnarray}
S:\tau \longrightarrow -\frac{1}{\tau}\ , \qquad\qquad
T:\tau \longrightarrow \tau + 1\ ,
\end{eqnarray}
which satisfy the following algebraic relations, 
\begin{equation}
S^2 =T^2 = (ST)^3 =\mathbb{1}\ .
\end{equation}

 We introduce the series of groups $\Gamma(N)~ (N=1,2,3,\dots)$ defined by
 \begin{align}
 \begin{aligned}
 \Gamma(N)= \left \{ 
 \begin{pmatrix}
 a & b  \\
 c & d  
 \end{pmatrix} \in SL(2,\mathbb{Z})~ ,
 ~~
 \begin{pmatrix}
  a & b  \\
 c & d  
 \end{pmatrix} =
  \begin{pmatrix}
  1 & 0  \\
  0 & 1  
  \end{pmatrix} ~~({\rm mod} N) \right \}
 \end{aligned} .
 \end{align}
 For $N=2$, we define $\bar\Gamma(2)\equiv \Gamma(2)/\{I,-I\}$.
Since the element $-I$ does not belong to $\Gamma(N)$
  for $N>2$, we have $\bar\Gamma(N)= \Gamma(N)$,
  which are infinite normal subgroup of $\bar \Gamma$, called principal congruence subgroups.
   The quotient groups defined as
   $\Gamma_N\equiv \bar \Gamma/\bar \Gamma(N)$
  are  finite modular groups.
In this finite groups $\Gamma_N$, $T^N=\mathbb{1}$  is imposed.
 The  groups $\Gamma_N$ with $N=2,3,4,5$ are isomorphic to
$S_3$, $A_4$, $S_4$ and $A_5$, respectively \cite{deAdelhartToorop:2011re}.

Modular forms of  level $N$ are 
holomorphic functions $f(\tau)$  transforming under the action of $\Gamma(N)$ as:
\begin{equation}
f(\gamma\tau)= (c\tau+d)^k f(\tau)~, ~~ \gamma \in \Gamma(N)~ ,
\end{equation}
where $k$ is the so-called as the  modular weight.

We discuss the modular symmetric theory without supersymmetry. 
In this paper, we fix the modular $S_3$ ($N=2$) group. 
Under the modular transformation of Eq.(\ref{eq:tau-SL2Z}), fields $\phi^{(I)}$ 
transform as 
\begin{equation}
\phi^{(I)} \to (c\tau+d)^{-k_I}\rho^{(I)}(\gamma)\phi^{(I)},
\end{equation}
where  $-k_I$ is the modular weight and $\rho^{(I)}(\gamma)$ denotes a unitary representation matrix of $\gamma\in\Gamma(2)$.

The kinetic terms of their scalar fields are {derived from the K$\ddot{\rm a}$hler potential invariant under the modular symmetry
up to K$\ddot{\rm a}$hler transformation given as,
\begin{eqnarray}
K=-h{\rm log}(-i\tau+i\bar{\tau})+\sum_I (-i\tau+i\bar{\tau})^{-k_I}|\phi^{(I)}|^2,
\end{eqnarray}
where $h$ is a positive constant and  the scalar $\phi^{(I)}$ has modular weight $k_I$.
Then, the  K$\ddot{\rm a}$hler metric  gives the kinetic terms of $\phi$ as
\begin{eqnarray}
{\cal L}_{\rm kin}=K_{\phi\bar{\phi}}\partial_{\mu}\bar{\phi}^{(I)} \partial^{\mu} \phi^{(I)}.
\end{eqnarray}
The final form of the kinetic terms of scalar field is given by}
\begin{equation}
\sum_I\frac{|\partial_\mu\phi^{(I)}|^2}{(-i\tau+i\bar{\tau})^{k_I}} ,~
\label{kinetic}
\end{equation}
which is invariant under the modular transformation.
Also, the Lagrangian such as $|\phi^{I}|^2$ should be invariant under the modular symmetry.
%%%
%

%\bibliographystyle{utphys}
%\bibliography{ref}

\begin{thebibliography}{99}
\bibitem{Kajita:2016cak}
T.~Kajita, %``Nobel Lecture: Discovery of atmospheric neutrino oscillations,''
Rev. Mod. Phys. \textbf{88}, no.3, 030501 (2016)
doi:10.1103/RevModPhys.88.030501
%295 citations counted in INSPIRE as of 14 Apr 2025

\bibitem{McDonald:2016ixn}
A.~B. McDonald,
%``Nobel Lecture: The Sudbury Neutrino Observatory: Observation of flavor change for solar neutrinos,''
Rev. Mod. Phys. \textbf{88}, no.3, 030502 (2016)
doi:10.1103/RevModPhys.88.030502
%281 citations counted in INSPIRE as of 14 Apr 2025

\bibitem{Bertone:2004pz}
G.~Bertone, D.~Hooper, and J.~Silk, 
%``Particle dark matter: Evidence, candidates and constraints,''
Phys. Rept. \textbf{405}, 279-390 (2005)
doi:10.1016/j.physrep.2004.08.031
[arXiv:hep-ph/0404175 [hep-ph]].
%5584 citations counted in INSPIRE as of 14 Apr 2025

\bibitem{Planck:2018vyg}
N.~Aghanim \textit{et al.} [Planck],
%``Planck 2018 results. VI. Cosmological parameters,''
Astron. Astrophys. \textbf{641}, A6 (2020)
[erratum: Astron. Astrophys. \textbf{652}, C4 (2021)]
doi:10.1051/0004-6361/201833910
[arXiv:1807.06209 [astro-ph.CO]].

\bibitem{PhysRevLett.38.1440}
R.~D.~Peccei and H.~R.~Quinn,
%``CP Conservation in the Presence of Instantons,''
Phys. Rev. Lett. \textbf{38}, 1440-1443 (1977)
doi:10.1103/PhysRevLett.38.1440
%8495 citations counted in INSPIRE as of 14 Apr 2025


\bibitem{PhysRevD.16.1791}
R.~D.~Peccei and H.~R.~Quinn,
%``Constraints Imposed by CP Conservation in the Presence of Instantons,''
Phys. Rev. D \textbf{16}, 1791-1797 (1977)
doi:10.1103/PhysRevD.16.1791
%4392 citations counted in INSPIRE as of 14 Apr 2025

\bibitem{PhysRevLett.40.223}
S.~Weinberg,
%``A New Light Boson?,''
Phys. Rev. Lett. \textbf{40}, 223-226 (1978)
doi:10.1103/PhysRevLett.40.223
%6068 citations counted in INSPIRE as of 14 Apr 2025

\bibitem{PhysRevLett.40.279}
F.~Wilczek,
%``Problem of Strong  $P$  and  $T$  Invariance in the Presence of Instantons,''
Phys. Rev. Lett. \textbf{40}, 279-282 (1978)
doi:10.1103/PhysRevLett.40.279
%5813 citations counted in INSPIRE as of 14 Apr 2025

\bibitem{Kim:1979if}
J.~E.~Kim,
%``Weak Interaction Singlet and Strong CP Invariance,''
Phys. Rev. Lett. \textbf{43}, 103 (1979)
doi:10.1103/PhysRevLett.43.103
%3294 citations counted in INSPIRE as of 14 Apr 2025

\bibitem{Shifman:1979if}
M.~A.~Shifman, A.~I.~Vainshtein and V.~I.~Zakharov,
%``Can Confinement Ensure Natural CP Invariance of Strong Interactions?,''
Nucl. Phys. B \textbf{166}, 493-506 (1980)
doi:10.1016/0550-3213(80)90209-6
%2966 citations counted in INSPIRE as of 14 Apr 2025

\bibitem{Zhitnitsky:1980tq}
A.~R.~Zhitnitsky,
%``On Possible Suppression of the Axion Hadron Interactions. (In Russian),''
Sov. J. Nucl. Phys. \textbf{31}, 260 (1980)
%2405 citations counted in INSPIRE as of 14 Apr 2025

\bibitem{Dine:1981rt}
M.~Dine, W.~Fischler and M.~Srednicki,
%``A Simple Solution to the Strong CP Problem with a Harmless Axion,''
Phys. Lett. B \textbf{104}, 199-202 (1981)
doi:10.1016/0370-2693(81)90590-6
%3712 citations counted in INSPIRE as of 14 Apr 2025

\bibitem{Tao:1996vb}
Z.~j.~Tao,
%``Radiative seesaw mechanism at weak scale,''
Phys. Rev. D \textbf{54}, 5693-5697 (1996)
doi:10.1103/PhysRevD.54.5693
[arXiv:hep-ph/9603309 [hep-ph]].
%56 citations counted in INSPIRE as of 14 Apr 2025

\bibitem{Ma:2006km}
E.~Ma,
%``Verifiable radiative seesaw mechanism of neutrino mass and dark matter,''
Phys. Rev. D \textbf{73}, 077301 (2006)
doi:10.1103/PhysRevD.73.077301
[arXiv:hep-ph/0601225 [hep-ph]].
%1589 citations counted in INSPIRE as of 14 Apr 2025


\bibitem{Hirsch:2013ola}
M.~Hirsch, R.~A.~Lineros, S.~Morisi, J.~Palacio, N.~Rojas and J.~W.~F.~Valle,
%``WIMP dark matter as radiative neutrino mass messenger,''
JHEP \textbf{10}, 149 (2013)
doi:10.1007/JHEP10(2013)149
[arXiv:1307.8134 [hep-ph]].
%99 citations counted in INSPIRE as of 14 Apr 2025


\bibitem{Preskill:1982cy}
J.~Preskill, M.~B.~Wise and F.~Wilczek,
%``Cosmology of the Invisible Axion,''
Phys. Lett. B \textbf{120}, 127-132 (1983)
doi:10.1016/0370-2693(83)90637-8
%3706 citations counted in INSPIRE as of 14 Apr 2025

\bibitem{Abbott:1982af}
L.~F.~Abbott and P.~Sikivie,
%``A Cosmological Bound on the Invisible Axion,''
Phys. Lett. B \textbf{120}, 133-136 (1983)
doi:10.1016/0370-2693(83)90638-X
%3428 citations counted in INSPIRE as of 14 Apr 2025

\bibitem{Dine:1982ah}
M.~Dine and W.~Fischler,
%``The Not So Harmless Axion,''
Phys. Lett. B \textbf{120}, 137-141 (1983)
doi:10.1016/0370-2693(83)90639-1
%3397 citations counted in INSPIRE as of 14 Apr 2025

\bibitem{DiLuzio:2020wdo}
L.~Di Luzio, M.~Giannotti, E.~Nardi and L.~Visinelli,
%``The landscape of QCD axion models,''
Phys. Rept. \textbf{870}, 1-117 (2020)
doi:10.1016/j.physrep.2020.06.002
[arXiv:2003.01100 [hep-ph]].
%828 citations counted in INSPIRE as of 14 Apr 2025

\bibitem{Adams:2022pbo}
C.~B.~Adams, N.~Aggarwal, A.~Agrawal, R.~Balafendiev, C.~Bartram, M.~Baryakhtar, H.~Bekker, P.~Belov, K.~K.~Berggren and A.~Berlin, \textit{et al.}
%``Axion Dark Matter,''
[arXiv:2203.14923 [hep-ex]].
%175 citations counted in INSPIRE as of 12 Aug 2024

\bibitem{Arcadi:2017kky}
G.~Arcadi, M.~Dutra, P.~Ghosh, M.~Lindner, Y.~Mambrini, M.~Pierre, S.~Profumo and F.~S.~Queiroz,
%``The waning of the WIMP? A review of models, searches, and constraints,''
Eur. Phys. J. C \textbf{78}, no.3, 203 (2018)
doi:10.1140/epjc/s10052-018-5662-y
[arXiv:1703.07364 [hep-ph]].
%923 citations counted in INSPIRE as of 14 Apr 2025

\bibitem{Mohapatra:1982tc}
R.~N.~Mohapatra and G.~Senjanovic,
%``The Superlight Axion and Neutrino Masses,''
Z. Phys. C \textbf{17}, 53-56 (1983)
doi:10.1007/BF01577819
%125 citations counted in INSPIRE as of 24 Aug 2024

\bibitem{Langacker:1986rj}
P.~Langacker, R.~D.~Peccei and T.~Yanagida,
%``Invisible Axions and Light Neutrinos: Are They Connected?,''
Mod. Phys. Lett. A \textbf{1}, 541 (1986)
doi:10.1142/S0217732386000683
%249 citations counted in INSPIRE as of 24 Aug 2024

\bibitem{Shin:1987xc}
M.~Shin,
%``Light Neutrino Masses and Strong {CP} Problem,''
Phys. Rev. Lett. \textbf{59}, 2515 (1987)
[erratum: Phys. Rev. Lett. \textbf{60}, 383 (1988)]
doi:10.1103/PhysRevLett.59.2515
%64 citations counted in INSPIRE as of 24 Aug 2024

\bibitem{He:1988dm}
X.~G.~He and R.~R.~Volkas,
%``Models Featuring Spontaneous {CP} Violation: An Invisible Axion and Light Neutrino Masses,''
Phys. Lett. B \textbf{208}, 261 (1988)
[erratum: Phys. Lett. B \textbf{218}, 508 (1989)]
doi:10.1016/0370-2693(88)90427-3
%30 citations counted in INSPIRE as of 24 Aug 2024

\bibitem{Batra:2023erw}
A.~Batra, H.~B.~C\^amara, F.~R.~Joaquim, R.~Srivastava and J.~W.~F.~Valle,
%``Axion Paradigm with Color-Mediated Neutrino Masses,''
Phys. Rev. Lett. \textbf{132}, no.5, 051801 (2024)
doi:10.1103/PhysRevLett.132.051801
[arXiv:2309.06473 [hep-ph]].
%1 citations counted in INSPIRE as of 24 Aug 2024

\bibitem{deAdelhartToorop:2011re}
R.~de Adelhart Toorop, F.~Feruglio and C.~Hagedorn,
%``Finite Modular Groups and Lepton Mixing,''
Nucl. Phys. B \textbf{858}, 437-467 (2012)
doi:10.1016/j.nuclphysb.2012.01.017
[arXiv:1112.1340 [hep-ph]].
%307 citations counted in INSPIRE as of 25 Aug 2024

\bibitem{Feruglio:2017spp}
F.~Feruglio,
%``Are neutrino masses modular forms?,''
doi:10.1142/9789813238053\_0012
[arXiv:1706.08749 [hep-ph]].
%259 citations counted in INSPIRE as of 25 Aug 2024


\bibitem{Ding:2023htn}
G.~J.~Ding and S.~F.~King,
%``Neutrino mass and mixing with modular symmetry,''
Rept. Prog. Phys. \textbf{87}, no.8, 084201 (2024)
doi:10.1088/1361-6633/ad52a3
[arXiv:2311.09282 [hep-ph]].
%38 citations counted in INSPIRE as of 25 Aug 2024

\bibitem{Kobayashi:2018vbk}
T.~Kobayashi, K.~Tanaka and T.~H.~Tatsuishi,
%``Neutrino mixing from finite modular groups,''
Phys. Rev. D \textbf{98}, no.1, 016004 (2018)
doi:10.1103/PhysRevD.98.016004
[arXiv:1803.10391 [hep-ph]].
%207 citations counted in INSPIRE as of 25 Aug 2024

\bibitem{Kobayashi:2018wkl}
T.~Kobayashi, Y.~Shimizu, K.~Takagi, M.~Tanimoto, T.~H.~Tatsuishi and H.~Uchida,
%``Finite modular subgroups for fermion mass matrices and baryon/lepton number violation,''
Phys. Lett. B \textbf{794}, 114-121 (2019)
doi:10.1016/j.physletb.2019.05.034
[arXiv:1812.11072 [hep-ph]].
%153 citations counted in INSPIRE as of 25 Aug 2024

\bibitem{Kobayashi:2019rzp}
T.~Kobayashi, Y.~Shimizu, K.~Takagi, M.~Tanimoto and T.~H.~Tatsuishi,
%``Modular $S_3$-invariant flavor model in SU(5) grand unified theory,''
PTEP \textbf{2020}, no.5, 053B05 (2020)
doi:10.1093/ptep/ptaa055
[arXiv:1906.10341 [hep-ph]].
%108 citations counted in INSPIRE as of 25 Aug 2024

\bibitem{Altarelli:2010gt}
G.~Altarelli and F.~Feruglio,
%``Discrete Flavor Symmetries and Models of Neutrino Mixing,''
Rev. Mod. Phys. \textbf{82}, 2701-2729 (2010)
doi:10.1103/RevModPhys.82.2701
[arXiv:1002.0211 [hep-ph]].
%1015 citations counted in INSPIRE as of 25 Aug 2024

\bibitem{Ishimori:2010au}
H.~Ishimori, T.~Kobayashi, H.~Ohki, Y.~Shimizu, H.~Okada and M.~Tanimoto,
%``Non-Abelian Discrete Symmetries in Particle Physics,''
Prog. Theor. Phys. Suppl. \textbf{183}, 1-163 (2010)
doi:10.1143/PTPS.183.1
[arXiv:1003.3552 [hep-th]].
%992 citations counted in INSPIRE as of 25 Aug 2024

\bibitem{Ishimori:2012zz}
H.~Ishimori, T.~Kobayashi, H.~Ohki, H.~Okada, Y.~Shimizu and M.~Tanimoto,
%``An introduction to non-Abelian discrete symmetries for particle physicists,''
Lect. Notes Phys. \textbf{858}, 1-227 (2012)
doi:10.1007/978-3-642-30805-5
%149 citations counted in INSPIRE as of 25 Aug 2024

\bibitem{Hernandez:2012ra}
D.~Hernandez and A.~Y.~Smirnov,
%``Lepton mixing and discrete symmetries,''
Phys. Rev. D \textbf{86}, 053014 (2012)
doi:10.1103/PhysRevD.86.053014
[arXiv:1204.0445 [hep-ph]].
%250 citations counted in INSPIRE as of 25 Aug 2024

\bibitem{King:2013eh}
S.~F.~King and C.~Luhn,
%``Neutrino Mass and Mixing with Discrete Symmetry,''
Rept. Prog. Phys. \textbf{76}, 056201 (2013)
doi:10.1088/0034-4885/76/5/056201
[arXiv:1301.1340 [hep-ph]].
%848 citations counted in INSPIRE as of 25 Aug 2024

\bibitem{Ding:2020zxw}
G.~J.~Ding, F.~Feruglio and X.~G.~Liu,
%``Automorphic Forms and Fermion Masses,''
JHEP \textbf{01}, 037 (2021)
doi:10.1007/JHEP01(2021)037
[arXiv:2010.07952 [hep-th]].
%77 citations counted in INSPIRE as of 14 Apr 2025



%\cite{Ding:2024inn}
\bibitem{Ding:2024inn}
G.~J.~Ding, J.~N.~Lu, S.~T.~Petcov and B.~Y.~Qu,
%``Non-holomorphic modular S$_{4}$ lepton flavour models,''
JHEP \textbf{01}, 191 (2025)
doi:10.1007/JHEP01(2025)191
[arXiv:2408.15988 [hep-ph]].
%8 citations counted in INSPIRE as of 31 Mar 2025

\bibitem{Qu:2024rns}
B.~Y.~Qu and G.~J.~Ding,
%``Non-holomorphic modular flavor symmetry,''
JHEP \textbf{08}, 136 (2024)
doi:10.1007/JHEP08(2024)136
[arXiv:2406.02527 [hep-ph]].
%20 citations counted in INSPIRE as of 14 Apr 2025

\bibitem{Criado:2018thu}
J.~C.~Criado and F.~Feruglio,
%``Modular Invariance Faces Precision Neutrino Data,''
SciPost Phys. \textbf{5}, no.5, 042 (2018)
doi:10.21468/SciPostPhys.5.5.042
[arXiv:1807.01125 [hep-ph]].
%209 citations counted in INSPIRE as of 14 Apr 2025


\bibitem{Novichkov:2019sqv}
P.~P.~Novichkov, J.~T.~Penedo, S.~T.~Petcov and A.~V.~Titov,
%``Generalised CP Symmetry in Modular-Invariant Models of Flavour,''
JHEP \textbf{07}, 165 (2019)
doi:10.1007/JHEP07(2019)165
[arXiv:1905.11970 [hep-ph]].
%168 citations counted in INSPIRE as of 25 Aug 2024

\bibitem{Aoki:2014cja}
M.~Aoki and T.~Toma,
%``Impact of semi-annihilation of $\mathbb{Z}_3$ symmetric dark matter with radiative neutrino masses,''
JCAP \textbf{09}, 016 (2014)
doi:10.1088/1475-7516/2014/09/016
[arXiv:1405.5870 [hep-ph]].
%70 citations counted in INSPIRE as of 25 Aug 2024


%\cite{Vagnozzi:2017ovm}
\bibitem{Vagnozzi:2017ovm}
S.~Vagnozzi, E.~Giusarma, O.~Mena, K.~Freese, M.~Gerbino, S.~Ho and M.~Lattanzi,
%``Unveiling $\nu$ secrets with cosmological data: neutrino masses and mass hierarchy,''
Phys. Rev. D \textbf{96} (2017) no.12, 123503
doi:10.1103/PhysRevD.96.123503
[arXiv:1701.08172 [astro-ph.CO]].
%373 citations counted in INSPIRE as of 20 Aug 2024

%\cite{DESI:2024mwx}
\bibitem{DESI:2024mwx}
A.~G.~Adame \textit{et al.} [DESI],
%``DESI 2024 VI: Cosmological Constraints from the Measurements of Baryon Acoustic Oscillations,''
[arXiv:2404.03002 [astro-ph.CO]].
%247 citations counted in INSPIRE as of 20 Aug 2024


%\cite{KamLAND-Zen:2016pfg}
\bibitem{KamLAND-Zen:2016pfg} 
  A.~Gando {\it et al.} [KamLAND-Zen Collaboration],
  %``Search for Majorana Neutrinos near the Inverted Mass Hierarchy Region with KamLAND-Zen,''
  Phys.\ Rev.\ Lett.\  {\bf 117}, no. 8, 082503 (2016)
  Addendum: [Phys.\ Rev.\ Lett.\  {\bf 117}, no. 10, 109903 (2016)]
  doi:10.1103/PhysRevLett.117.109903, 10.1103/PhysRevLett.117.082503
  [arXiv:1605.02889 [hep-ex]]; Phys. Rev. Lett. \textbf{130}, no.5, 051801 (2023)
doi:10.1103/PhysRevLett.130.051801
[arXiv:2203.02139 [hep-ex]].
  %%CITATION = doi:10.1103/PhysRevLett.117.109903, 10.1103/PhysRevLett.117.082503;%%
  %395 citations counted in INSPIRE as of 20 Jun 2019

%\cite{Esteban:2020cvm}
\bibitem{Esteban:2020cvm}
I.~Esteban, M.~C.~Gonzalez-Garcia, M.~Maltoni, T.~Schwetz and A.~Zhou,
%``The fate of hints: updated global analysis of three-flavor neutrino oscillations,''
JHEP \textbf{09}, 178 (2020)
doi:10.1007/JHEP09(2020)178
[arXiv:2007.14792 [hep-ph]].
%1460 citations counted in INSPIRE as of 05 May 2025


%\cite{Baek:2016kud}
\bibitem{Baek:2016kud} 
  S.~Baek, T.~Nomura and H.~Okada,
  %``An explanation of one-loop induced h → μτ decay,''
  Phys.\ Lett.\ B {\bf 759}, 91 (2016)
  doi:10.1016/j.physletb.2016.05.055
  [arXiv:1604.03738 [hep-ph]].
  %%CITATION = doi:10.1016/j.physletb.2016.05.055;%%
  %27 citations counted in INSPIRE as of 09 Jul 2019

\bibitem{Lindner:2016bgg} 
  M.~Lindner, M.~Platscher and F.~S.~Queiroz,
  %``A Call for New Physics : The Muon Anomalous Magnetic Moment and Lepton Flavor Violation,''
  Phys.\ Rept.\  {\bf 731}, 1 (2018)
  doi:10.1016/j.physrep.2017.12.001
  [arXiv:1610.06587 [hep-ph]].
  %%CITATION = doi:10.1016/j.physrep.2017.12.001;%%
  %141 citations counted in INSPIRE as of 17 Jul 2019


%\cite{TheMEG:2016wtm}
\bibitem{TheMEG:2016wtm} 
  A.~M.~Baldini {\it et al.} [MEG Collaboration],
  %``Search for the lepton flavour violating decay $\mu ^+ \rightarrow \mathrm {e}^+ \gamma $ with the full dataset of the MEG experiment,''
  Eur.\ Phys.\ J.\ C {\bf 76}, no. 8, 434 (2016)
%  doi:10.1140/epjc/s10052-016-4271-x
  [arXiv:1605.05081 [hep-ex]].
  %%CITATION = doi:10.1140/epjc/s10052-016-4271-x;%%
  %124 citations counted in INSPIRE as of 02 Oct 2017


  %\cite{Renga:2018fpd}
\bibitem{Renga:2018fpd} 
  F.~Renga [MEG Collaboration],
  %``The quest for $\mu \to e \gamma$: present and future,''
  Hyperfine Interact.\  {\bf 239}, no. 1, 58 (2018)
%  doi:10.1007/s10751-018-1534-y
  [arXiv:1811.05921 [hep-ex]].
  %%CITATION = doi:10.1007/s10751-018-1534-y;%%
  %1 citations counted in INSPIRE as of 31 Jan 2019

  
  %\cite{Aubert:2009ag}
\bibitem{Aubert:2009ag}
  B.~Aubert {\it et al.} [BaBar Collaboration],
  %``Searches for Lepton Flavor Violation in the Decays tau+- ---> e+- gamma and tau+- ---> mu+- gamma,''
  Phys.\ Rev.\ Lett.\  {\bf 104} (2010) 021802
%  doi:10.1103/PhysRevLett.104.021802
  [arXiv:0908.2381 [hep-ex]].
  %%CITATION = doi:10.1103/PhysRevLett.104.021802;%%
  %333 citations counted in INSPIRE as of 02 Oct 2017

\bibitem{GrillidiCortona:2015jxo}
G.~Grilli di Cortona, E.~Hardy, J.~Pardo Vega and G.~Villadoro,
%``The QCD axion, precisely,''
JHEP \textbf{01}, 034 (2016)
doi:10.1007/JHEP01(2016)034
[arXiv:1511.02867 [hep-ph]].
%699 citations counted in INSPIRE as of 11 Aug 2024

\bibitem{Visinelli:2009zm}
L.~Visinelli and P.~Gondolo,
%``Dark Matter Axions Revisited,''
Phys. Rev. D \textbf{80}, 035024 (2009)
doi:10.1103/PhysRevD.80.035024
[arXiv:0903.4377 [astro-ph.CO]].
%273 citations counted in INSPIRE as of 11 Jun 2025

%\cite{Kawasaki:2014sqa}
\bibitem{Kawasaki:2014sqa}
M.~Kawasaki, K.~Saikawa and T.~Sekiguchi,
%``Axion dark matter from topological defects,''
Phys. Rev. D \textbf{91} (2015) no.6, 065014
doi:10.1103/PhysRevD.91.065014
[arXiv:1412.0789 [hep-ph]].
%381 citations counted in INSPIRE as of 03 Jul 2025]



\bibitem{deSalas:2015glj}
P.~F.~de Salas, M.~Lattanzi, G.~Mangano, G.~Miele, S.~Pastor and O.~Pisanti,
%``Bounds on very low reheating scenarios after Planck,''
Phys. Rev. D \textbf{92}, no.12, 123534 (2015)
doi:10.1103/PhysRevD.92.123534
[arXiv:1511.00672 [astro-ph.CO]].
%255 citations counted in INSPIRE as of 12 Aug 2024

\bibitem{Beltran:2006sq}
M.~Beltran, J.~Garcia-Bellido and J.~Lesgourgues,
%``Isocurvature bounds on axions revisited,''
Phys. Rev. D \textbf{75}, 103507 (2007)
doi:10.1103/PhysRevD.75.103507
[arXiv:hep-ph/0606107 [hep-ph]].
%144 citations counted in INSPIRE as of 12 Aug 2024

\bibitem{DiLuzio:2016sbl}
L.~Di Luzio, F.~Mescia and E.~Nardi,
%``Redefining the Axion Window,''
Phys. Rev. Lett. \textbf{118}, no.3, 031801 (2017)
doi:10.1103/PhysRevLett.118.031801
[arXiv:1610.07593 [hep-ph]].
%211 citations counted in INSPIRE as of 12 Aug 2024

\bibitem{Borsanyi:2016ksw}
S.~Borsanyi, Z.~Fodor, J.~Guenther, K.~H.~Kampert, S.~D.~Katz, T.~Kawanai, T.~G.~Kovacs, S.~W.~Mages, A.~Pasztor and F.~Pittler, \textit{et al.}
%``Calculation of the axion mass based on high-temperature lattice quantum chromodynamics,''
Nature \textbf{539}, no.7627, 69-71 (2016)
doi:10.1038/nature20115
[arXiv:1606.07494 [hep-lat]].
%694 citations counted in INSPIRE as of 12 Aug 2024

\bibitem{Co:2019jts}
R.~T.~Co, L.~J.~Hall and K.~Harigaya,
%``Axion Kinetic Misalignment Mechanism,''
Phys. Rev. Lett. \textbf{124}, no.25, 251802 (2020)
doi:10.1103/PhysRevLett.124.251802
[arXiv:1910.14152 [hep-ph]].
%181 citations counted in INSPIRE as of 25 Aug 2024

\bibitem{Co:2020dya}
R.~T.~Co, L.~J.~Hall, K.~Harigaya, K.~A.~Olive and S.~Verner,
%``Axion Kinetic Misalignment and Parametric Resonance from Inflation,''
JCAP \textbf{08}, 036 (2020)
doi:10.1088/1475-7516/2020/08/036
[arXiv:2004.00629 [hep-ph]].
%73 citations counted in INSPIRE as of 26 Aug 2024

\bibitem{Armengaud:2014gea}
E.~Armengaud, F.~T.~Avignone, M.~Betz, P.~Brax, P.~Brun, G.~Cantatore, J.~M.~Carmona, G.~P.~Carosi, F.~Caspers and S.~Caspi, \textit{et al.}
%``Conceptual Design of the International Axion Observatory (IAXO),''
JINST \textbf{9}, T05002 (2014)
doi:10.1088/1748-0221/9/05/T05002
[arXiv:1401.3233 [physics.ins-det]].
%361 citations counted in INSPIRE as of 26 Aug 2024

\bibitem{Stern:2016bbw}
I.~Stern,
%``ADMX Status,''
PoS \textbf{ICHEP2016}, 198 (2016)
doi:10.22323/1.282.0198
[arXiv:1612.08296 [physics.ins-det]].
%89 citations counted in INSPIRE as of 26 Aug 2024

\bibitem{Beurthey:2020yuq}
S.~Beurthey, N.~B\"ohmer, P.~Brun, A.~Caldwell, L.~Chevalier, C.~Diaconu, G.~Dvali, P.~Freire, E.~Garutti and C.~Gooch, \textit{et al.}
%``MADMAX Status Report,''
[arXiv:2003.10894 [physics.ins-det]].
%77 citations counted in INSPIRE as of 26 Aug 2024

\bibitem{Semertzidis:2019gkj}
Y.~K.~Semertzidis, J.~E.~Kim, S.~Youn, J.~Choi, W.~Chung, S.~Haciomeroglu, D.~Kim, J.~Kim, B.~Ko and O.~Kwon, \textit{et al.}
%``Axion Dark Matter Research with IBS/CAPP,''
[arXiv:1910.11591 [physics.ins-det]].
%37 citations counted in INSPIRE as of 26 Aug 

\bibitem{OHare:2024nmr}
C.~A.~J.~O'Hare,
%``Cosmology of axion dark matter,''
PoS \textbf{COSMICWISPers}, 040 (2024)
doi:10.22323/1.454.0040
[arXiv:2403.17697 [hep-ph]].
%98 citations counted in INSPIRE as of 11 Jun 2025

\bibitem{Foster:2022fxn}
J.~W.~Foster, S.~J.~Witte, M.~Lawson, T.~Linden, V.~Gajjar, C.~Weniger and B.~R.~Safdi,
%``Extraterrestrial Axion Search with the Breakthrough Listen Galactic Center Survey,''
Phys. Rev. Lett. \textbf{129}, no.25, 251102 (2022)
doi:10.1103/PhysRevLett.129.251102
[arXiv:2202.08274 [astro-ph.CO]].
%70 citations counted in INSPIRE as of 11 Jun 2025

\bibitem{Chan:2021gjl}
M.~H.~Chan,
%``Constraining the axion\textendash{}photon coupling using radio data of the Bullet cluster,''
Sci. Rep. \textbf{11}, no.1, 20087 (2021)
doi:10.1038/s41598-021-99495-3
[arXiv:2109.11734 [astro-ph.CO]].
%18 citations counted in INSPIRE as of 10 Jun 2025

\bibitem{Ayala:2014pea}
A.~Ayala, I.~Dom\'\i{}nguez, M.~Giannotti, A.~Mirizzi and O.~Straniero,
%``Revisiting the bound on axion-photon coupling from Globular Clusters,''
Phys. Rev. Lett. \textbf{113}, no.19, 191302 (2014)
doi:10.1103/PhysRevLett.113.191302
[arXiv:1406.6053 [astro-ph.SR]].
%522 citations counted in INSPIRE as of 10 Jun 2025


\bibitem{CAST:2017uph}
V.~Anastassopoulos \textit{et al.} [CAST],
%``New CAST Limit on the Axion-Photon Interaction,''
Nature Phys. \textbf{13}, 584-590 (2017)
doi:10.1038/nphys4109
[arXiv:1705.02290 [hep-ex]].
%1043 citations counted in INSPIRE as of 11 Jun 2025

\bibitem{Manzari:2024jns}
C.~A.~Manzari, Y.~Park, B.~R.~Safdi and I.~Savoray,
%``Supernova Axions Convert to Gamma Rays in Magnetic Fields of Progenitor Stars,''
Phys. Rev. Lett. \textbf{133}, no.21, 211002 (2024)
doi:10.1103/PhysRevLett.133.211002
[arXiv:2405.19393 [hep-ph]].
%25 citations counted in INSPIRE as of 11 Jun 2025

\end{thebibliography}
\end{document}